\documentclass[12pt]{article}
\pdfoutput=1
\usepackage{jheppub}
\usepackage{amsmath}
\usepackage{bm}
\usepackage{slashed}
\usepackage{graphicx}

\usepackage{cancel}

\newlength{\dummysp}
\newcommand{\beq}{\begin{eqnarray}}
\newcommand{\eeq}{\end{eqnarray}}

\newcommand{\gappeq}{\mathrel{\rlap {\raise.5ex\hbox{$>$}}
{\lower.5ex\hbox{$\sim$}}}}
\newcommand{\lappeq}{\mathrel{\rlap{\raise.5ex\hbox{$<$}}
{\lower.5ex\hbox{$\sim$}}}}

\newcommand{\ben}{\begin{enumerate}}
\newcommand{\een}{\end{enumerate}}

\newcommand{\bit}{\begin{itemize}}
\newcommand{\eit}{\end{itemize}}

\def\[{\left [}
\def\]{\right ]}
\def\({\left (}
\def\){\right )}

\def\Z{{\mathbb Z}}

\title{Multi-fractional instantons in $SU(N)$ Yang-Mills theory on the twisted $\mathbb T^4$}

 \author[a]{Mohamed M. Anber,}\author[b]{Erich Poppitz} 
  
\affiliation[a]{Centre for Particle Theory, Department of Mathematical Sciences, Durham University, South Road, Durham DH1 3LE, UK}
\affiliation[b]{Department of Physics,   University of Toronto, 60 St George St., 
Toronto, ON M5S 1A7, Canada}
\emailAdd{mohamed.anber@durham.ac.uk}\emailAdd{poppitz@physics.utoronto.ca}   
   
\abstract{
We construct analytical self-dual Yang-Mills fractional instanton solutions on a four-torus $\mathbb{T}^4$ with 't Hooft twisted boundary conditions. These instantons possess topological charge $Q=\frac{r}{N}$, where $1\leq r< N$. To implement the twist, we employ $SU(N)$ transition functions that satisfy periodicity conditions up to center elements and are embedded into $SU(k)\times SU(\ell)\times U(1)\subset SU(N)$, where $\ell+k=N$. The self-duality requirement imposes a condition, $k L_1L_2=r\ell L_3L_4$, on the lengths of the periods of $\mathbb{T}^4$ and yields solutions with abelian field strengths. However, by introducing a detuning parameter $\Delta\equiv (r\ell L_3L_4-k L_1 L_2)/\sqrt{L_1 L_2L_3L_4}$, we generate self-dual nonabelian solutions on a general $\mathbb{T}^4$ as an expansion in powers of $\Delta$. We explore the moduli spaces associated with these solutions and find that they exhibit intricate structures. Solutions with topological charges greater than $\frac{1}{N}$ and $k\neq r $ possess non-compact moduli spaces, along which the  ${\cal O}(\Delta)$ gauge-invariant densities exhibit runaway behavior. On the other hand, solutions with $Q=\frac{r}{N}$ and $k=r$ have compact moduli spaces, whose coordinates correspond to the allowed holonomies in the $SU(r)$ color space. These solutions can be represented as a sum over $r$ lumps centered around the $r$ distinct holonomies, thus resembling a liquid of instantons. In addition, we show that each lump supports $2$ adjoint fermion zero modes.}

\begin{document}

\maketitle

\flushbottom

%%%%%%%%%%%%%%%%%%%%%%%%%%%%
\section{Introduction, summary, and outlook}
%%%%%%%%%%%%%%%%%%%%%%%%%%%%
 
Instantons are prominent in studying many nonperturbative phenomena in Yang-Mills theory, including the vacuum structure, condensates, and confinement. One of the least-explored instantons are {\em't Hooft fluxes} of $SU(N)$ gauge theory on the $4$-torus $\mathbb T^4$ with twisted boundary conditions \cite{tHooft:1981nnx}.  Such solutions, found by 't Hooft, carry fractional topological charges and have constant abelian field strength. While the field strength is abelian, for a general number of colors $N$, the boundary conditions on $\mathbb T^4$  are implemented via non-abelian transition functions (i.e.~there exists no gauge where all transition functions commute).

Although 't Hooft's solutions have been known since the 1980s,  relatively little attention has been devoted to their study since  \cite{vanBaal:1984ar}. The notable exception is the work of the Madrid group over many years, reviewed in  \cite{Gonzalez-Arroyo:2023kqv}. The recent development of generalized global symmetries \cite{Gaiotto:2014kfa} resurrected the interest in this subject. It was shown in \cite{Gaiotto:2017yup} that introducing background fields for the $1$-form $\mathbb Z_N^{(1)}$ center symmetry of Yang-Mills theory can lead to new 't Hooft anomalies, restricting the symmetry realizations and thus the infrared dynamics. 

The gauge field of the $1$-form symmetry is a $2$-form field whose nonvanishing holonomies implement the 't Hooft twist of the boundary conditions on $\mathbb T^4$. The fractional $2$-form flux is merely an external field that imposes kinematical constraints. On the other hand, finding the field configurations which minimize the action (or energy) in the presence of twists requires dynamical considerations. Recently, the authors questioned the role instantons in the presence of twists could play in determining the dynamics of the theory \cite{Anber:2022qsz}. In particular, we examined the gaugino condensate in $SU(2)$ super Yang-Mills theory with twists on $\mathbb T^4$. The fractional topological charge $Q=\frac{1}{2}$ of the $SU(2)$ solution supports two gaugino zero modes and yields a non-vanishing condensate, which was found to be independent of the torus size. The computations were carried within the limit of the small-torus size, taken to be much smaller than the inverse strong scale, so we remained in the semi-classical domain. Thus, we could perform reliable computations and, thanks to supersymmetry, extract the numerical coefficient of the condensate.  However, our computations gave twice the condensate's numerical value on $\mathbb R^4$. Thus, our results warrant further examination of the situation for $SU(2)$ and for a general number of colors. 

The current work is a continuation of the efforts in this direction. One of the crucial conditions for studying the dynamics is the self-duality of the fractional instantons. A non-self dual solution is not a minimum of the action; it has negative fluctuation modes and hence, is unstable. Insisting on the abelian solutions found by 't Hooft \cite{tHooft:1981nnx}, the ratio between the periods of $\mathbb T^4$ needs to satisfy a specific condition to respect the self-duality of the solutions. We call such $\mathbb T^4$ a self-dual torus. However, in \cite{Anber:2022qsz}, it was found that instantons on the self-dual torus support extra fermion zero modes, more than needed to support the bilinear gaugino condensate.

A way to lift the extra zero modes is to deform the self-dual $\mathbb T^4$. The price to pay, insisting on the self-duality of the instantons, is to depart from the simple abelian solutions found by 't Hooft. 
One is then faced with the fact that a non-abelian analytical solution on a generic $\mathbb T^4$ with general 't Hooft twists is not currently known. Furthermore, even a description of the moduli space and of its metric\footnote{These data alone  suffice to perform certain instanton computations in supersymmetric theories.} of such self-dual solutions is not available.
Fortunately, the authors of \cite{GarciaPerez:2000aiw}  developed a systematic approach to obtaining approximate $SU(2)$ nonabelian self-dual solutions as expansion in a small parameter $\Delta$, measuring the deviation from the self-dual torus.\footnote{This is the solution used in \cite{Anber:2022qsz}, which, at $\Delta >0$, supports exactly two zero modes needed to give rise to the bilinear condensate.}
The approach in \cite{GarciaPerez:2000aiw}  was generalized in \cite{Gonzalez-Arroyo:2019wpu} to the case of $SU(N)$. Nevertheless, it was only used to obtain solutions with minimal topological charge $Q=\frac{1}{N}$.

 In this paper, we carry out a systematic analysis to obtain self-dual solutions with generic topological charge $Q=\frac{r}{N}$, with integer $N>r>1$, on a non-self dual torus. The main effort of the present work is directed at exploring the structure of the bosonic moduli space of the solutions as well as the fermion zero modes in these backgrounds. 

\bigskip

{\flushleft{ \bf Summary.} }
The main findings of this rather technical paper are described below:

\smallskip

{W}e let $L_{1},L_{2},L_{3},L_{4}$ be the lengths of the periods of $\mathbb T^4$. Following 't Hooft  \cite{tHooft:1981nnx}, we embed the $SU(N)$ transition functions and gauge fields in $SU(k)\times SU(\ell)\times U(1)\subset SU(N)$, such that $k$ and $\ell$ are positive integers and $k+\ell=N$. We choose the transition functions to give rise to  't Hooft twists on $\mathbb T^4$ corresponding to topological charge $Q=\frac{r}{N}$ (Section \ref{sec:reviewsolution}).  
 Even though the transition functions are fully non-abelian,  the original 't Hooft solution with topological charge $Q=\frac{r}{N}$ has only an abelian gauge field $A_\mu$ along the $U(1)$ generator.\footnote{See   Section \ref{sec:reviewsolution1}: the $Q=\frac{r}{N}$ transition functions are in (\ref{the set of transition functions for Q equal r over N, general solution}) and the abelian solution is in (\ref{r over N abelian sol}).}
 The self-duality of this solution imposes the condition $k L_1L_2= r\ell L_3L_4$. As already mentioned, a $\mathbb T^4$ that satisfies this condition is said to be self-dual.
 
\begin{figure}[t] %  figure placement: here, top, bottom, or page
   \centering
   \includegraphics[width=5in]{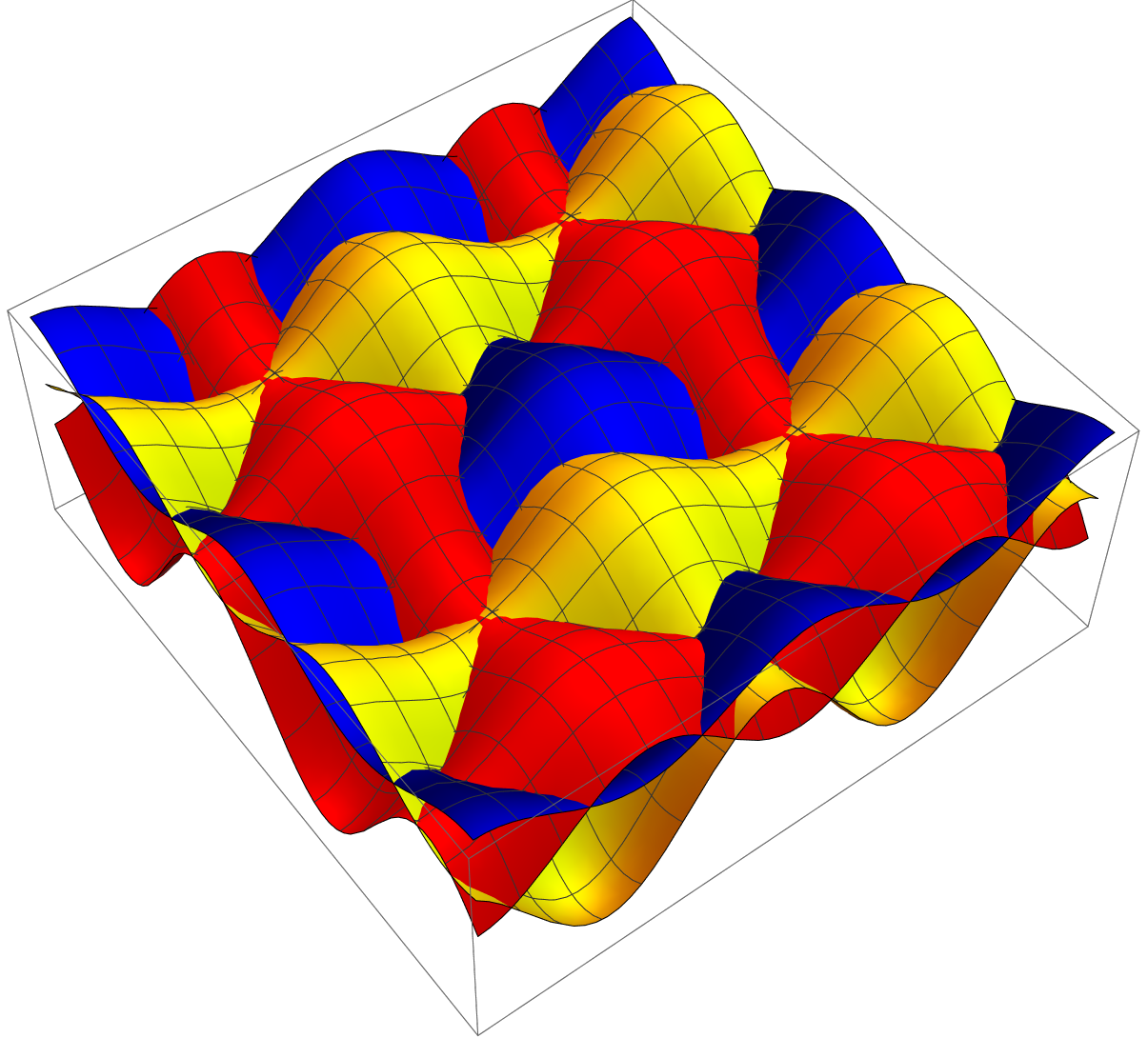} 
   \caption{A $3D$ plot of the profile given by Eq. (\ref{fractional1}), with $r=3$, as a function of $(x_1,x_2)$, for fixed $(x_3,x_4)$. For better visualization, we show double the periods in $x_1$ and $x_2$. We see three solutions, in red, yellow, and blue,  lumped around three distinct centers.   These lumps, however, are not well-separated, comprising a liquid rather than a dilute gas. Earlier \cite{Gonzalez-Arroyo:1996eos},  similar configurations were constructed numerically and used to study confinement, see \cite{Gonzalez-Arroyo:2023kqv}. }
   \label{visual of liquid}
\end{figure}

 Next, we define a {\em detuning parameter} $\Delta$, that measures the deviation from the self-dual $\mathbb T^4$, as $\Delta\equiv (r\ell L_3L_4-k L_1L_2)/\sqrt{L_1L_2L_3L_4}$. Then, the self-dual non-abelian solution is obtained as an expansion in $\Delta$, similar to \cite{GarciaPerez:2000aiw,Gonzalez-Arroyo:2019wpu}. The solution now has nontrivial components along the abelian $U(1)$ generator as well as the nonabelian subgroups $SU(k)\times SU(\ell)$. We carry out our analysis to the leading order in $\Delta$, from which we observe the following:
 \begin{enumerate}
 
 \item  To the leading order in $\Delta$, the solution of the self-dual Yang-Mills equations is in one-to-one correspondence with the solution to the Dirac equation of the gaugino zero modes on the self-dual $\mathbb T^4$ (Section \ref{sec:fermionzeromodes}). Thus, one can borrow the latter's solutions and show that they satisfy the self-dual Yang-Mills equations to the leading order (Section \ref{sec:deforming}). 
 
 \item Among all solutions with $Q = \frac{r}{N}$, the ones with  $k=r$ stand out.  For this case, we find $4r$ arbitrary physical parameters that label the self-dual Yang-Mills solutions, in accordance with the index theorem. We interpret these parameters as the coordinates on the compact moduli space: these are the $r \; (=k)$ holonomies in the $SU(k)$ color space in each of the $4$ spacetime directions (Section \ref{sec:compactvsnoncompact}).
 
\item In addition, we find that gauge-invariant densities for the $k=r$ solutions can be cast into the form of a sum over $r$ identical lumps centered about the values taken by the $r \; (=k)$ different holonomies.
This indicates that a solution with topological charge $Q=\frac{r}{N}$ can be thought of as composed of $r$  ``elementary,''  yet strongly overlapping  ones---thus, resembling a liquid, rather than a dilute gas \cite{Gonzalez-Arroyo:2023kqv} (Section \ref{sec:fractionalbose}). See Figure \ref{visual of liquid} for an illustration. 

Further support for this interpretation follows from solving the Dirac equation in the background of the full non-abelian solution, showing that $2$ fermion zero modes are centered about each of the $r$ holonomies, giving a total of $2r$ fermion zero modes as required by the index theorem (Section \ref{sec:fractionalfermion}).  
 
 \item We also study the  $\Delta$-expansion around the other $Q = \frac{r}{N}$ solutions, the ones with $k \ne r$  (Section \ref{sec:compactvsnoncompact}). Here, we find that the moduli space becomes non-compact. To further understand the significance of this finding, we show that gauge-invariant local densities grow without limit in the noncompact moduli directions, clashing with the spirit of the $\Delta$ expansion for $k \ne r$ (Section \ref{sec:compactvsnoncompact} and Appendix \ref{appx:blowup}).  
 This blow-up leads us to conjecture that the only self-dual $Q=\frac{r}{N}$ solutions, obtained via the $\Delta$-expansion,  are the ones with $k=r$.
 \end{enumerate}

%%%%%%%%%%%%%%%%%%%%%%%%%%%%%%%%%%%%%%%%%%%%%%%%%%%%%%%%%%%%%%%%%%%%%%
%%%%%%%%%%%%%%%%%%%%%%%%%%%%%%%%%%%%%%%%%%%%%%%%%%%%%%%%%%%%%%%%%%%%%%%%%%%%

\smallskip

{\flushleft{ \bf Outlook.}}  There are  several directions where this work can be applied to or extended:

\smallskip 
 
  The study of the present paper sets the stage for a forthcoming paper to shed light on a few dynamical and kinematical aspects of supersymmetric and non-supersymmetric $SU(N)$ gauge theories. This includes the higher-order condensates, cluster decomposition principle, and exactness/holomorphy of supersymmetric results. 
 
We have yet to achieve a deeper understanding of the apparent failure of the $\Delta$ expansion for $k \ne r$ that we observed in the leading order. This may be require better control of the higher orders in the  $\Delta$-expansion.
Numerical studies of instantons on the twisted torus can also be used to study the convergence of the expansion as well as the approach to various large volume limits.

%%%%%%%%%%%%%%%%%%%%%%%%%%%%%%%%%%%%%%%%%%%%%%%%%%%%%%%%
\section{Review of 't Hooft's constant-flux solutions on $\mathbb T^4$}
\label{sec:reviewsolution}
%%%%%%%%%%%%%%%%%%%%%%%%%%%%%%%%%%%%%%%%%%%%%%%%%%%%%%%%
This section quickly reviews $SU(N)$ 't Hooft twisted solution on the four-torus $\mathbb T^4$.  We take the torus to have periods of length $L_\mu$, $\mu=1,2,3,4$, where  $\mu,\nu$ runs over the spacetime dimensions. The gauge fields $A_\mu$ are Hermitian traceless $N\times N$ matrices,  and taken to obey the boundary conditions
\begin{eqnarray}
A_\nu(x+L_\mu \hat e_\mu)=\Omega_\mu(x) A_\nu(x) \Omega_\mu^{-1}(x)-i \Omega_\mu(x) \partial_\nu \Omega_\mu^{-1}(x)\,,
\label{conditions on gauge field}
\end{eqnarray}
upon traversing $\mathbb T^4$ in each direction. The transition functions $\Omega_\mu$ are $N\times N$ unitary matrices, and  $\hat e_\nu$ are unit vectors in the $x_\nu$ direction. The subscript $\mu$ in $\Omega_\mu$ means that the function $\Omega_\mu$ does not depend on the coordinate $x_\mu$. The boundary condition (\ref{conditions on gauge field}) means that the gauge fields $A_\mu$ are periodic up to a gauge transformation. Let us for the moment use the short-hand-notation $[\Omega_\mu] A_\nu$ to denote $\Omega_\mu A_\nu \Omega_\mu^{-1}-i \Omega_\mu \partial_\nu \Omega_\mu^{-1}$. Then, the compatibility of (\ref{conditions on gauge field}) at the corners of the $x_\mu-x_\nu$ plane of $\mathbb T^4$ gives: 
\begin{eqnarray}
\nonumber
A_\lambda(x+L_\mu \hat e_\mu+L_\nu \hat e_\nu)&=&[\Omega_\mu(x+L_\nu \hat e_\nu)][\Omega_\nu(x+L_\mu \hat e_\mu)]A_\lambda(x)\\
&=&[\Omega_\nu(x+L_\mu \hat e_\mu)][\Omega_\mu(x+L_\nu \hat e_\nu)]A_\lambda(x)\,,
\label{compatibility on the corners}
\end{eqnarray}
from which we obtain the periodicity conditions on the transition functions $\Omega_\mu$ (now giving up the short-hand notation and going back to the original $\Omega_\mu$ that appears in (\ref{conditions on gauge field}))
\begin{eqnarray}\label{cocycle}
\Omega_\mu (x + \hat{e}_\nu L_\nu) \; \Omega_\nu (x) = e^{i {2 \pi n_{\mu\nu} \over N}} \Omega_\nu (x+ \hat{e}_\mu L_\mu) \; \Omega_\mu (x)~.
\end{eqnarray}
Equation (\ref{cocycle}) is the cocycle conditions on the transition functions $\Omega_\mu$.  The exponent $e^{i {2 \pi n_{\mu\nu} \over N}}$, with integers $n_{\mu\nu}=-n_{\nu\mu}$, is the $\mathbb Z_N$ center of $SU(N)$. The freedom to introduce the center stems from the fact that both the transition function and its inverse appear in (\ref{conditions on gauge field}).

't Hooft found a solution to the consistency conditions (\ref{cocycle}) carrying a fractional topological charge by embedding the $SU(N)$ transition functions $\Omega_\mu(x)$ in $SU(k)\times SU(\ell)\times U(1)\subset SU(N)$, such that $N=k+\ell$ and  writing them in the form
\begin{eqnarray}\label{transitionfunctions}
\Omega_\mu(x)=P_k^{s_\mu}Q_k^{t_\mu}\otimes P_\ell^{u_\mu}Q_\ell^{v_\mu}\; e^{i\omega \frac{\alpha_{\mu\lambda} x_{\lambda}}{L_\lambda}}\, .
\label{transition functions}
\end{eqnarray} 
Here, $s_\mu,t_\mu,u_\mu,v_\mu$ are integers,   a sum over $\lambda$ is implied in the exponent, and $\alpha_{\mu\lambda}$ is a real matrix with vanishing diagonal components without any (anti-)symmetry properties. The matrices $P_k$ and $Q_k$ (similarly the matrices $P_\ell$ and $Q_\ell$) are the $k\times k$ (similarly $\ell\times\ell$) shift and clock matrices:
\begin{eqnarray}\label{pandq}
P_k=\gamma_k\left[\begin{array}{cccc} 0& 1&0&...\\ 0&0&1&...\\... \\ ...&  &0&1 \\1&0&...&0\end{array}\right]\,,\quad Q_k=\gamma_k\; \mbox{diag}\left[1, e^{\frac{i 2\pi}{k}}, e^{2\frac{i 2\pi}{k}},...\right]\,,
\end{eqnarray}
which satisfy the relation $P_kQ_k=e^{i\frac{2\pi}{k}}Q_kP_k$. The factor $\gamma_k\equiv e^{\frac{i\pi (1-k)}{k}}$ ensures that $\det Q_k=1$ and $\det P_k=1$.

 In the rest of this paper, we take primed upper-case Latin letters to denote elements of $k\times k$ matrices: $C', D'=1,2,...,k$, and the unprimed upper-case Latin letters to denote $\ell\times \ell$ matrices: $C,D=1,2,..,\ell$.  The matrices $P_k$ and $Q_k$ can then be written as $(P_k)_{B'C'}=\delta_{B',C'-1 \; (\text{mod} k)}$ and $(Q_k)_{C'B'}=\gamma_k \; e^{i2\pi \frac{C'-1}{k}}\delta_{C'B'}$. The matrix $\omega$ is the $U(1)$ generator. It is  given by
\begin{eqnarray}\label{omega}
\omega=2\pi\mbox{diag}\left[\underbrace{\ell, \ell,...,\ell}_{k\, \mbox{times}},\underbrace{ -k,-k,...,-k}_{\ell\,\mbox{times}}\right]\,,
\end{eqnarray}
and clearly commutes with $P_k,P_\ell, Q_k,Q_\ell$. 

Writing the twist matrix $n_{\mu\nu}$  appearing in the cocycle condition (\ref{cocycle}) as $n_{\mu\nu}=n_{\mu\nu}^{(1)}+n_{\mu\nu}^{(2)}$, the antisymmetric part of the coefficients $\alpha_{\mu\nu}$ are taken to be
\begin{eqnarray}\label{alphaandn}
\alpha_{\mu\nu}-\alpha_{\nu\mu}=\frac{n_{\mu\nu}^{(2)}}{N\ell}-\frac{n_{\mu\nu}^{(1)}}{Nk}\,.
\end{eqnarray}
Recall that $\alpha_{\mu\nu}$  have vanishing diagonal elements; it is convenient, see Section \ref{sec:reviewsolution1}, to choose a particular form for their symmetric part, which amounts to a gauge choice.

A  solution of the transition functions (\ref{transitionfunctions}) obeying the cocycle conditions (\ref{cocycle})  with $\alpha_{\mu\nu}$ and $n_{\mu\nu}$  related as in (\ref{alphaandn}) can be obtained provided that $s_\mu,t_\mu,u_\mu,v_\mu\in \mathbb Z$ can be found such that
\begin{eqnarray}\label{ens1}
n_{\mu\nu}^{(1)}=s_\mu t_\nu-s_\nu t_\mu +k A_{\mu\nu}\,,\quad n_{\mu\nu}^{(2)}=u_\mu v_\nu-v_\nu u_\mu +\ell  B_{\mu\nu}\,,
\end{eqnarray} 
where $A_{\mu\nu}$ and $B_{\mu\nu}$ are integers, and
\begin{eqnarray}\label{ens2}
n_{\mu\nu}^{(1)}\tilde n_{\mu\nu}^{(1)}=0 \; (\mbox{mod}\, k)\,,\quad n_{\mu\nu}^{(2)}\tilde n_{\mu\nu}^{(2)}=0 \; (\mbox{mod}\, \ell)\,, 
\end{eqnarray}
and $\tilde n_{\mu\nu}=\frac{1}{2}\epsilon_{\mu\nu\alpha\beta}n_{\alpha\beta}$.

While the details of the derivation are not shown here (see \cite{tHooft:1981nnx}), the data we have given above suffice to check that upon plugging (\ref{ens2}, \ref{ens1},  \ref{alphaandn}) into (\ref{transitionfunctions}) one finds, using (\ref{omega}) and (\ref{pandq}), that the cocycle conditions (\ref{cocycle}) are obeyed, with twist matrices $n_{\mu\nu}  = n_{\mu\nu}^{(1)} + n_{\mu\nu}^{(2)}$.

An abelian gauge field configuration along the $U(1)$ generator $\omega$, which obeys the boundary conditions specified by the $\Omega_\mu$ thus constructed,  is given by the expression
\begin{eqnarray}
A_\lambda=-\omega\left( \frac{\alpha_{\mu\lambda}x_\mu}{L_\mu L_\lambda}+ \frac{z_\lambda}{L_\lambda}\right)\,.
\end{eqnarray}
The corresponding field strength $F_{\mu\nu}=\partial_\mu A_\nu-\partial_\nu A_\mu +i[A_\mu, A_\nu]$ is constant everywhere on $\mathbb T^4$:
\begin{eqnarray}
F_{\mu\nu}=-\omega \frac{\alpha_{\mu\nu}-\alpha_{\nu\mu}}{L_\mu L_\lambda}\,.
\end{eqnarray}
 The constants $z_\mu$ label the holonomies along the $U(1)$ generator, which are translational moduli. This solution carries a fractional topological charge:
\begin{eqnarray} \label{Q of n}
Q=-\frac{1}{4N}n_{\mu\nu}\tilde n_{\mu\nu}=-\frac{n_{12}n_{34}+n_{13}n_{42}+n_{14}n_{23}}{N}\,.
\end{eqnarray}
Without loss of generality, we can always assume $n_{13}=n_{42}=n_{14}=n_{23}=0$. Thus, we only consider fluxes in the 1-2 and 3-4 planes. Then, a self-dual solution satisfies the relation $F_{12}=F_{34}$, from which one can find the ratio $\frac{L_1L_2}{L_3L_4}$ that defines the self-dual torus.  The action of the self-dual solution is
\begin{eqnarray}
S_{0}=\frac{1}{2g^2}\int_{\mathbb T^4}\mbox{tr}\left[F_{\mu\nu}F_{\mu\nu}\right]=\frac{8\pi^2|Q|}{g^2}\,.
\end{eqnarray}

%%%%%%%%%%%%%%%%%%%%%%%%%%%%%%%%%%%

%
\section{Fermion zero modes in the $Q = {r \over N}$ constant-flux background}
\label{sec:fermionzeromodes}

In this Section, we study the zero modes of the adjoint fermions in the constant-flux abelian background with topological charge $r\over N$, described in Section \ref{sec:reviewsolution1} (see eqn.~(\ref{r over N abelian sol})). These results are useful when constructing the nonabelian self-dual solution with $Q = {r\over N}$ on the deformed $\mathbb T^4$. 

We find that there are  $2 {\rm gcd}(k,r)$ dotted (Section \ref{sec:dottedzeromodes}) and $2 {\rm gcd}(k,r)$ undotted (Section \ref{sec:undotteddiagonal}) 
{\it constant} fermion zero modes. We also find $2 r$ undotted adjoint fermion zero modes with nontrivial $x$-dependence (Section \ref{sec:offdiagonalzeromodes}, see eqns.~(\ref{lambdazeromode1}--\ref{form of Phi}) for the explicit solution  and Appendix \ref{appx:offdiagonalfermion} for the rather technical derivation). The latter are the ones determining the bosonic nonabelian self-dual background on the deformed torus in the $\Delta$-expansion.

\subsection{The solution with topological charge $Q=\frac{r}{N}$}
\label{sec:reviewsolution1}
%%%%%%%%%%%%%%%%%%%%%%%%%%%%%%%%%%%

A solution with  topological charge $Q=\frac{r}{N}$ is obtained from Section \ref{sec:reviewsolution} by taking $n_{12}^{(1)}=-r, n_{12}^{(2)}=0, n_{34}^{(1)}=0,n_{34}^{(2)}=1$, and, hence $n_{12}=-r, n_{34}=1$. We also take $s_1=-r,t_2=1,u_3=v_4=1$ and set $A_{\mu\nu}=B_{\mu\nu}=0$ and the rest of $s_\mu$, $t_\mu$, $u_\mu$, and $v_\mu$ to zero. Thus, without loss of generality, we take $
\alpha_{12}=\frac{r}{Nk}\,,  \alpha_{21}=0\,, \alpha_{34}=\frac{1}{N\ell}\,,\alpha_{43}=0$.

 The upshot is that  the transition functions (\ref{transitionfunctions}) now read
\begin{eqnarray}
\nonumber
\Omega_1&=&P_k^{-r}\otimes I_\ell e^{i \omega \frac{r x_2}{Nk L_2}} = \left[\begin{array}{cc}P_k^{-r}e^{i2\pi \ell r \frac{x_2}{Nk L_2}}&0\\0& e^{-i 2\pi r\frac{x_2}{NL_2}}I_\ell\end{array}\right],\\
\nonumber
\Omega_2&=&Q_k\otimes I_\ell = \left[\begin{array}{cc}Q_k&0\\0& I_\ell\end{array}\right],\\
\nonumber
\Omega_3&=&I_k\otimes P_\ell e^{i \omega \frac{x_4}{N\ell L_4}} = \left[\begin{array}{cc} e^{i2\pi  \frac{x_4}{N L_4}} I_k&0\\0& e^{-i 2\pi k\frac{x_4}{N \ell L_4}}P_\ell\end{array}\right],\\
\Omega_4&=&I_k\otimes Q_\ell = \left[\begin{array}{cc}I_k&0\\0& Q_\ell\end{array}\right].
\label{the set of transition functions for Q equal r over N, general solution}
\end{eqnarray}
where we recall that  $\omega$ is given by (\ref{omega}), $P$ and $Q$ in (\ref{pandq}), and $I_k$ ($I_\ell$) denote $k\times k$ ($\ell\times \ell$) unit matrices. Above, we introduced our $k \times \ell$ block-matrix notation, to be used further in this paper.

The reader can use (\ref{the set of transition functions for Q equal r over N, general solution}), recalling that $k + \ell =N$, with $k$, $\ell$ being positive integers, and that $P$ and $Q$ are the clock and shift matrices (\ref{pandq}), to explicitly check that $\Omega_\mu$  obey the cocycle conditions (\ref{cocycle}) with only $n_{12}=-r$ and $n_{34}=1$ being nonzero, and that these  hold for any $1 \le r \le N$. 
Likewise, it is easy to check that the abelian gauge field and the field strength of the constant flux background%
\begin{eqnarray}
\nonumber
A_1&=&-\omega \frac{z_1}{L_1}\,, ~A_2=-\omega\left(\frac{ r x_1}{Nk L_1L_2}+\frac{z_2}{L_2}\right)\,, ~ A_3=-\omega \frac{z_3}{L_3}\,, ~ A_4=-\omega\left(\frac{x_3}{N\ell L_3L_4}+\frac{z_4}{L_4}\right)\,\\
F_{12}&=&-\omega\frac{r}{Nk L_1L_2}\,,F_{34}=-\omega\frac{1}{N\ell L_3L_4}\,.
\label{r over N abelian sol}
\end{eqnarray}
 obey the boundary conditions (\ref{conditions on gauge field}) with transition functions (\ref{the set of transition functions for Q equal r over N, general solution}).\footnote{If one of $k$ or $\ell$ is unity, the cocycle conditions with $n_{12}=-r$, $n_{34}=1$ and the corresponding boundary conditions  (\ref{conditions on gauge field}) are obeyed with the corresponding $P$ and $Q$ in $\Omega_\mu$ replaced by unity.}

If we require the self-duality of the solution $F_{12}=F_{34}$, we find that a self-dual torus sides have to obey the relation
\begin{eqnarray}\label{selfdualtorus1}
\frac{L_1L_2}{L_3L_4}=\frac{r\ell}{k}\,.
\end{eqnarray}

\subsection{Boundary conditions for the adjoint fermions}
\label{sec:boundarycondadjoint}

In the rest of Section \ref{sec:fermionzeromodes}, we solve the Weyl equations $D_\mu\bar\sigma_\mu\lambda=0$ and $D_\mu\sigma_\mu \bar \lambda=0$ for massless adjoint fermions in the background (\ref{r over N abelian sol}).\footnote{\label{footnote:notation0}Here, $\sigma_\mu \equiv(i\vec\sigma,1)$, $\bar\sigma_\mu \equiv(-i\vec\sigma,1)$, $\vec \sigma$ are the Pauli matrices which determine the $\mu={1,2,3}$ components of the four-vectors $\sigma_\mu, \bar\sigma_\mu$. The Euclidean action for fermions and the matrices $\sigma_\mu$, $\bar\sigma_\mu$, $\sigma_{\mu\nu}$, $\bar\sigma_{\mu\nu}$, are  as in \cite{Dorey:2002ik}, except   that we use hermitean gauge fields, necessitating the replacement $A^{\text{that ref.}} = i A^{\text{this paper}}$. } This will enable us to understand the fermionic zero modes in the background with topological charge $Q=\frac{r}{N}$ on the self-dual torus. In subsequent sections, the results help the construction of the self-dual bosonic background on the deformed torus in the small-$\Delta$ expansion. 

Before we begin, let us discuss the moduli of the solution. We first note that 
 the constant holonomies $z_\mu$ in the $U(1)$ direction $\omega$, appearing in (\ref{r over N abelian sol}),  are the most general ones commuting with the transition functions (\ref{the set of transition functions for Q equal r over N, general solution}), provided gcd$(k,r)=1$ (that this is so follows from the discussion below). 
 
However, when gcd$(k,r)>1$, there are  gcd$(k,r)$ different holonomies permitted for each $\mu$. To work them out for future use, we first note that   the holonomies have to be in the Cartan subalgebra, because they have to commute with $Q_{k}$ and $Q_l$ from (\ref{the set of transition functions for Q equal r over N, general solution}) in order that (\ref{conditions on gauge field}) be obeyed. Thus,  the additional (to $z_\mu$ from (\ref{r over N abelian sol})) holonomies would add, to the background (\ref{r over N abelian sol}), $\delta A_\mu = H^{a'}  \phi^{a'}_\mu + H^a  \phi^a_\mu$, with constant $\phi$'s, where $H^{a'}$ ($a'=1,...,k-1$) and $H^a$ ($a=1,...l-1$) are the $SU(k)$ and $SU(l)$ Cartan generators, respectively. The generators $H^{a'}$, $H^a$ are extended to have zero entries in their respective complement to $SU(N)$.
In addition, $H^{a'}$ and $H^a$ have to commute with the transition functions (\ref{the set of transition functions for Q equal r over N, general solution}), which means that $P_k^{-r} H^{a'} P_k^r = H^{a'}$ and $P_l H^a P_l^{-1} = H^a$. Clearly, there are no nonzero $SU(\ell)$ generators $H^a$ allowed, thus we set the corresponding holonomies to zero $\phi^a_\mu=0$. The condition for $H^{a'}$ only allows nonzero $\phi^{a'}_\mu$ if gcd$(k,r)>1$. If gcd$(k,r)=k$, any Cartan generator obeys $P_k^{-r} H^{a'} P_k^r = H^{a'}$ and so there are $k-1$ $\phi^{a'}_\mu$'s allowed (for reasons that become clear later, we shall study this case in great detail in what follows). For generic values of gcd$(k,r)$, $1 < {\rm gcd}(k,r) \le k$,  there are only gcd$(k,r)$ holonomies along the $SU(k)$ Cartan generators allowed. Let us now describe them in a manner useful for the future.

 For general values of gcd$(r,k)$, we combine the allowed holonomies in the $SU(k)$ part of $SU(N)$ with the $z_\mu$ holonomies (the ones proportional to  $\omega$, see (\ref{r over N abelian sol})). We use primed indices $C', B' = 1...k$ to denote the $k\times k$ part of the components of the $SU(N)$ gauge field and unprimed $C,B = 1,...\ell$ to denote the $SU(\ell)$ components. Thus, we describe the general abelian background (\ref{r over N abelian sol}) as 
 \begin{equation}\label{gaugewithholonomies}
\hat A_\mu = (A_\mu)\vert_{{\rm of\; eqn.~(\ref{r over N abelian sol}) \; with} \; z_\mu = 0} + \left[\begin{array}{cc}|| \delta A_{\mu \; C' B'}||&0\\0& ||\delta A_{\mu \; C B}|| \end{array}\right]~,
 \end{equation}
 using the same block-matrix form as in (\ref{the set of transition functions for Q equal r over N, general solution}), with, e.g. $|| \delta A_{\mu \; C' B'}||$ denoting a $k\times k$ matrix with components $\delta A_{\mu \; C' B'}$, etc. 
 All holonomies (including $z_\mu$) are now included in the second term and are given by  \begin{eqnarray}\label{allowedholonomies}
 \delta A_{\mu \; C D} &=& \delta_{CD} \; 2 \pi k {z_\mu \over L_\mu}, \\
 \delta A_{\mu \; C' D'} &=& \delta_{C'D'}(- 2 \pi \ell {z_\mu \over L_\mu} + \phi_\mu^{C'}),  \nonumber \\
{\rm where} ~ \phi_\mu^{C'} &=& \phi_\mu^{C'-r ({\rm mod} \; k)} \equiv \phi_\mu^{\[C'-r\]_k} \; {\rm and} \; \sum_{C'=1}^{k}\phi_\mu^{C'} = 0.\nonumber
\end{eqnarray}
The $SU(k)$ holonomies, denoted by $\phi_\mu^{C'}$, must obey   the condition from the last line to ensure commutativity with $P_k^r$. 
In (\ref{allowedholonomies}) we also introduced the short-hand notation that we shall often use in this paper:\footnote{Notice that, to conform to (\ref{defofmod}), in (\ref{allowedholonomies}) and further, since $q$(mod$q) = 0$, we take the range of the $SU(k)$ index $C'$ to be  $0...k-1$ instead of $1...k$. Likewise, we take the range of the unprimed $SU(\ell)$ indices $0...\ell-1$.}
\begin{equation}\label{defofmod}
\[x\]_q \equiv x ({\rm mod} \; q)~.
\end{equation}

We now turn to the adjoint fermions (gauginos), which obey the boundary conditions (\ref{conditions on gauge field}) without the inhomogeneous term
\begin{eqnarray} \label{boundaryconditions}
\lambda(x+L_\mu \hat e_\mu)=\Omega_{\mu}\lambda(x)\Omega_{\mu}^{-1}\,,
\end{eqnarray}
with $\Omega_\mu$ from (\ref{the set of transition functions for Q equal r over N, general solution}).  Omitting the spinor index, we write the gaugino field, an $N \times N$ traceless matrix, as a block of $k \times k$, $k \times \ell$, $\ell \times k$ and $\ell \times \ell$ matrices (recall $N = k + \ell$):
\begin{eqnarray} \label{blockform}
\lambda &=& \left[\begin{array}{cc}||\lambda_{C'B'}|| & ||\lambda_{C'B}||\\  ||\lambda_{CB'}||& ||\lambda_{CB}||\end{array}\right]~, ~ C',B' \in \{0,...k-1\}, ~ C,B  \in \{0,...\ell-1\}~, 
\end{eqnarray}
obeying the tracelessness condition
\begin{eqnarray}\label{tracelesslambda}
 \sum\limits_{C'=0}^{k-1} \lambda_{C'C'} +  \sum\limits_{C=0}^{\ell-1} \lambda_{CC} =0~.
\end{eqnarray}

The explicit form of the boundary conditions follows from  (\ref{boundaryconditions}) and (\ref{blockform}).  For $\lambda_{C'B'}$, they are
\begin{eqnarray}
\nonumber
\lambda_{C' B'}(x+L_1\hat e_1)&=&\lambda_{[C'-r]_k \; [B'-r]_k}(x)\,,\\
\nonumber
\lambda_{C' B'}(x+L_2\hat e_2)&=& e^{i2\pi\frac{C'-B'}{k}}\lambda_{C' B'}(x)\,,\\
\nonumber
\lambda_{C' B'}(x+L_3\hat e_3)&=&\lambda_{C' B'}(x)\,,\\
\lambda_{C' B'}(x+L_4\hat e_4)&=&\lambda_{C' B'}(x)\,,
\label{BCS lambda A}
\end{eqnarray}
while  $\lambda_{CB}$ obeys
\begin{eqnarray}
\nonumber
\lambda_{CB}(x+L_1\hat e_1)&=&\lambda_{C B}(x)\,,\\
\nonumber
\lambda_{CB}(x+L_2\hat e_2)&=&\lambda_{CB}(x)\,,\\
\nonumber
\lambda_{CB}(x+L_3\hat e_3)&=&\lambda_{[C+1]_\ell \; [B+1]_\ell}(x)\,,\\
\lambda_{CB}(x+L_4\hat e_4)&=& e^{i2\pi\frac{C-B}{\ell}}\; \lambda_{CB}(x)\,,\label{BCS lambda a}
\end{eqnarray}
and  $\lambda_{C'B}$:
\begin{eqnarray}
\nonumber
\lambda_{C' B}(x+L_1\hat e_1)&=&\gamma_k^{-r}e^{i2\pi \frac{rx_2}{kL_2}} \;\lambda_{[C'-r]_k\;B}(x)\,,\\
\nonumber
\lambda_{C' B}(x+L_2\hat e_2)&=&\gamma_k e^{i2\pi\frac{(C'-1)}{k}} \;\lambda_{C'  B}(x)\,,\\
\nonumber
\lambda_{C' B}(x+L_3\hat e_3)&=&\gamma_\ell^{-1}e^{i2\pi \frac{x_4}{\ell L_4}}\; \lambda_{C' [B+1]_\ell}(x)\,,\\
\lambda_{C' B}(x+L_4\hat e_4)&=&\gamma_\ell^{-1} e^{-i2\pi\frac{(B-1)}{\ell}}\; \lambda_{C' B}(x)\,.
\label{BCS lambda beta}
\end{eqnarray}
We also note that $\lambda_{C B'}$ obeys the h.c. conditions to (\ref{BCS lambda beta}).
In addition,  the dotted fermions $\bar\lambda$ obey boundary conditions equal to the ones given above, written in terms of a decomposition of $\bar\lambda$ in terms of $\bar\lambda_{C'B'}$, $\bar\lambda_{C'B}$, $\bar\lambda_{CB}$ and $\bar\lambda_{CB'}$,  identical  to the one in (\ref{blockform}). 

We can now solve  the Weyl equations $D_\mu \bar \sigma_\mu\lambda=0$ and $D_\mu \sigma_\mu \bar\lambda = 0$ with the above boundary conditions. The covariant derivative is given by $D_\mu=\partial_\mu+i [A_\mu,\,]$ with 
$A_\mu$ already given in (\ref{gaugewithholonomies}) and (\ref{allowedholonomies}). We solve for the zero modes of the Weyl equation in the abelian background, beginning with the simplest cases.

\subsection{Dotted-fermion zero modes}
\label{sec:dottedzeromodes}

First, we solve the Weyl equation for the dotted fermions, $D_\mu\sigma^\mu \bar \lambda=0$. Here, we ignore the allowed nonzero holonomies from (\ref{allowedholonomies}), since  (as we shall see later) they do not affect the solution in an interesting way. We find, keeping in mind the tracelessness condition (\ref{tracelesslambda}),
\begin{eqnarray}
\nonumber
\partial_\mu\sigma^\mu \bar\lambda_{C B \; \dot\alpha}=0\,, \quad \partial_\mu\sigma^\mu \bar\lambda_{C' B'\; \dot\alpha}&=&0, \; \; \text{with} \; \dot\alpha = \dot{1}, \dot{2},\\
\nonumber
\left(\partial_3-i\partial_4-\frac{2\pi x_3}{\ell L_3 L_4}\right)\bar \lambda_{C' B \; \dot 1}+\left(\partial_1-i\partial_2-\frac{2\pi r x_1}{k L_1L_2}\right)\bar \lambda_{C' B\;  \dot2} &=&0\,,\\
\left(\partial_1+i\partial_2+\frac{2\pi r x_1}{k L_1 L_2}\right)\bar \lambda_{C' B \;\dot 1}+\left(-\partial_3-i\partial_4-\frac{2\pi x_3}{\ell L_3L_4}\right)\bar \lambda_{C' B \;\dot2}&=&0\,, \label{barlambdaequations}
\end{eqnarray}
 and similar equations for $\bar\lambda_{C B' \;\dot\alpha}$.
One can convince themselves that there exist no normalizable solutions for $\bar\lambda_{C' B \; \dot\alpha}$ and $\bar\lambda_{C B' \; \dot\alpha}$ obeying the boundary conditions. We shall not repeat the details here but only note that this follows from the analysis of \cite{Anber:2022qsz} and the realization that normalizability of the  solutions on the four torus (after expanding in eigenmodes) ends up requiring normalizability of  simple-harmonic oscillator wavefunctions, the solutions of (\ref{barlambdaequations}), in the infinite $x_1$-$x_3$ plane (the two oscillators being in the $x_1$ and $x_3$ directions).

The only normalizable solution involves the diagonal components  $\bar\lambda_{C C \; \dot\alpha}$ and $\bar\lambda_{C' C' \; \dot\alpha}$ and is constant. This is because the boundary conditions (\ref{BCS lambda a}, \ref{BCS lambda A}) only allow for constant diagonal solutions and also further restrict the solutions as we now discuss. 
The boundary conditions for the $
\ell \times \ell$-components  only permit the solution
\begin{eqnarray}\label{barlambda1}
\bar\lambda_{CC \; \dot\alpha} =  \bar \vartheta_{\dot{\alpha}}, ~ \forall C = 0,..., \ell-1,
\end{eqnarray}
with equal diagonal entries. Here $\bar\vartheta_{\dot\alpha}$ are two Grassmann variables.
The   $k \times k$ part of the dotted fermions, $\bar\lambda_{C'C' \; \dot\alpha}$ allows for gcd$(k,r)$ such solutions (due to the first boundary condition in (\ref{BCS lambda A})), which can be written as
\begin{eqnarray}\label{barlambda2}
\bar \lambda_{C' C' \; \dot\alpha} = \bar \vartheta_{\dot{\alpha}}^{\[C'-r\]_k},
\end{eqnarray}
for arbitrary Grassmann $\bar \vartheta_{\dot{\alpha}}^{\[C'-r\]_k}$. Clearly, for every value of $\dot\alpha$, there are gcd$(k,r)$ such different $\bar \vartheta_{\dot{\alpha}}^{\[C'-r\]_k}$, which one can label 
$\bar\vartheta_{\dot\alpha}^{0}$, $\bar\vartheta_{\dot\alpha}^{1}$ to $...\bar\vartheta_{\dot\alpha}^{{\rm gcd}(k,r)-1}$. The tracelessness condition (\ref{tracelesslambda}), however, determines the $SU(\ell)$ Grassmann variables  (\ref{barlambda1}) in terms of  the $SU(k)$ ones, (\ref{barlambda2}).

In conclusion, there are 
a total of $2 {\rm{gcd}} (k,r)$ dotted-fermion zero modes in the constant-flux instanton background.

\subsection{Undotted-fermion zero modes}
\label{sec:undottedzeromodes}

\subsubsection{The ``diagonal'': $U(1)$, $SU(\ell)$ and $SU(k)$ undotted zero modes} 
\label{sec:undotteddiagonal} 

Now, we continue with the undotted fermions $\lambda_{BC}$ and $\lambda_{B'C'}$, i.e. their componets in the $U(1)$, $SU(k)$ and $SU(\ell)$ directions. Because the abelian  background  (\ref{gaugewithholonomies}, \ref{allowedholonomies}) commutes with the $U(1)$, $SU(k)$ and $SU(\ell)$ generators, these ``diagonal'' components satisfy a free Dirac equation:
\begin{eqnarray}
\partial_\mu\bar\sigma_\mu \lambda_{C'B'}&=&0, \nonumber\\
\partial_\mu\bar\sigma_\mu \lambda_{CB}&=&0, ~ \text{with} ~~ \sum\limits_{C'=0}^{k-1} \lambda_{C'C'} +  \sum\limits_{C=0}^{\ell-1} \lambda_{CC} =0~,
\label{EOM diagonal}
\end{eqnarray}
along with the $SU(N)$ tracelessness condition (\ref{tracelesslambda}).

One needs to solve these equations with the boundary conditions (\ref{BCS lambda A}) and (\ref{BCS lambda a}). We now state the results, since the analysis is similar to that in \cite{Tanizaki:2022plm,Anber:2022qsz}.
The first remark is that, following the steps outlined for the dotted zero modes, one finds that there are no normalizable solutions for the components of $\lambda_{C'B'}$ and $\lambda_{CB}$ 
with $C' \ne B'$ and $C \ne B$ obeying the boundary conditions. 

Next, we note that the only  solution for $\lambda_{CC}$ is the one where $\lambda_{CC \;\alpha} = \eta_\alpha$, with a constant spinor $\eta_\alpha$, for all $C$ (this is needed to satisfy (\ref{BCS lambda a})). The tracelessness condition in (\ref{EOM diagonal}), however, relates this to the $\lambda_{B'B'}$ solutions on which we now focus.
The boundary conditions (\ref{BCS lambda A}) are satisfied by the constant solutions
\begin{eqnarray} \label{undotteddiagonalzeromodes}
\lambda_{B' C' \; \alpha}= \delta_{B'C'} \sum\limits_{j = 0}^{\rm{gcd}(k,r)-1} \; \vartheta_\alpha^{(j)} \; \sum\limits_{n=0}^{\frac{k}{{\rm gcd}(k,r)}-1} \delta_{B',[j+nr]_k},
\end{eqnarray}
with gcd$(k,r)$ arbitrary constant Grassmann spinors  $\vartheta_{\alpha}^{(j)}$. 
We conclude that there are $2\mbox{gcd}(k,r)$ independent zero modes of $\lambda_{B'C'}$ and, from the above remarks, of the all ``diagonal" components of the undotted fermions considered in this Section. 

Note that the number of diagonal undotted zero modes is precisely the same  as the number of the dotted fermion zero modes of Section \ref{sec:dottedzeromodes}. In particular, the  contribution of the zero modes of Sections \ref{sec:dottedzeromodes} and \ref{sec:undotteddiagonal} to the index cancels out.

\subsubsection{The ``off-diagonal'' $k \times \ell$ and $\ell \times k$ undotted zero modes.}
\label{sec:offdiagonalzeromodes}

 The zero modes most worthy of our attention, the ones which determine the nonabelian instanton solution to leading order in  $\Delta$, are the ones considered in this Section. 
Finding the off-diagonal undotted zero modes, the ones for $\lambda_{C'B}$ ($k \times \ell$) and $\lambda_{CB'}$ ($\ell \times k$),  is the most important and least trivial part of our study. We  find that there are $r$ zero modes for $\lambda_{C'B}$ and $r$ zero modes for $\lambda_{CB'}$, in agreement with the index theorem which requires that the number of undotted minus the number of dotted zero modes be $2r$. 

The derivation of the results quoted in this Section is technically involved and the details are relegated to Appendix \ref{appx:offdiagonalfermion}. Here, we simply give  the explicit formulae for the zero modes for $\lambda_{C'B}$, the $k \times \ell$ ones.\footnote{Noting that the $\ell \times k$  zero modes (which come with their own Grassmann parameters) are obtained by hermitean conjugation of  $\Phi^{(p)}$ in (\ref{lambdazeromode1}), as per the remark after (\ref{BCS lambda beta}).} We find that in the background (\ref{gaugewithholonomies}, \ref{allowedholonomies}), only one spinor component has $r$ normalizable zero modes
\begin{eqnarray}
\label{lambdazeromode1}
\lambda_{C'B\; 1} &=& \sum\limits_{p=0}^{{r \over {\rm gcd}(k,r)} - 1} \eta^{[C' + p k]_r}  \; \Phi^{(p)}_{C'B}(x, \hat \phi),\nonumber \\
\lambda_{C'B \; 2} &=& 0~.
\end{eqnarray}
Here, $\eta^j$, $j=0,...,r-1$, are $r$ Grassmann parameters associated with the zero modes (clearly, $[C'+pk]_r$ takes $r$ values).
Notice that a given zero mode, proportional to $\eta^j$ with some $j \in \{0,...,r-1\}$,  nontrivially intertwines  the gauge indices in (\ref{lambdazeromode1}).  

Before giving the form of the functions $\Phi^{(p)}$ governing the $x$-dependence of the zero modes (\ref{lambdazeromode1}), we introduce the notation   
$\hat \phi_\mu^{C'}$ to denote the way various gauge field holonomies appear in the equations governing the off diagonal zero modes. These combine the $U(1)$-holonomy $z_\mu$ with the extra ones allowed when gcd$(k,r)>1$, as per the discussion around (\ref{allowedholonomies}):\footnote{The reason that $2 \pi N$ (and not $2 \pi \ell$) appears here is that $\hat \phi^{C'}$ encodes the action of the commutator on the off diagonal components $\lambda_{C'B}$.}
 \begin{equation}\label{holonomy1}
\hat\phi_\mu^{C'} \equiv \phi_\mu^{C'} - 2 \pi N {z_\mu \over L_\mu}, ~ {\rm{with}}~ \hat\phi_\mu^{C'} =  \hat\phi_\mu^{[C'-r]_k}. 
\end{equation} 
The explicit solution for $\hat\phi^{C'}$ obeying the relations above (and from (\ref{allowedholonomies})) can be written out in a somewhat unwieldy form (which, however, serves to show that there are gcd$(k,r)$ independent holonomies for each $\mu$)\footnote{We note that this is similar to eqn.~(\ref{undotteddiagonalzeromodes}) for the undotted diagonal zero modes of the next Section.}   
\begin{eqnarray}\label{explicit holonomies}
\hat\phi^{C'}_\mu = \sum\limits_{j=0}^{{\rm gcd}(k,r)-1} \varphi^j_\mu \sum\limits_{n=0}^{{k \over {\rm gcd}(k,r)}-1} \delta^{C',[j+nr]_k}~.
\end{eqnarray}
Here, we use the notation (\ref{defofmod}), taking the range of $C'$ to be $0...k-1$. The sum over $n$ for each $j$ simply incorporates the fact that the index $C'$  takes values an ``orbit'' of size $k \over \text{gcd}(k,r)$. Each of the gcd$(k,r)$ ``orbits,'' labelled by $j$, has the same holonomy $\phi_\mu^j$ and contains   values of $C'$ jumping by $r$ units, as required by commutativity of the holonomy with $P_k$.
Although (\ref{explicit holonomies}) explicitly shows that, for each $\mu$, there are gcd$(k,r)$ independent holonomies $\varphi^j_\mu$, we prefer to further denote them as $\hat\phi_\mu^{C'}$, remembering the relations they obey. However, we make explicit use of (\ref{explicit holonomies}) later on, see Section \ref{sec:compactvsnoncompact}.

 The zero modes $\lambda_{C'B \; 1}$  of (\ref{lambdazeromode1}) depend on $(x, \hat\phi^{C'}, \eta^j)$. Their $x$- and $\hat\phi$-dependence is through the ${r \over {\rm{gcd}}(k,r)}$ functions $\Phi^{(p)}$, given by (for derivation, see Appendix \ref{appx:offdiagonalfermion}):\begin{eqnarray}
\nonumber
\Phi^{(p)}_{C' B}(x,\hat\phi)&=& \sum_{\scriptsize m=p+\frac{rm'}{\mbox{gcd}(k,r)},\, m'\in \mathbb Z}~~\sum_{n'\in \mathbb Z}e^{\frac{i2\pi x_2 }{L_2}(m+\frac{2	C'-1-k}{2k})}e^{\frac{i2\pi x_4 }{L_4}(n'-\frac{2	B-1-\ell}{2\ell})}\\
\nonumber
&&~~\times e^{-i\frac{\pi (1-k)}{k}\left(C'-\frac{1+k(1-2m)}{2}\right)} e^{i\frac{\pi(1-\ell)}{\ell}\left(B-\frac{1+\ell(2n'+1)}{2}\right)}\\
\nonumber
&&~~\times \; e^{-\frac{\pi r}{k L_1 L_2}\left[x_1-\frac{k L_1 L_2}{2\pi r}(\hat \phi_2^{[C']_r}-i\hat \phi_1^{[C']_r})-\frac{L_1}{r}\left(km +\frac{2C'-1-k}{2}\right)\right]^2}\\
&&~~\times \; e^{-\frac{\pi }{\ell L_3 L_4}\left[x_3-\frac{\ell L_3 L_4}{2\pi }(\hat \phi_4^{[C']_r}-i\hat \phi_3^{[C']_r})-L_3\left(\ell n' -\frac{2B-1-\ell}{2}\right)\right]^2}\,.
\label{form of Phi}
\end{eqnarray}
The explicit form of the functions $\Phi^{(p)}$ will be useful later, in our study of the properties  of the self-dual fractional instantons on the deformed torus. 
Eqns.~(\ref{lambdazeromode1}, \ref{holonomy1}, \ref{form of Phi}) give the general normalizable solution of the massless undotted Weyl equation $D_\mu\bar\sigma_\mu\lambda=0$ for $\lambda_{C'B \; \alpha}$ in the abelian constant-field strength background (\ref{gaugewithholonomies},\ref{allowedholonomies}) of topological charge $Q={r \over N}$.

In summary of  Section \ref{sec:fermionzeromodes}, we found that there is a number of dotted and undotted zero modes in the abelian background of topological charge $r \over N$. The total number is consistent with the index theorem. The solutions for the non-constant fermion zero modes will be used to construct the nonabelian self-dual solution of charge $r \over N$ on the deformed torus.

\section{Deforming the self-dual torus: small-$\Delta$ expansion for the bosonic background with $Q= {r \over N}$}
\label{sec:deforming}

To remedy the zero modes problem we saw in the previous section, i.e.,  to lift the dotted zero modes, we now depart from the self-dual torus and search for a self-dual instanton solution with topological charge $Q=\frac{r}{N}$ on a deformed $\mathbb T^4$, following the strategy of  \cite{GarciaPerez:2000aiw,Gonzalez-Arroyo:2019wpu}. We write the general gauge field  on the non-self-dual torus in the form
\begin{eqnarray}
A_\mu(x)= \hat A_\mu +{\cal S}_\mu^\omega(x) \;\omega +\delta_\mu(x)\,.
\label{gauge field in general}
\end{eqnarray}
Here, $\omega$ is the $U(1)$ generator (\ref{omega}), $\hat A_\mu$ is the abelian gauge field with constant field strength defined previously in  (\ref{gaugewithholonomies}) and ${\cal S}_\mu^\omega(x)$ is the nonconstant field component along the $U(1)$ generator. The non-abelian part $\delta_\mu(x)$ is given by the $N\times N$ matrix, which, as earlier in (\ref{the set of transition functions for Q equal r over N, general solution}), (\ref{gaugewithholonomies}), (\ref{blockform}), is decomposed in a block form:\footnote{\label{footnote:notation}Here ${\cal S}_\mu^{k}$ and ${\cal S}_\mu^l$ are traceless $su(k)$- and $su(l)$-algebra elements, respectively, while  ${\cal W}_\mu^{k\times \ell}$ is a complex $k \times \ell$ matrix with ${\cal W}_\mu^{\dagger \ell \times k}$ its hermitean conjugate. In the second (bracketed) term in (\ref{fields1})  we have indicated the index notation used earlier in describing the zero modes of the adjoint fermions, recall (\ref{lambdazeromode1}). Here, we find it convenient to use the block matrix notation $S^k, S^\ell, W^{k \times \ell}, W^{\dagger \; \ell \times k}$ and will revert to using indices $B'C',B'C$, etc., when needed.}
\begin{eqnarray}\label{fields1}
\delta_\mu=\left[\begin{array}{cc} {\cal S}_{\mu }^k & {\cal W}_\mu^{k\times \ell}\\ {\cal W}_\mu^{\dagger \ell \times k}& {\cal S}_\mu^\ell\end{array} \right] \; \; \;\; \left(  \equiv \left[\begin{array}{cc} ||{\cal S}_{\mu \; B' C'}^k|| & ||{\cal W}_{\mu \; B' C}|| \\ ||{\cal (W}_{\mu}^\dagger)_{C B'}||& ||{\cal S}_{\mu \; B C}^\ell||\end{array} \right] \right)\,.
\end{eqnarray}

The boundary conditions  (\ref{conditions on gauge field}) with transition functions (\ref{the set of transition functions for Q equal r over N, general solution}) imply  that ${\cal S}^\omega_\mu$ satisfy periodic boundary conditions in all directions (because $\hat A_\mu$ absorbs the inhomogenous part of (\ref{conditions on gauge field})):
\begin{eqnarray}
{\cal S}^\omega_\mu(x+\hat e_\nu L_\nu)={\cal S}^\omega_\mu(x)\,.
\label{periodic BCS for S omega}
\end{eqnarray}
On the other hand,  $ {\cal S}_\mu^k$, ${\cal S}_\mu^\ell$,  ${\cal W}_\mu^{k\times \ell}$, and ${\cal W}_\mu^{\dagger \ell \times k}$ satisfy  exactly the same gaugino-field boundary conditions we discussed in the previous section, and we refrain from repeating (thus, the boundary conditions are given by equations (\ref{BCS lambda A}), (\ref{BCS lambda a}), (\ref{BCS lambda beta}), respectively, for $ {\cal S}_\mu^k$, ${\cal S}_\mu^\ell$,  ${\cal W}_\mu^{k\times \ell}$, recalling (\ref{fields1}) and Footnote \ref{footnote:notation}).

The field strength of (\ref{gauge field in general}), $F_{\mu\nu}=\partial_\mu A_\nu-\partial_\nu A_\mu +i[A_\mu, A_\nu]$, is  given by
\begin{eqnarray}
\nonumber
F_{\mu\nu}&=&\hat F_{\mu\nu}+F_{\mu\nu}^s \omega+\hat D_\mu\delta_\nu-\hat D_\nu \delta_\mu+i\left[{\cal S}_\mu^\omega\omega, \delta_\nu\right]+i\left[\delta_\mu,{\cal S}_\nu^\omega\omega\right]+i[\delta_\mu,\delta_\nu]\,,\\
&\equiv&\hat F_{\mu\nu}+F_{\mu\nu}^s\omega+\left[\begin{array}{cc} F_{\mu\nu}^k & {\cal F}_{\mu\nu}^{k\times\ell}\\ {\cal F}_{\mu\nu}^{\dagger \ell\times k} & F_{\mu\nu}^{\ell} \end{array}\right]\,,
\end{eqnarray}
where $\hat D_\mu=\partial_\mu+i[\hat A_\mu,\,]$ is the covariant derivative w.r.t. the gauge field $\hat A_\mu$. Using (\ref{gauge field in general}, \ref{fields1}), we obtain:
\begin{eqnarray}
\nonumber
F_{\mu\nu}^s&=&\partial_\mu {\cal S}_\nu^\omega-\partial_\nu {\cal S}_\mu^\omega\,,\\
\nonumber
F_{\mu\nu}^{k}&=&\partial_\mu {\cal S}_\nu^k-\partial_\nu {\cal S}_\mu^k+i[{\cal S}_\mu^k, {\cal S}_\nu^k]+i{\cal W}_\mu^{k \times \ell} {\cal W}_\nu^{\dagger\ell \times k} -i{\cal W}_\nu^{k \times \ell} {\cal W}_\mu^{\dagger \ell \times k}\,, \\
\nonumber
F_{\mu\nu}^{\ell}&=&\partial_\mu {\cal S}_\nu^\ell-\partial_\nu {\cal S}_\mu^\ell+i[{\cal S}_\mu^\ell, {\cal S}_\nu^\ell]+i{\cal W}_\mu^{\dagger \ell \times k}{\cal W}_\nu^{k \times \ell}  -i{\cal W}_\nu^{\dagger \ell \times k}{\cal W}_\mu^{k \times \ell} \,,\\ 
\nonumber
{\cal F}_{\mu\nu}^{k \times \ell}&=&\hat D_\mu {\cal W}_\nu^{k \times \ell}-\hat D_\nu {\cal W}_\mu^{k \times \ell}+i{\cal S}_\mu^k {\cal W}_\nu^{k\times \ell}-i{\cal S}_\nu^k {\cal W}_\mu^{k\times \ell}+i{\cal W}_\mu^{k\times \ell}{\cal S}_\nu^{\ell}-i{\cal W}_\nu^{k\times \ell}{\cal S}_\mu^{\ell}\\
&&+i2\pi N \left({\cal S}_\mu^\omega {\cal W}_\nu^{k\times \ell}-{\cal S}_\nu^\omega {\cal W}_\mu^{k\times\ell}\right)\,,
\label{components of F}
\end{eqnarray}
where $\hat D_\mu {\cal W}_\nu^{k \times \ell}$ is understood as
\begin{eqnarray}\label{cov1}
\hat D_\mu {\cal W}_\nu^{k \times \ell}=\left[\partial_\mu +i 2\pi N\hat A_\mu^\omega\right]{\cal W}_\nu^{k \times \ell}\,,
\end{eqnarray}
and we have written $\hat A_\mu =\hat A_\mu^\omega \omega$, for $\hat A_\mu$ 
from (\ref{gaugewithholonomies}).\footnote{For brevity,   the nontrivial holonomies' (allowed when gcd$(k,r)>1$) are not explicitly shown here. They should, however, be included in the covariant derivatives in (\ref{cov1},\ref{cov2}) and our final solution (\ref{expressions of W2 and W4 with holonomies})  does take these into account.}
Similarly,
\begin{eqnarray}\label{cov2}
\hat D_\mu {\cal W}_\nu^{\dagger  \ell\times k}=\left[\partial_\mu -i 2\pi N\hat A_\mu^\omega\right]{\cal W}_\nu^{\dagger \ell\times k}\,.
\end{eqnarray}

Next, we  impose self duality on the background (\ref{gauge field in general}) on the deformed $\mathbb T^4$. 
Imposing self-duality is equivalent (see e.g. \cite{Dorey:2002ik}) to imposing the constraint on the field strength \begin{eqnarray}
\bar \sigma_{\mu\nu}F_{\mu\nu}=0\,.\label{selfduality1}
\end{eqnarray}
where\footnote{Recall that the matrices $\sigma_\mu$, $\bar\sigma_\mu$ were defined in Footnote \ref{footnote:notation0}.} 
  $\bar\sigma_{\mu\nu}=\frac{1}{2}(\bar\sigma_\mu\sigma_\nu-\bar\sigma_\nu\sigma_\mu)$. Now, we recall $\hat F_{\mu\nu} = \hat F_{\mu\nu}^\omega \omega$, and use (\ref{r over N abelian sol}) to find $\hat F_{12}^\omega=-\frac{r}{Nk L_1L_2}$ and $\hat F_{34}^\omega=-\frac{1}{N\ell L_3L_4}$. Recalling the properties of the self-dual $\mathbb T^4$, eqn.~(\ref{selfdualtorus1}), we also define the parameter $\Delta$, which parametrizes the deviation from the self-dual torus:
\begin{eqnarray}\label{deltadef}
\Delta\equiv \frac{r\ell L_3L_4-k L_1 L_2}{\sqrt{V}}\,.
\end{eqnarray}
We assume, without loss of generality, $\Delta\ge0$. Thus, we find that%
\begin{eqnarray}\label{fdelta}
\hat F_{\mu\nu}^\omega\bar\sigma_{\mu\nu}=-\frac{2i \Delta}{Nk \ell \sqrt{V}}\sigma_3\,.
\end{eqnarray}

To continue, for every  four-vector ${\cal V}_\mu$, we define the quaternions  ${\cal V}\equiv \sigma_{\mu}{\cal V}_\mu$ and $\bar{\cal V}\equiv \bar\sigma_\mu{\cal V}_\mu$. 
Then, using (\ref{components of F}) and (\ref{fdelta}), we find  that self-duality (\ref{selfduality1}) implies that
\begin{eqnarray}
\frac{1}{2}\bar \sigma_{\mu\nu}F_{\mu\nu}=\left(-\frac{i \Delta}{Nk \ell \sqrt{V}}\sigma_3+\bar\partial {\cal S}^\omega-\partial_\mu S_\mu^\omega\right)\omega
+\left[\begin{array}{cc} {\cal A}^k &{\cal A}^{k\times \ell}\\{\cal A}^{\dagger \ell\times k}& {\cal A}^\ell\end{array}\right]=0\,,
\label{slef duality constraint matrix}
\end{eqnarray}
where\footnote{Here and below, the terms that have sums over $\mu$ should be multiplied by unit quaternion   $\sigma_4$, which we have omitted for brevity. Thus, temporarily not denoting explicitly that these are $k \times \ell$ matrices, we warn the reader to keep in mind the difference between the quaternions, ${\cal W} \equiv {\cal W}_\mu \sigma_\mu$, $\bar{\cal W}= \bar\sigma_\mu W_\mu$, and the four-vector ${\cal W}_\mu$ and, furthermore, note that  ${\cal W}^\dagger = \sigma_\mu W^\dagger_\mu$ and $\bar{\cal W}^\dagger = \bar\sigma_\mu W^\dagger_\mu$. }
\begin{eqnarray}
\nonumber
{\cal A}^k&=&\bar\partial {\cal S}^k-\partial_\mu{\cal S}_\mu^k-i\bar{\cal S}^k{\cal S}^k+i {\cal S}_\mu^k{\cal S}_\mu^k+i \bar {\cal W}^{k\times \ell}{\cal W}^{\dagger\ell\times k}-i{\cal W}_\mu^{k\times \ell}{\cal W}_\mu^{\dagger\ell\times k}\,,\\
\nonumber
{\cal A}^{k\times\ell}&=&\bar {\hat D} {\cal W}^{k\times \ell}-\hat D_\mu {\cal W}_\mu^{k\times\ell}+i\bar{\cal S}^{k}{\cal W}^{k\times \ell}-i {\cal S}_\mu^k{\cal W}_\mu^{k\times \ell}+i\bar {\cal W}^{k\times \ell}{\cal S}^\ell-i{\cal W}_\mu^{k\times \ell} {\cal S}_\mu^\ell\\
\nonumber
&&+i 2\pi N\left(\bar {\cal S}^\omega{\cal W}^{k\times \ell}-{\cal S}_\mu^\omega {\cal W}_\mu^{k\times \ell}\right)\,,\\
{\cal A}^\ell&=&\bar\partial {\cal S}^\ell-\partial_\mu{\cal S}_\mu^\ell-i\bar{\cal S}^\ell{\cal S}^\ell+i {\cal S}_\mu^\ell{\cal S}_\mu^\ell+i \bar {\cal W}^{\dagger \ell\times k}{\cal W}^{k\times \ell}-i{\cal W}_\mu^{\dagger \ell\times k}{\cal W}_\mu^{  k\times \ell}\,.
\end{eqnarray}
In order to remove gauge redundancies, we impose the background gauge condition with respect to the field $\hat A_\mu$:
\begin{eqnarray}
\hat D_\mu A_\mu=0
\end{eqnarray}
which in components reads:
\begin{eqnarray}
\partial_\mu {\cal S}_\mu^\omega=0\,,\partial_\mu {\cal S}_\mu^k=0\,, \partial_\mu {\cal S}_\mu^\ell=0\,,  {\hat D}_\mu {\cal W}_\mu^{k\times \ell}=0\,,{\hat D}_\mu {\cal W}_\mu^{\dagger  \ell \times k}=0\,.
\label{BGC}
\end{eqnarray}
Using (\ref{BGC}) in (\ref{slef duality constraint matrix}), we find the set of equations imposing the self-duality condition on the background (\ref{gauge field in general}):
\begin{eqnarray}
\nonumber
\left(-\frac{i 2\pi \Delta}{Nk \sqrt{V}}\sigma_3+2\pi\ell\bar\partial {\cal S}^\omega\right) I_k+\bar\partial {\cal S}^k-i\bar{\cal S}^k{\cal S}^k+i {\cal S}_\mu^k{\cal S}_\mu^k+i \bar {\cal W}^{k\times \ell}{\cal W}^{\dagger\ell\times k}-i{\cal W}_\mu^{k\times \ell}{\cal W}_\mu^{\dagger\ell\times k}&=&0\,,\\
\nonumber
\left(\frac{i 2\pi \Delta}{N\ell \sqrt{V}}\sigma_3-2\pi k\bar\partial {\cal S}^\omega\right) I_\ell +\bar\partial {\cal S}^\ell-i\bar{\cal S}^\ell{\cal S}^\ell+i {\cal S}_\mu^\ell{\cal S}_\mu^\ell+i \bar {\cal W}^{\dagger \ell\times k}{\cal W}^{k\times \ell}-i{\cal W}_\mu^{\dagger \ell\times k}{\cal W}_\mu^{  k\times \ell}&=&0\,,\\
\nonumber
\bar {\hat D} {\cal W}^{k\times \ell}+i\bar{\cal S}^{k}{\cal W}^{k\times \ell}-i {\cal S}_\mu^k{\cal W}_\mu^{k\times \ell}+i\bar {\cal W}^{k\times \ell}{\cal S}^\ell-i{\cal W}_\mu^{k\times\ell} {\cal S}_\mu^\ell+i 2\pi N\left(\bar {\cal S}^\omega{\cal W}^{k\times \ell}-{\cal S}_\mu^\omega {\cal W}_\mu^{k\times \ell}\right)&=&0\,.\\
\label{main equations we need to solve}  
\end{eqnarray}
We note that here $\bar {\hat D} \equiv \bar\sigma_\mu \hat D_\mu$, precisely the Weyl operator for the undotted fermions, whose zero modes were studied in Section \ref{sec:undottedzeromodes}.

The idea of    the method introduced in \cite{GarciaPerez:2000aiw} is that a solution of the self-duality conditions (\ref{main equations we need to solve}) can be obtained via series expansions in the deformation parameter $\Delta$ of (\ref{deltadef}). 
The approximate solution of the self-duality equations thus obtained is then also an approximation to the minimal action solution of the equations of motion, i.e. a fractional instanton with $Q = {r \over N}$. 
Comparing the $\Delta$ scaling of the various terms in (\ref{main equations we need to solve}), the $\Delta$-expansion is found to have the following form
\begin{eqnarray}
\nonumber
{\cal W}^{k\times \ell}&=&\sqrt{\Delta} \sum_{a=0}^\infty \Delta^a {\cal W}^{(a)k\times \ell}\, ,\\
{\cal S}&=& \Delta \sum_{a=0}^\infty \Delta^a{\cal S}^{(a)}\,, \label{expansion1}
\end{eqnarray}
where ${\cal S}$ accounts for ${\cal S}^\omega$, ${\cal S}^k$, and ${\cal S}^\ell$. 

We proceed to leading order\footnote{The $\Delta$ expansion was tested to high orders, and found to converge (even to the infinite volume limit) in the two dimensional abelian Higgs model in \cite{Gonzalez-Arroyo:2004fmv}. Convergence is not well  understood for the general case of $SU(N)$ in four dimensions. For $SU(2)$, the  comparisons with the exact numerical solution (obtained by minimizing the lattice Yang-Mills action) of \cite{GarciaPerez:2000aiw} give evidence for the convergence of the expansion for small $\Delta$. It should be possible to analytically study the properties of higher orders in the expansion (\ref{expansion1}) of the solutions of (\ref{main equations we need to solve}); however, this rather formidable task is left for the future.}  in $\Delta$ by considering solutions of ${\cal W}^{k\times \ell}$ to order $\sqrt {\Delta}$ and ${\cal S}$ to order $\Delta$, thus keeping only the terms ${\cal S}^{(0)}$ and ${\cal W}^{(0)}$ in (\ref{expansion1}). Then, to this order, (\ref{main equations we need to solve}) gives
\begin{eqnarray}
\nonumber
\left(-\frac{i 2\pi }{Nk \sqrt{V}}\sigma_3+2\pi\ell\bar\partial {\cal S}^{(0)\omega}\right) I_k+\bar\partial {\cal S}^{(0)k}+i \bar {\cal W}^{(0)k\times \ell}{\cal W}^{\dagger (0)\ell\times k}-i{\cal W}_\mu^{(0)k\times \ell}{\cal W}_\mu^{\dagger(0)\ell\times k}&=&0\,,\\\nonumber
\left(\frac{i 2\pi }{N\ell \sqrt{V}}\sigma_3-2\pi k\bar\partial {\cal S}^{(0)\omega}\right) I_\ell +\bar\partial {\cal S}^{(0)\ell}+i \bar {\cal W}^{\dagger(0) \ell\times k}{\cal W}^{(0)k\times \ell}-i{\cal W}_\mu^{\dagger(0) \ell\times k}{\cal W}_\mu^{ (0) k\times \ell}&=&0\,,\\
&& \label{Sequations}
\end{eqnarray}
and
\begin{eqnarray}
\bar {\hat D} {\cal W}^{(0)k\times \ell}&=&0\,.
\label{Wequation}
\end{eqnarray}
The strategy of solving the leading-order equations (\ref{Sequations}, \ref{Wequation}) is as follows:
\begin{enumerate}
 \item Solve (\ref{Wequation}) for the quaternions ${\cal W}^{(0)k\times \ell}$. This equation has the form of two copies of the undotted fermion zero-mode equation, whose general normalizable solutions were already found in Section \ref{sec:offdiagonalzeromodes}, recall (\ref{lambdazeromode1}).
 \item Next, plug the general solution of (\ref{Wequation}) into  (\ref{Sequations}). 
The result is a set of first-order differential equations for the quaternions ${\cal{S}}^{(0)}$, with periodic boundary conditions for ${\cal{S}}^{(0) \omega}$ and with ${ \cal{S}}^{(0) k}, {\cal{S}}^{(0) \ell}$, obeying (\ref{BCS lambda A}), (\ref{BCS lambda a}), respectively. 
The resulting equations for ${\cal S}^{(0)}$  have  nonvanishing source terms, comprised of a constant piece (the one proportional to $\sigma_3$ in (\ref{Sequations})) and of terms quadratic in the just-found general solution of (\ref{Wequation}), ${\cal W}^{(0) k \times \ell}$. Consistency of these equations requires that the source term be orthogonal to the zero modes of the differential operator acting on the various components of ${\cal S}^{(0)}$. 
\item One then needs to determine the zero modes of $\bar \partial$, the operator acting on $\cal S^{(0)}$, obeying the appropriate boundary conditions. This task was already accomplished in Section \ref{sec:undotteddiagonal}, since $\bar\partial$ is simply the undotted diagonal Weyl operator. We then require  orthogonality of these zero modes to the source terms in (\ref{Sequations}). 
On one hand, this will be shown to provide restrictions on the arbitrary coefficients appearing in the general  solution of  (\ref{Wequation}), ${\cal W}^{(0) k\times \ell}$. The  coefficients left arbitrary determine the moduli space of the multi-fractional instanton. On the other hand, imposing consistency of (\ref{Sequations}) allows one to determine ${\cal S}^{(0)}$ by expanding both sides in a chosen basis of functions and equating the coefficients on both sides. \end{enumerate}
The   procedure outlined above can be, in principle, iterated to higher orders. The way this procedure works to higher orders was, in principle,  studied in \cite{Gonzalez-Arroyo:2004fmv}. However, implementing it to determine the higher-order solution becomes technically challenging. Here, we shall only study the leading-order  and determine the constraints of the arbitrary coefficients in ${\cal W}^{(0)\; k \times \ell}$, which restrict the moduli space of the multi-fractional instantons. 

To begin implementing the above steps, we start with (\ref{Wequation}), written explicitly as
\begin{eqnarray}
\bar \sigma_\mu \hat D_\mu \left[\begin{array}{cccc} {\cal W}_4^{(0)k\times \ell}+i {\cal W}_3^{(0)k\times \ell} &&& {\cal W}_2^{(0)k\times \ell}+i {\cal W}_1^{(0)k\times \ell} \\ -{\cal W}_2^{(0)k\times \ell}+i {\cal W}_1^{(0)k\times \ell} &&& {\cal W}_4^{(0)k\times \ell}-i {\cal W}_3^{(0)k\times \ell} \end{array}\right]=0\,,
\label{semifinal equation}
\end{eqnarray}
where $\hat D_\mu = \partial_\mu + i \;[\hat A_\mu, ]$ is the covariant derivative in the background (\ref{gaugewithholonomies}). 
As already stated, (\ref{semifinal equation}) represent two copies of the undotted gaugino zero mode equations in the $\Delta=0$ background  $A^\omega$, one for each column of the $\cal{W}$-quaternion given above. Further, as for the gauginos, one can show that normalizability on $\mathbb T^4$ requires normalizability in the infinite $x_1, x_3$ plane of the simple harmonic oscillator wave functions, the solutions of (\ref{semifinal equation}).  
Thus, we borrow the solutions for the gauginos from Section \ref{sec:offdiagonalzeromodes}, we find that equations (\ref{semifinal equation}) have normalizable solutions if and only if 
\begin{eqnarray}\label{assertion1}
{\cal W}_4^{(0)k\times \ell}=i{\cal W}_3^{(0)k\times \ell}\,, \quad {\cal W}_2^{(0)k\times \ell}=i{\cal W}_1^{(0)k\times \ell}\,
\end{eqnarray}
noting that these are nothing but the conditions of vanishing of $\lambda_{C'B\; 2}$, recall (\ref{lambdazeromode1}).
The  solutions for ${\cal W}_4^{(0)k\times \ell}, {\cal W}_2^{(0)k\times \ell}$ are then  borrowed from (\ref{lambdazeromode1}):\footnote{For further use, in (\ref{expressions of W2 and W4 with holonomies}), we also introduced the short-hand notation ${ W}_{2 \; C'C}$ and ${ W}_{4 \; C'C}$ for the general solutions of (\ref{Wequation}).}
\begin{eqnarray}
\nonumber
\left({\cal W}_2^{(0)k\times \ell}\right)_{C'C}=V^{-1/4} \sum_{p=0}^{\scriptsize \frac{r}{\mbox{gcd}(k,r)}-1}{\cal C}^{[C'+pk]_r}_2\Phi^{(p)}_{C'C}(x,\hat\phi) =: {W}_{2 \; C'C}\,,\\
\left({\cal W}_4^{(0) k\times \ell}\right)_{C'C}=V^{-1/4} \sum_{p=0}^{\scriptsize \frac{r}{\mbox{gcd}(k,r)}-1}{\cal C}^{[C'+pk]_r}_4\Phi^{(p)}_{C'C}(x,\hat\phi)  =:{ W}_{4 \; C'C}\,,
\label{expressions of W2 and W4 with holonomies}
\end{eqnarray}
where $\Phi^{(p)}_{C'C}(x,\hat\phi)$ are given by (\ref{form of Phi}) and the volume factor is included for future convenience. 
 Thus, there are $2r$ arbitrary coefficients ${\cal C}^{[C'+pk]_r}_2$ and ${\cal C}^{[C'+pk]_r}_4$, which are now complex bosonic variables. In the following, we shall discuss the physical significance of ${\cal C}_{2, 4}$. 

We now continue with the next step: imposing orthogonality to the various zero modes of $\bar \partial = \bar\sigma_\mu \partial_\mu$, the solutions of the equation $\bar \partial {\cal S}^{(0)} =0$. Notice that $\bar \partial$ is precisely the Weyl operator for the diagonal undotted fermions discussed in Section \ref{sec:undotteddiagonal} and that we shall borrow our results from that Section shortly. To continue, however, it is convenient to rewrite (\ref{Sequations}) using the index notation, recalling eqn.~(\ref{fields1}) and Footnote \ref{footnote:notation}. This necessitates using   (\ref{assertion1}) and the definition of the quaternions, in order to express everything through the general solutions of (\ref{Wequation}), denoted by ${W}_{4 \; (\text{or} 2) \; C'C}$ of (\ref{expressions of W2 and W4 with holonomies}). This produces, from the first equation of (\ref{Sequations}),
an equation determining ${\cal S}_{C'B'}$ (which includes the component ${\cal S}^\omega \omega$ from (\ref{gauge field in general})):
\begin{eqnarray}\label{equationforSk}
&&\bar \partial {\cal S}_{C'B'} = \\
&&i \left(\begin{array}{cc}  {2 \pi \over N k \sqrt{V}} \delta_{C'B'} - 2 \;(W_{2} W^*_2 -W_{4 } W^*_{4})_{C'B'} &
 4 \;(W_{2} W^*_{4})_{C'B'}  \cr
 4 \;(W_{2} W^*_{4})_{C'B'}   
  &- {2 \pi \over N k \sqrt{V}} \delta_{C'B'} + 2\; (W_{2} W^*_{2 } -W_{4 } W^*_{4})_{C'B'}   \end{array}\right)~, \nonumber
\end{eqnarray}
where we introduced the shorthand notation, $(W_{2 } W^*_{4 })_{C'B'} \equiv W_{2 \; C' D} W^*_{4 \; B' D}$, with a sum over $D$ implied, and similar for the other contractions. Likewise, the equation for ${\cal S}_{CB}$ obtained from the second of eqns.~(\ref{Sequations})
 reads:
\begin{eqnarray} \label{equationforSell}
&& \bar\partial {\cal S}_{CB} = \\
&&i \left(\begin{array}{cc}  -{2 \pi \over N \ell \sqrt{V}} \delta_{CB} + 2  (W_{2}^* W_{2 } -W_{4 }^* W_{4})_{CB} &
- 4 ( W^*_{4} W_{2})_{CB}  \cr
 -4 (W_{2}^* W_{4})_{CB}   
  &  {2 \pi \over N \ell \sqrt{V}} \delta_{CB} - 2  (W_{2}^* W_{2 } -W_{4 }^* W_{4})_{CB}   \end{array}\right)~,  \nonumber
\end{eqnarray}
using a similar shorthand (e.g. $(W_{2}^* W_2)_{CB} \equiv W^*_{2 \; D'C} W_{2 \; D' B}$ with a sum over $D'$).

 Next, we recall that the operator $\bar \partial$ is the Weyl operator for the diagonal undotted fermions, whose zero modes were determined in Section \ref{sec:undotteddiagonal}. We also recall that $\cal S$ is a quaternion, hence (similar to 
  (\ref{Wequation})), we can think of $\cal S$ as of two sets of Weyl fermions, one for each column of the quaternion matrix.
 We can thus borrow the results for the zero modes, recalling
 (\ref{EOM diagonal}) and (\ref{undotteddiagonalzeromodes}), and then impose  their orthogonality of the r.h.s. of (\ref{equationforSk}, \ref{equationforSell}). 
 As shown there, undotted fermions have  $2$gcd$(k,r)$ constant zero modes. This implies that there are $4$gcd$(k,r)$ zero modes of $\cal S$, which we label by an arbitrary {\it quaternionic} coefficient $\epsilon^{(j)}$, $j = 0,...,{\rm gcd}(k,r)-1$.
  The corresponding wave functions, which we denote $s_{B'C'}$ and $s_{BC}$, have only diagonal entries
  \begin{eqnarray}\label{szeromodes}
s_{B' C' }&=& \delta_{B'C'} \sum\limits_{j = 0}^{{\rm{gcd}}(k,r)-1} \; \epsilon^{(j)} \; \sum\limits_{n=0}^{\frac{k}{{\rm gcd}(k,r)}-1} \delta_{B',[j+nr]_k}\, , \nonumber \\
s_{B C}&=& - {\delta_{BC} \over \ell} \sum_{B' = 0}^{k-1} s_{B'B'}~, ~ \forall B = 0,..., \ell-1~.
\end{eqnarray}

The simplest condition is the orthogonality of $s_{BC}$ (which is simply a constant quaternionic mode) to the source term in the equation for ${\cal S}_{CB}$. Multiplying the source term by the $s_{BC}$ zero mode, taking the trace,  and integrating over $\mathbb T^4$, we find  that orthogonality implies that the integral of the trace of the r.h.s. over $\mathbb T^4$ should vanish for every entry in the quaternion source on the r.h.s. of (\ref{equationforSell}). Explicitly, this gives the conditions
\begin{eqnarray}\label{diagonalconsistencycondition}
\int_{\mathbb T^4}  (W_{2 \;  B'C}^* W_{2 \; B' C } -W_{4 \;  B'C }^* W_{4 \; B' C}) &=& { \pi \over N} \sqrt{V}~, \nonumber \\
\int_{\mathbb T^4} W_{4 \; B'C}^* W_{2 \; B' C }&=&0~,
\end{eqnarray}
 with a sum over the full range of repeated indices implied.

 However, the conditions imposed by orthogonality to the $4$gcd$(k,r)$ zero modes $s_{B'B'}$ labelled by $\epsilon^{(j)}$ are more detailed than (\ref{diagonalconsistencycondition}). Proceeding similar to the above, we find the gcd$(k,r)$  conditions:
\begin{eqnarray}
\label{gcdconsistencycondition}
\nonumber
&&\sum\limits_{B=0}^{\ell-1}  \sum\limits_{C'=0}^{k-1} \sum\limits_{n=0}^{\frac{k}{{\rm gcd}(k,r)}-1} \delta_{C', [j + n r]_k}\int_{\mathbb T^4}   (W_{2 \; C' B} W^*_{2 \; C'B  } -W_{4 \; C' B } W^*_{4 \; C'B}) = { \pi \over N \text{gcd}(k,r)} \sqrt{V}  \\
\nonumber
&&\sum\limits_{B=0}^{\ell-1}  \sum\limits_{C'=0}^{k-1} \sum\limits_{n=0}^{\frac{k}{{\rm gcd}(k,r)}-1}  \delta_{C', [j + n r]_k}\int_{\mathbb T^4}W_{4 \; C' B}^* W_{2 \;  C'B }=0,\quad j=0,..., \mbox{gcd}(k,r)-1\,.\\
\end{eqnarray}
That the above  gcd$(k,r)$ conditions are more general than (\ref{diagonalconsistencycondition}) follows by observing that summing up the gcd$(k,r)$ conditions in each line of  (\ref{gcdconsistencycondition}) (i.e., summing over $j$) we obtain (\ref{diagonalconsistencycondition}).

The importance of the conditions (\ref{gcdconsistencycondition}) is that they restrict the $2 r$ complex coefficients ${\cal C}_2$ and ${\cal C}_4$, and thus determine the moduli space of the multifractional instanton. Studying this is the subject of  the next Section.

\section{The moduli of the $Q= {r \over N}$ bosonic solution: compact vs. noncompact}
\label{sec:compactvsnoncompact}

To study the constraints (\ref{diagonalconsistencycondition}, \ref{gcdconsistencycondition}) with  $W_2$ and $W_4$ from (\ref{expressions of W2 and W4 with holonomies}), we now define, for each $j=0,..., {\rm gcd}(k,r)-1$ and $a,b \in \{2,4\}$:
\begin{eqnarray}\label{Iab1}
I^{ab}_j &=& \sum\limits_{C'=0}^{k-1} \sum\limits_{n=0}^{\frac{k}{{\rm gcd}(k,r)}-1}   \delta_{C', [j+nr]_k} \sum\limits_{p,p'=0}^{{r\over {\rm gcd}(k,r)}-1}
{{\cal C}_a^{[C'+ pk]_r} \; {\cal C}_b^{* \; [C'+ p'k]_r} \over \sqrt{V}} \int_{\mathbb T^4} \sum\limits_{B=0}^{\ell-1} \Phi^{(p)}_{C'B} \Phi^{(p') \; *}_{C'B}. ~ \end{eqnarray}
In terms of $I_j^{ab}$, the constraints (\ref{diagonalconsistencycondition}, \ref{gcdconsistencycondition}) take the form:
\begin{eqnarray}\label{constraints in terms of I}
I_j^{22} - I_j^{44} &=& {\pi \sqrt{V} \over {\rm gcd}(k,r) N}, \\
I_j^{42} &=& 0, ~~~{\rm where} ~~ j = 0,...,{\rm gcd}(k,r)-1~.\nonumber
\end{eqnarray}

The expressions (\ref{Iab1}) are evaluated in Appendix \ref{appx:identity1}. Substituting $I^{ab}_j$ from (\ref{IabFinal}) in, we find the constraints  (\ref{diagonalconsistencycondition}, \ref{gcdconsistencycondition}) expressed in terms of  the undetermined complex coefficients ${\cal C}_2^A$ and ${\cal C}_4^A$ from the solution of the equations for ${\cal W}_\mu$ (\ref{expressions of W2 and W4 with holonomies}):\footnote{We also note that the origin of the $(\varphi^j_{1,3})^2$-terms on the r.h.s. is in the imaginary $\hat\phi_1, \hat\phi_3$-terms appearing in the last two lines in $\Phi^{(p)}$ from  (\ref{form of Phi}). One can show that they can be absorbed in the definition of the coefficients ${\cal C}^j$ (or $\eta^j$).}
\begin{eqnarray}\label{constraintssubbed}
\sum\limits_{A_j \in S_j} {\cal C}_2^{A_j} \;{\cal C}_2^{* \; A_j} - {\cal C}_4^{A_j} \;{\cal C}_4^{* \; A_j} &=& {2 \pi \over {\rm gcd}(k,r) N}\sqrt{ r L_1 L_3 \over \ell k L_2 L_4}    \; {e^{- {L_1 L_2 k \over 2 \pi r} ( \varphi_1^{j})^2}}  \; {e^{- {L_3 L_4 \ell \over 2 \pi} (\varphi_3^j)^2}} ,\nonumber  \\
\sum\limits_{A_j \in S_j} {\cal C}_2^{ A_j}\; {\cal C}_4^{* \; A_j} &=&0~.
\end{eqnarray}
Here, $S_j$ are  ${\rm gcd}(k,r)$ sets of integers ($\in \{0,...,r-1\}$), defined in (\ref{setSj}) and repeated here for convenience: $$S_j = \bigg\{ [[j+nr]_k+ pk]_r, \text{for} \; n = 0,...\frac{k}{{\rm gcd}(k,r)}-1,  \text{and} \; p = 0,...,{r\over {\rm gcd}(k,r)}-1 \bigg\}. $$ Repeated entries in $S_j$ are identified so that each set has  $r \over {\rm gcd}(k,r)$ elements. The union of all sets $S_j$ is the set $ \{0,...,r-1\}$.

As we shall shortly see, the structure of the ``moduli space'' of ${\cal C}_{2,4}^A$ defined by (\ref{constraintssubbed}) is quite rich. 
Let us, however, first  count the number of moduli for general $k$ and $r > 1$, taking into account the constraints (\ref{constraintssubbed}). First, there are $4$ gcd$(k,r)$ Wilson lines $\varphi_\mu^j$, as per (\ref{explicit holonomies}). 
Then, there are $2 r$ real components of ${\cal C}_2^A$ and $2 r$ real components of ${\cal C}_4^A$. Thus the total number of real moduli is $4 r + 4 {\rm gcd}(k,r)$. These are subject to the constraints of eqn.~(\ref{constraintssubbed}): the gcd$(k,r)$ real constraints on the first line and $2$gcd$(k,r)$ real constraints on the second line.
Thus, it would appear that the number of moduli minus the number of constraints is $4 r + {\rm gcd}(k,r)$. 
We notice, however, that  the gauge conditions (\ref{BGC}) are invariant under   constant gauge transformations  in the gcd$(k,r)$ Cartan directions, the ones along the allowed holonomies (\ref{explicit holonomies}) (i.e. ones that commute with the transition functions).\footnote{In the next Section, we shall  explicitly see that  no gauge invariant characterizing the instanton depends on these phases.}  Thus, the total number of bosonic moduli for $k \ne r > 1$ is  $4 r$, as required by the index theorem for a selfdual solution.

{\flushleft{W}}e now consider the various cases in detail:   
\begin{enumerate}
\item {\bf The case $\mathbf{k=r}$}. This case is singled out by the fact that there are $k$ complex coefficients ${\cal C}_2^A$ (and $k$ ${\cal C}_4^A$). In addition, the $r$ sets $S_j$ are such that each contains a single element, one of the $r$ allowed values of $A$. Thus the $r (=k)$ constraints become, with $c$  a real number, determined by the r.h.s. of (\ref{constraintssubbed}):
\begin{eqnarray} \label{requalkconstraints}
{\cal C}_2^{A} \;{\cal C}_2^{* \; A} - {\cal C}_4^{A} \;{\cal C}_4^{* \; A} &=& c^2  ~(\text{no sum over $A$})\,,  \\
{\cal C}_2^{ A}\; {\cal C}_4^{* \; A} &=&0  \qquad  \implies {\cal C}_4^A = 0, ~ {\cal C}_2^A = e^{i \alpha_A} c, ~  \forall ~ A \in \{0,...,r-1\} .\nonumber 
\end{eqnarray} 
 
Thus, all ``moduli'' ${\cal C}_{2,4}^A$ are fixed up to $r$ undetermined phases $\alpha_A$. These phases are unphysical and correspond to the already mentioned ability to perform $r$ ($=$gcd($k,r$)) constant gauge transformations preserving the gauge conditions  (\ref{BGC}).  Thus, the only moduli left are the $r$ phases $\varphi^j_\mu$, $j=0,...,r$, recall (\ref{explicit holonomies}). 
 
 Thus the multifractional instanton obtained for $k=r$, with $Q = {r \over N}$, has $4 r$ compact moduli, as expected from the index theorem. Further studies of the instantons for $k=r$ and the interpretation of these moduli will be discussed in the next Section.

\item {\bf 
The case $\mathbf{ k \ne r ,  r>1}$.}\footnote{We do not consider $r=1$ here, as it was studied in detail before \cite{Gonzalez-Arroyo:2019wpu}.  As is also easy to see from our formulae, for $r=1$, the moduli ${\cal C}_{2,4}$ are also fixed.}  This case is quite different.
Here the  $r$ sets $S_j$ contain more than a single number each. Thus, the second equation in (\ref{requalkconstraints}) does not set any modulus to zero (recall that it required that all ${\cal C}_4^A$ vanish for $k=r$). Instead,  as we argue below, the constraints permit the moduli ${\cal C}_{2,4}$ to grow without bound, thus making the ``moduli'' space noncompact. 

To illustrate the noncompactness for $k \ne r >1$, we abandon generality and focus on a simple example $r=2, k=3$, a case with gcd$(k,r)=1$ (we shall further use this example in the following). Here, there is only a single set $S_j$, $S_0 =\{0,1\}$ and after the following relabeling, with all $x$'s and $y$'s real,\footnote{A trivial rescaling setting the r.h.s. of the first equation in (\ref{constraintssubbed}) to unity is not explicitly shown.} 
\begin{eqnarray}
{\cal C}_2^0 \rightarrow x_1+i y_1\,,\quad {\cal C}_4^0\rightarrow x_2+i y_2\,,\quad {\cal C}_2^1\rightarrow x_3+iy_3\,,\quad {\cal C}_4^1 \rightarrow x_4+i y_4\,,
\end{eqnarray}
we obtain for eqns.~(\ref{constraintssubbed}):
 \begin{eqnarray}
 \nonumber
 x_1^2+y_1^2+x_3^2+y_3^2 -x_2^2-y_2^2 -x_4^2-y_4^2&=&1\,,\\
 \nonumber
x_1x_2+y_1y_2+x_3x_4+y_3y_4&=&0\,,\\
x_2y_1-x_1y_2+y_3x_4-x_3y_4&=&0\,.
 \label{first condition k 3 r 2}
 \end{eqnarray}
Conditions (\ref{first condition k 3 r 2}) eliminate $3$ out of $8$ real parameters, leaving $4$ physical parameters that parameterize the moduli space in addition to the single arbitrary unphysical phase mentioned above (recall that here gcd$(k,r)$=1).  

The moduli space spanned by the hypersurface given by the constraints (\ref{first condition k 3 r 2}) is non-compact. To see this, we set for simplicity $x_2=y_1=y_3=x_4=0$. Then, the constraints become
 \begin{eqnarray}
x_1y_2=-x_3y_4\,, \quad x_1^2-y_2^2+x_3^2-y_4^2=1\,.
 \end{eqnarray}
For every $x_3=y_4 \in (-\infty,\infty)$ we find
\begin{eqnarray}
x_1^2=\frac{x_3^4}{x_1^2}+1\,,
\end{eqnarray} 
 which has  at least  two real solutions of $x_1$. We also find that $x_1\rightarrow \infty$ as $x_3=y_4\rightarrow \infty$. We conclude that the moduli space is non-compact. For a later convenience, we parametrize the asymptotic region ($u \rightarrow \infty$) of this noncompact direction of the moduli space as
 \begin{eqnarray} \label{noncompact}
 {\cal C}_2^0 \sim \pm u\,,\quad  {\cal C}_2^1 \sim  u\,, \quad  {\cal C}_4^0 \sim \mp iu\,, \quad  {\cal C}_4^1 \sim  iu\,.
 \end{eqnarray}
It is easy to see, even without following the derivation, that (\ref{noncompact}) obey (\ref{constraintssubbed}) with vanishing r.h.s., i.e. at $u \rightarrow \infty$
\end{enumerate}
 The presence of  noncompact moduli for the $k \ne r$ instantons is difficult to interpret  in a $\mathbb T^4$ geometry.  In the later Sections, we shall see that on this noncompact moduli space, ${\cal O}(\Delta)$ gauge invariants characterizing the multifractional instanton grow without bounds---see the end of Section \ref{sec:local} for a brief discussion of the blowup and Appendix \ref{appx:blowup} for details of its derivation.  
  This blow up clashes with the spirit of the $\Delta$ expansion.   As we mentioned in the Introduction, it would be nice to achieve a deeper understanding of this finding.

\section{Local gauge invariants of the $Q={r\over N}$ solution and its ``dissociation"}

\label{sec:dissociation}

In this Section, we give  expressions for local gauge invariant densities characterizing the multifractional instanton to order $\Delta$. These expressions are evaluated in the Appendices. We use the results to, first, show that ${\cal O}(\Delta)$ local gauge invariants grow without bound along the noncompact moduli directions found for $k \ne r$, and, second, to argue for the fractionalization of the $k=r$ multifractional instanton into $r$ identical lumps located at positions on $\mathbb T^4$ determined by the $r$ distinct holonomies/moduli.

%%%%%%%%%%%%%%%%%%%%%%%
\subsection{Gauge-invariant local densities to order $\Delta$ and their blow up for $k \ne r$}
\label{sec:local}
%%%%%%%%%%%%%%%%%%%%%%%

The gauge-invariant local density of the lowest scaling dimension is
\begin{eqnarray} \label{local1}
\mbox{tr}\left[F_{\mu_1\nu_1}F_{\mu_2\nu_2}\right]\,,
\end{eqnarray}
where 
\begin{eqnarray} \label{eff1}
F_{\mu\nu}=\left(F_{\mu\nu}^\omega+F_{\mu\nu}^s\right)\omega +\left[\begin{array}{cc}F_{\mu\nu}^k & {\cal F}_{\mu\nu}\\ {\cal F}_{\mu\nu}^\dagger & F_{\mu\nu}^\ell\end{array}\right]\,,
\end{eqnarray}
and we recall that   the components of (\ref{eff1})  were already defined in  (\ref{components of F}).\footnote{For brevity, we have omitted the $k \times \ell$ and $\ell \times k$ superscripts  in writing (\ref{eff1}).}

 In  Appendix \ref{appx:strength}, we compute the various field strength components appearing in (\ref{eff1}) to order $\Delta$ (shown in eqn.~(\ref{leading order in field strength}))  as well as the action density and action. Then, following the same steps used in deriving the action density there,  we obtain for eqn.~(\ref{local1}) to order $\Delta$
\begin{eqnarray}
\nonumber
&&\mbox{tr}\left[F_{\mu_1\nu_1}F_{\mu_2\nu_2}\right]=\\
\nonumber
&&\mbox{tr}[\omega^2]\left\{ \hat F_{\mu_1\nu_1}^\omega \hat F_{\mu_2\nu_2}^\omega+\Delta \hat F_{\mu_1\nu_1}^\omega\left(\partial_{\mu_2}{\cal S}^{(0)\omega }_{\nu_2}-\partial_{\nu_2}{\cal S}^{(0)\omega }_{\mu_2}\right)+\Delta \hat F_{\mu_2\nu_2}^\omega\left(\partial_{\mu_1}{\cal S}^{(0)\omega }_{\nu_1}-\partial_{\nu_1}{\cal S}^{(0)\omega }_{\mu_1}\right) \right\}\\
\nonumber
&&+2\pi \ell \Delta \hat F_{\mu_1\nu_1}^\omega\mbox{tr}_k \left[\partial_{\mu_2}{\cal S}^{(0)k }_{\nu_2}-\partial_{\nu_2}{\cal S}^{(0)k }_{\mu_2}\right]+2\pi \ell \Delta \hat F_{\mu_2\nu_2}^\omega\mbox{tr}_k \left[\partial_{\mu_1}{\cal S}^{(0)k }_{\nu_1}-\partial_{\nu_1}{\cal S}^{(0)k }_{\mu_1}\right]\\
\nonumber
&&-2\pi k \Delta \hat F_{\mu_1\nu_1}^\omega\mbox{tr}_\ell \left[\partial_{\mu_2}{\cal S}^{(0)\ell }_{\nu_2}-\partial_{\nu_2}{\cal S}^{(0)\ell }_{\mu_2}\right]-2\pi k \Delta \hat F_{\mu_2\nu_2}^\omega\mbox{tr}_\ell \left[\partial_{\mu_1}{\cal S}^{(0)\ell }_{\nu_1}-\partial_{\nu_1}{\cal S}^{(0)\ell }_{\mu_1}\right]\\
\nonumber
&&+i 2\pi N \Delta \hat F_{\mu_1\nu_1}^\omega\mbox{tr}_k\left[{\cal W}_{\mu_2}{\cal W}_{\nu_2}^\dagger-{\cal W}_{\nu_2}{\cal W}_{\mu_2}^\dagger\right]+i 2\pi N \Delta \hat F_{\mu_2\nu_2}^\omega\mbox{tr}_k\left[{\cal W}_{\mu_1}{\cal W}_{\nu_1}^\dagger-{\cal W}_{\nu_1}{\cal W}_{\mu_1}^\dagger\right]\\
&&+\Delta \mbox{tr}_k\left({\cal F}_{\mu_1\nu_1}{\cal F}_{\mu_2\nu_2}^\dagger\right)+\Delta \mbox{tr}_\ell\left({\cal F}_{\mu_1\nu_1}^\dagger{\cal F}\,_{\mu_2\nu_2}\right)\,.
\end{eqnarray}
Using $\mbox{tr}_\ell {\cal S}^{(0\ell)}_\mu=\mbox{tr}_k {\cal S}^{(0k)}_\mu=0$, we obtain
\begin{eqnarray}
\nonumber
&&\mbox{tr}\left[F_{\mu_1\nu_1}F_{\mu_2\nu_2}\right]=\\
\nonumber
&&\mbox{tr}[\omega^2]\left\{ \hat F_{\mu_1\nu_1}^\omega \hat F_{\mu_2\nu_2}^\omega+\Delta \hat F_{\mu_1\nu_1}^\omega\left(\partial_{\mu_2}{\cal S}^{(0)\omega }_{\nu_2}-\partial_{\nu_2}{\cal S}^{(0)\omega }_{\mu_2}\right)+\Delta \hat F_{\mu_2\nu_2}^\omega\left(\partial_{\mu_1}{\cal S}^{(0)\omega }_{\nu_1}-\partial_{\nu_1}{\cal S}^{(0)\omega }_{\mu_1}\right) \right\}\\
\nonumber
&&+i 2\pi N \Delta \hat F_{\mu_1\nu_1}^\omega\mbox{tr}_k\left[{\cal W}_{\mu_2}{\cal W}_{\nu_2}^\dagger-{\cal W}_{\nu_2}{\cal W}_{\mu_2}^\dagger\right]+i 2\pi N \Delta \hat F_{\mu_2\nu_2}^\omega\mbox{tr}_k\left[{\cal W}_{\mu_1}{\cal W}_{\nu_1}^\dagger-{\cal W}_{\nu_1}{\cal W}_{\mu_1}^\dagger\right]\\
&&+\Delta \mbox{tr}_k\left({\cal F}_{\mu_1\nu_1}{\cal F}_{\mu_2\nu_2}^\dagger\right)+\Delta \mbox{tr}_\ell\left({\cal F}_{\mu_1\nu_1}^\dagger{\cal F}\,_{\mu_2\nu_2}\right)\,.
\label{local gauge invariant density}
\end{eqnarray}

  In Appendix \ref{appx:blowup}, we compute (for definiteness) the gauge invariant density $\mbox{tr}\left[F_{34}F_{34}\right]$ for the $k \ne r$ solution  and show that it grows without bounds along the noncompact moduli direction of  (\ref{noncompact}). This local gauge invariant, from (\ref{local gauge invariant density}), is given by
\begin{eqnarray}
\nonumber
&&\mbox{tr}\left[F_{34}F_{34}\right]=\\
\nonumber
&&\mbox{tr}[\omega^2]\left\{ \hat F_{34}^\omega \hat F_{34}^\omega+2\Delta \hat F_{34}^\omega\left(\partial_{3}{\cal S}^{(0)\omega }_{4}-\partial_{4}{\cal S}^{(0)\omega }_{3}\right) \right\}+i 4\pi N \Delta \hat F_{34}^\omega\mbox{tr}_k\left[{\cal W}_{3}{\cal W}_{4}^\dagger-{\cal W}_{4}{\cal W}_{3}^\dagger\right]=\\
&&\mbox{tr}[\omega^2]\left\{ \hat F_{34}^\omega \hat F_{34}^\omega+2\Delta \hat F_{34}^\omega\left(\partial_{3}{\cal S}^{(0)\omega }_{4}-\partial_{4}{\cal S}^{(0)\omega }_{3}\right) \right\}+ 8\pi N \Delta \hat F_{34}^\omega\mbox{tr}_k\left[{\cal W}_{4}{\cal W}_{4}^\dagger\right]\,,
\label{field strength F34F34}
\end{eqnarray}
and we used ${\cal W}_3=-i{\cal W}_4$. 

To show the blow up, we use the example $r=2$, $k=3$ studied in Section \ref{sec:compactvsnoncompact}. In Appendix \ref{appx:blowup}, we show that in the noncompact direction (\ref{noncompact}) the ${\cal O}(\Delta)$ gauge invariant blows up as $u \rightarrow \infty$. This runaway behaviour of local gauge invariant densities along the noncompact moduli space runs counter the spirit of the $\Delta$-expansion. Thus, in what follows, we concentrate on the properties of the $k=r$ solutions with compact moduli space. 

%%%%%%%%%%%%%%%%%%%%%%%%%%%%%%%%%%%%%%%%%
 \subsection{Fractionalization of solutions with topological charges $r>1$}
%%%%%%%%%%%%%%%%%%%%%%%%%%%%%%%%%%%%%%%%%
\label{sec:fractional}

\subsubsection{Bosonic gauge invariant densities}
\label{sec:fractionalbose}

In this section, we use the results for the local gauge invariants to argue that instantons with topological charges $r>1$ dissociate into $r$ identical components. It is clear from the discussion in the previous section that unless one takes $k=r$, one faces the undesired runaway behavior of the gauge-invariant densities. Thus, we limit our discussion to the case $k=r$, where we show that the gauge-invariant densities take the form of a sum over $r$ independent lumps centered around $r$ distinct holonomies. 

To this end, consider (\ref{local gauge invariant density}) taking $\mu_1=\mu_3=1, \mu_2=\mu_4=2$.  Thus, one obtains
\begin{eqnarray}
\nonumber
\mbox{tr}\left[F_{12}F_{12} \right]=\mbox{tr}[\omega^2]\left\{ \hat F_{12}^\omega \hat F_{12}^\omega+2\Delta \hat F_{12}^\omega\left(\partial_{1}{\cal S}^{(0)\omega }_{2}-\partial_{2}{\cal S}^{(0)\omega }_{1}\right) \right\}+ 8\pi N \Delta \hat F_{12}^\omega\mbox{tr}_k\left[{\cal W}_{2}{\cal W}_{2}^\dagger\right]\,,\\
\label{field strength F12F12}
\end{eqnarray}
where, using (\ref{the long expressions of different S}), we find
\begin{eqnarray}
\nonumber
\left(\partial_1{\cal S}_2^{(0)\omega}-\partial_2{\cal S}_1^{(0)\omega} \right)&=&-\left(\pi \ell k \Box\right)^{-1}\left(\partial_1^2+\partial_2^2\right)\mbox{tr}_k\left[{\cal W}_2^{(0)}{\cal W}_2^{\dagger(0)}\right]\,. \\
\label{final expression of superpositions of S in the 12 direction}
\end{eqnarray}
Here, 
\begin{eqnarray}
{\cal W}^{(0)}_{2\,C',C}(x)&=&V^{-1/4}{\cal C}_{2}^{C'}\Phi^{(0)}_{C',C}(x,\hat\phi)\,, \quad C'=1,2,...,k=r\,, \quad C=1,2,..,\ell\,.
\end{eqnarray}
It is more convenient to express $\Phi^{(0)}_{C',C}(x,\hat\phi)$ in the form given in (\ref{expression of Phi that shows fractionalization}) 
\begin{eqnarray}
\nonumber
 \Phi^{(0)}_{C',C}(x,\hat\phi)&=& e^{\frac{k L_1 L_2}{2\pi r}\hat \phi_1^{C'}\left(i\hat \phi_2^{C'}+\hat \phi_1^{C'}/2\right)}  e^{\frac{\ell L_3 L_4}{2\pi }\hat \phi_3^{C'}\left(i\hat \phi_4^{C'}+\hat \phi_3^{C'}/2\right)}  e^{-i\hat \phi_1^{C'} x_1}  e^{-i\hat \phi_3^{C'} x_3}\\
\nonumber
&&\times  \sum_{m'\in \mathbb Z}\sum_{n'\in \mathbb Z} e^{i \left(\frac{ 2\pi x_2 }{L_2}+L_1\hat \phi_1^{C'}\right)(m'+\frac{2	C'-1-k}{2k})}e^{i\left(\frac{2\pi x_4 }{L_4}+\ell L_3\hat \phi_3^{C'}\right)(n'-\frac{2	C-1-\ell}{2\ell})}\\
\nonumber
&&\times e^{-i\frac{\pi (1-k)}{k}\left(C'-\frac{1+k(1-2m)}{2}\right)} e^{i\frac{\pi(1-\ell)}{\ell}\left(C-\frac{1+\ell(2n'+1)}{2}\right)}\\
\nonumber
&&\times e^{-\frac{\pi r}{k L_1 L_2}\left[x_1-\frac{L_1 L_2}{2\pi }\hat \phi_2^{C'}-\frac{L_1}{k}\left(km' +\frac{2C'-1-k}{2}\right)\right]^2}\\
&&\times e^{-\frac{\pi }{\ell L_3 L_4}\left[x_3-\frac{\ell L_3 L_4}{2\pi }\hat \phi_4^{C'}-L_3\left(\ell n' -\frac{2C-1-\ell}{2}\right)\right]^2}\,.
\label{expression of Phi that shows fractionalization, p=0}
\end{eqnarray}
The above eqns.~(\ref{final expression of superpositions of S in the 12 direction}, \ref{field strength F12F12}) imply that  the computation of the gauge-invariant density $\mbox{tr}\left[F_{12}F_{12} \right]$ requires finding the quantity
\begin{eqnarray}\label{quantity1}
\mbox{tr}_k\left[{\cal W}_2^{(0)}{\cal W}_2^{\dagger(0)}\right]=\sum_{C'=1}^{r}\left( \sum_{C=1}^{\ell}  |{\cal C}_2^{C'}|^2| \Phi^{(0)}_{C',C}(x,\hat\phi)|^2\right)\,.
\end{eqnarray}

To further study (\ref{quantity1}), we need to take into account the fact that the $r$ coefficients ${\cal C}_2$ are determined by the top equation in (\ref{constraintssubbed}), as described in (\ref{requalkconstraints}). It is important  that ${\cal C}_2$ do depend on the holonomies, which were absorbed into the coefficient $c$ in (\ref{requalkconstraints}). Taking this into account,\footnote{The $\hat\phi_{1,3}$-dependence of ${\cal C}_2$ cancels the $(\hat\phi_1)^2$ and $(\hat\phi_3)^2$ terms in the exponent on the first line  of (\ref{expression of Phi that shows fractionalization, p=0}). This  ensures that gauge invariant quantities have periodic dependence on the holonomies.} we find, after some rearrangement,  that the expression (\ref{quantity1}), which determines $\mbox{tr}\left[F_{12}F_{12} \right]$ to order $\Delta$ has the following form:\footnote{Up to an inessential $L_\mu, r, \ell, N$-dependent constant which can be easily determined.}
\begin{eqnarray}\label{fractional1}
&&\mbox{tr}_k\left[{\cal W}_2^{(0)}{\cal W}_2^{\dagger(0)}\right]  \sim \nonumber \\
&  & \sum\limits_{C'=1}^r\bigg|  \sum_{m'\in \mathbb Z}\;\;e^{i \left(\frac{ 2\pi x_2 }{L_2}+L_1\hat \phi_1^{C'}\right) m'  -\frac{\pi  }{  L_1 L_2}\left[x_1-\frac{L_1 L_2}{2\pi }\hat \phi_2^{C'}- \frac{L_1 C'}{r} - L_1 (m' -\frac{1+r}{2r}) \right]^2}\bigg|^2 \nonumber \\
 &&\;\;\; \times \;
\bigg|  \sum_{n'\in \mathbb Z} \;e^{i\left(\frac{2\pi x_4 }{\ell L_4}+ L_3\hat \phi_3^{C'}\right) n'   -\frac{\pi }{\ell L_3 L_4}\left[x_3-\frac{\ell L_3 L_4}{2\pi }\hat \phi_4^{C'}-L_3\left(\ell n' +\frac{ 1+\ell}{2}\right)\right]^2}\, \bigg|^2\nonumber  \\
&=:& \sum\limits_{C'=1}^r  F (x_1 - {L_1 L_2 \over 2 \pi} \hat\phi_2^{C'} - {L_1 C' \over r},\; x_2 +  {L_1 L_2 \over 2 \pi} \hat\phi_1^{C'},\; x_3 -  {\ell L_3 L_4 \over 2 \pi} \hat\phi_4^{C'},\; x_4+  { \ell L_3 L_4 \over 2 \pi} \hat\phi_3^{C'})\,. \nonumber \\
\end{eqnarray}
As indicated on the last line above, for every $C'=1,2,..,r$, the summand is given by the same function $F(x_1,x_2, x_3,x_4)$, implicitly defined above, but centered at a different point $x_\mu$ on $\mathbb T^4$. The position of each lump is determined by the moduli $\hat \phi_\mu^{C'}$, $\mu=1,2,3,4$, $C'=1,...,r$. 
 The size of the lumps is, of course, set by  the size of $\mathbb T^4$, the only scale of the problem. Thus, the ``lumps'' we find are not well isolated, but strongly overlapping, rather like a liquid than a dilute gas (see Figure \ref{visual of liquid} for an illustration).

\subsubsection{Fermionic zero modes and their localization }
\label{sec:fractionalfermion}

The conclusion of the above analysis is that the local gauge invariant density of the multifractional instanton, $\mbox{tr}\left[F_{12}F_{12} \right]$, takes the form of a  sum of $r $ identical lumps, each centered at  $r$ distinct holonomies. 
Thus, the solution of topological charge $r/N$ can be thought of as composed of $r$ distinct lumps. Each lump is expected to contribute  $1/N$-th of the total topological charge. 

This expectation is strengthened by considering the fermion zero modes in the $Q={r \over N}$ self-dual solution. In Appendix \ref{appx:fermiondelta}, we show that there are $2r$ zero modes, labeled by a $2$-spinor $\bar\eta^{C'}_\alpha$, with $C'=1,...r$. 
 To order ${\cal O}(\sqrt{\Delta})$, the $x$-dependence of the zero modes appears in the off-diagonal components: \begin{eqnarray}\label{lambdazeromode13}
 \nonumber
\lambda_{1 \; C'D} &\sim & \bar\eta_2^{C'} (\partial_3 + i \hat\phi_3^{C'}) \Phi^{(0)}_{C',C}(x, \hat\phi)) \equiv \bar\eta_2^{C'} {\cal G}_{3 \; C'D}^{(0)}(x, \hat\phi^{C'}), \\
\lambda_{2 \; C'D} &=& 0~.
\end{eqnarray}
with the expression for ${\cal G}_{3 \; C'D}^{(0)}(x, \hat\phi^{C'})$ given in  Appendix \ref{appx:strength}, see  (\ref{the G3 function}). Likewise, the zero mode wave function in  the other off-diagonal component  is
\begin{eqnarray}\label{lambdazeromode23}\nonumber
\lambda_{1 \;  D C'} &= & 0 , \\
\lambda_{2 \; D C'} &\sim& \bar\eta_1^{C'}   {\cal G}_{3 \; C'D}^{* \; (0)}(x, \hat\phi^{C'}). 
\end{eqnarray}
Even without consulting the explicit expression, it is clear that the $C'$-th zero mode only depends on $\hat\phi^{C'}_\mu$, which, therefore, governs its $x_\mu$-dependence, similar to (\ref{fractional1}) above. 

 Explicitly, one can construct ${\cal O}(\Delta)$ gauge invariants formed from the zero modes, to find, as for the bosonic invariants, that they are governed by a ``lumpy'' structure, with each of the $r$  lumps supporting $2$ zero modes located at a position governed by the moduli $\hat\phi^{C'}_\mu$. Explicitly, we find that the    order-$\Delta$ gauge invariants built from the fermion zero modes contain terms like
\begin{eqnarray} \label{fermionzeromodedensity}\nonumber
&& \sum_{C',D} \lambda_{1 \; C'D} \lambda_{2 \; DC'}  \sim  \\
&& \sum_{C'} \bar\eta_1^{C'} \bar\eta_2^{C'}  \bigg| \sum\limits_{m} e^{i\frac{2 \pi m}{L_2} (x_2 + {L_1 L_2 \over 2 \pi} \hat\phi_1^{C'})   -\frac{  \pi  }{  L_1 L_2}\left[x_1-\frac{  L_1 L_2}{2\pi  }\hat \phi_2^{C'}  -\frac{L_1 C'}{r} + L_1 \frac{1+r}{2r}- L_1 m \right]^2} \bigg|^2\times \nonumber \\
\nonumber
&&   \bigg| \sum\limits_{n}   \left( x_3-\frac{\ell L_3 L_4}{2\pi } \hat \phi_4^{C'} -L_3 \ell  n - L_3 \frac{  1+\ell}{2} \right) e^{i \frac{2\pi n}{\ell L_4} (x_4 + {\ell L_3 L_4 \over 2 \pi} \hat\phi_3^{C'}) -\frac{\pi }{\ell L_3 L_4}\left[x_3-\frac{\ell L_3 L_4}{2\pi } \hat \phi_4^{C'} -L_3\left( \ell n  +\frac{1+\ell}{2}\right)\right]^2  } \bigg|^2\,.\\ \end{eqnarray}
This expression shows the same ``localization'' properties (determined by the holonomies $\hat\phi^{C'}$) of the fermion zeromodes that were made evident for the bosonic solution in (\ref{fractional1}). It is also clear that the $C^{' \text{th}}$ fermion zero mode vanishes at the position determined by the $C^{' \text{th}}$ holonomy.

\bigskip
 
{\bf{\flushleft Acknowledgements:}} We would like to thank  F. David Wandler for comments on the manuscript. M.A. acknowledges the hospitality
of the University of Toronto, where this work was completed.  M.A. is supported by STFC through grant ST/T000708/1. E.P. is supported by a Discovery Grant from NSERC.

\appendix

\section{Derivation of the off-diagonal fermion zero modes}
\label{appx:offdiagonalfermion}

\subsection{The zero modes at zero holonomy}

Within this appendix, we present the derivation of one of the main results in the main text, denoted as Eq. (\ref{lambdazeromode1}). Our objective revolves around solving the off-diagonal fermion zero modes of the Dirac equation $D_\mu\bar\sigma^\mu \lambda=0$. This equation pertains to the 't Hooft flux background, wherein the covariant derivative takes the form $D_\mu=\partial_\mu+i[A_\mu,\,]$. To streamline our approach, we commence by deactivating the holonomies. Subsequently, we can reintroduce them once we have obtained a general solution.

Using  (\ref{r over N abelian sol}) and writing $A_\mu\equiv A_\mu^\omega\omega$, we find the commutator
\begin{eqnarray}
[A_\mu, \lambda]=2\pi A_\mu^\omega \left[\begin{array}{cc}0& N||\lambda_{C'C}||\\-N||\lambda_{CC'}||&0\end{array}\right]\,,
\label{the first commutator}
\end{eqnarray}
In this appendix we take the range of $C$ and $C'$ to be $C=1,2,...,\ell$ and $C'=1,2,...,k$.
Substituting the above result into the Dirac equation, $D_\mu\bar\sigma^\mu \lambda=0$, we obtain for $\lambda_{C'C}$ (and similarly for $\lambda_{CC'}$ after replacing $+ i 2\pi N \rightarrow - i 2\pi N $):
\begin{eqnarray}
\bar \sigma^\mu\left[\partial_\mu\lambda_{C'C}+ i 2\pi N A_\mu^\omega \lambda_{C'C}\right]=0\,.
\end{eqnarray}
Writing $\lambda_{C'C}$ in terms of its two spinor components $\lambda_{C'C\;1}$ and $\lambda_{C'C\;2}$, the Dirac equation reads:
\begin{eqnarray}
\nonumber
\left(\partial_1-i\partial_2-\frac{2\pi r x_1}{k L_1L_2}\right)\lambda_{C'C\;2}+\left(\partial_3+i\partial_4+\frac{2\pi  x_3}{\ell L_3L_4}\right)\lambda_{C'C\;1}&=&0\,,\\
\left(\partial_1+i\partial_2+\frac{2\pi r x_1}{k L_1L_2}\right)\lambda_{C'C\;1}+\left(-\partial_3+i\partial_4+\frac{2\pi  x_3}{\ell L_3L_4}\right)\lambda^{\bm\beta}_{C'C\;2}&=&0\,.
\label{semi-final beta lambda}
\end{eqnarray}
A normalizable solution to the above equations can be found provided that we set $\lambda_{C'C\;2}=0$; an assertion that will be revisited below in the light of the most general normalizable solution on $\mathbb T^4$ we shall construct.   

We proceed further by defining the functions $U_{C'C}$ via:
\begin{eqnarray}
\lambda_{C'C\;1}\equiv e^{-\frac{\pi r x_1^2}{k L_1L_2}}e^{-\frac{\pi x_3^2}{\ell L_3L_4}}U_{C'C}\,,
\end{eqnarray}
which reduces (\ref{semi-final beta lambda}) to the two simple equations
\begin{eqnarray}
\left(\partial_1+i\partial_2\right)U_{C'C}=0\,,\quad \left(\partial_3+i\partial_4\right)U_{C'C}=0\,.
\label{the first two equations}
\end{eqnarray}
By defnining the complex variables $w_1\equiv x_1+ix_2$ and $w_2\equiv x_3+ix_4$, we can cast (\ref{the first two equations}) in the form
\begin{eqnarray}
\frac{\partial U_{C'C}}{\partial\bar w_1}=0\,,\quad \frac{\partial U_{C'C}}{\partial\bar w_2}=0\,,
\end{eqnarray}
and, thus, we conclude that $U_{C'C}$ is an analytic function of $w_1$ and $w_2$:
\begin{eqnarray}
U_{C'C}=U_{C'C}(w_1,w_2)\,.
\end{eqnarray}

We can also write the boundary conditions (\ref{BCS lambda beta}) as (the cyclic nature of the matrix elements, i.e., $U_{C'C}\equiv U_{C'+k; C+\ell}$ will be imposed below):
\begin{eqnarray}
\nonumber
U_{C'C}(w_1+L_1,w_2)&=&\gamma_k^{-r}e^{\frac{\pi r L_1}{k L_2}+\frac{2\pi r w_1}{k L_2}}U_{C'-r\;C}(w_1,w_2)\,,\\
\nonumber
U_{C'C}(w_1+iL_2,w_2)&=&\gamma_k e^{i\frac{2\pi (C'-1)}{k}}U_{C'C}(w_1,w_2)\,,\\
\nonumber
U_{C'C}(w_1,w_2+L_3)&=&\gamma_\ell^{-1}e^{\frac{\pi L_3}{\ell L_4}+\frac{2\pi  w_2}{\ell L_4}}U_{C'\;C+1}(w_1,w_2)\,,\\
U_{C'C}(w_1,w_2+iL_4)&=&\gamma_\ell^{-1} e^{-i\frac{2\pi (C-1)}{\ell}}U_{C'C}(w_1,w_2)\,.
\label{BCS in w1 and w2}
\end{eqnarray}
We notice that the transformation properties under imaginary shifts of $w_1$ by $iL_2$ and $w_2$  by $i L _4$ are satisfied by writing $U_{C'C}(w_1, w_2)$ as the phase factor \begin{equation}e^{ {w_1 \over L_2}{\pi \over k}(2 C' - 1 - k)  -  {w_2 \over L_4}{\pi \over \ell}(2 C - \ell - 1)  }\end{equation} times an analytic function which is periodic w.r.t. these imaginary shifts, i.e., is a linear combination of functions $e^{ 2 \pi n {w_1 \over L_2} + 2 \pi m {w_2 \over L_4}}$ where $n, m \in \Z$.\footnote{The periodicity in imaginary shifts requires the exponential dependence, while the rest follows by holomorphy. The functions $e^{2\pi n \frac{w_2}{L_4}}$  are normalizable on $\mathbb T^2$, and the ones with different $n$'s are orthogonal, as enforced by the imaginary part of integrals over $x_2$.} Thus, the expression for $U_{C'C}$ has the form
%To proceed, we recall $\gamma_k=e^{i\frac{\pi (1-k)}{k}}$ and $\gamma_\ell=e^{i\frac{\pi (1-\ell)}{\ell}}$ and write down the most general series expansion of $U_{C'C}$ that is analytic in $w_1,w_2$ and satisfies the second and fourth BCs in (\ref{BCS in w1 and w2}):
%
\begin{eqnarray}
U_{C'C}(w_1,w_2)=e^{\frac{\pi w_1 (2C'-1-k)}{k L_2}}e^{-\frac{\pi w_2(2C-1-\ell)}{\ell L_4}}\sum_{m,n \in \mathbb Z}d_{C',C,m,n}e^{2\pi m\frac{w_1}{L_2}+2\pi n \frac{w_2}{L_4}}\,.
\label{series for U}
\end{eqnarray}

Our next task is determining the coefficients $d_{C',C,\,m,n}$. Using the first and third BCs in (\ref{BCS in w1 and w2}), we obtain the recurrence relations
\begin{eqnarray}
d_{C',C,m,n}&=&e^{-i\frac{\pi r(1-k)}{k}}e^{-\frac{\pi L_1 (2C'-1-k)}{k L_2}-\frac{2\pi m L_1}{L_2}+\frac{\pi r L_1}{k L_2}}d_{C'-r,C,m,n}\,,
\label{general relation Cp}
\end{eqnarray}
and
\begin{eqnarray}
d_{C',C+1,m,n}&=&e^{i\frac{\pi(1-\ell)}{\ell}}e^{\frac{\pi (-2C+(2n+1)\ell)L_3}{\ell L_4}}d_{C',C,m,n}\,.
\label{general relation C}
\end{eqnarray}

These recurrence relations must be supplemented with boundary conditions that need to be satisfied to guarantee the cyclic nature of the solution, i.e., $U_{C'C}(w_1,w_2)=U_{C'+k\;C}(w_1,w_2)=U_{C'\;C+\ell}(w_1,w_2)$. First, using $U_{C'1}(w_1,w_2)=U_{C'\;1+\ell}(w_1,w_2)$ along with the third equation in  (\ref{BCS in w1 and w2}), we  obtain the following relationship between the elements $C=1$ and $C=\ell$ in $SU(\ell)$:
\begin{eqnarray}
U_{C'\;C=\ell}(w_1,w_2+L_3)=\gamma_\ell^{-1}e^{\frac{\pi L_3}{\ell L_4}}e^{\frac{2\pi w_2}{\ell L_4}}U_{C'\; C=1}(w_1,w_2)\,,
\end{eqnarray}
which yields via (\ref{series for U}):
\begin{eqnarray}
d_{C',\ell,m,n}=e^{-i\frac{\pi(\ell-1)}{\ell}}e^{\frac{\pi (1-2n)L_3}{L_4}}d_{C',1,m,n-1}\,.
\label{n relation}
\end{eqnarray}
We can generalize (\ref{general relation C}) and (\ref{n relation}) to
\begin{eqnarray}
\nonumber
d_{C',C,m,n}&=&e^{-i\frac{\pi(1-\ell)}{\ell}}e^{\frac{-\pi (-2C+(2n+1)\ell)L_3}{\ell L_4}}d_{C',C+1,m,n}\,,\quad \mbox{if}\,\, C+1<\ell\,\\
\nonumber
d_{C',C,m,n}&=&e^{-i\frac{\pi(1-\ell)}{\ell}}e^{\frac{-\pi (-2C+(2n+1)\ell)L_3}{\ell L_4}}d_{C',C_{\scriptsize \mbox{new}},m,n-1}\,,\quad C_{\scriptsize \mbox{new}}=C+1-\ell\quad \mbox{if}\,\, C+1>\ell\,.\\
\label{combined relation C}
\end{eqnarray}

We must also find the boundary condition for the recurrence relation (\ref{general relation Cp}). Using $U_{1C}(w_1,w_2)=U_{1+k\; C}(w_1,w_2)$ along with the first equation in (\ref{BCS in w1 and w2}),  we obtain the following relationship between the elements $C'=1$ and $C'=k-(r-1)$ in $SU(k)$:
\begin{eqnarray}
U_{C'=1\;C}(w_1+L_1,w_2)=\gamma_k^{-r}e^{\frac{\pi r L_1}{k L_2}}e^{\frac{2\pi r}{k L_2}w_1}U_{C'=k-(r-1)\;C}(w_1,w_2)\,,
\end{eqnarray}
 which yields via (\ref{series for U}):
\begin{eqnarray}
d_{1,C,m,n}=e^{-i\frac{\pi r(1-k)}{k}}e^{\frac{\pi(r-1+k-2mk)}{k}\frac{L_1}{L_2}}d_{k-(r-1),C,m-1,n}\,.
\label{m relation}
\end{eqnarray}
Notice that we had to shift $m$ by one unit since, according to the first equation in (\ref{BCS in w1 and w2}),  a shift in the $L_1$ direction relates the element $C'=1$ to the element $C'=1-r$.  However, since $1-r\leq0$, we needed to replace $C'=1-r$ by a new $C'_{\scriptsize\mbox{new}}=k-(r-1)$. This replacement forces us to shift $m\rightarrow m-1$ to obey the boundary condition (\ref{BCS in w1 and w2}) in the $L_1$ direction.  This shift in $m$ always happens whenever the matrix elements have  $C'-r\leq 0$. We may generalize (\ref{m relation}) for any $C'$ whenever the first condition (\ref{general relation Cp}) yields $d_{C'=C-r,C,m,n}$ with $C'<0$. Demanding the cyclicity $U_{C'+k\;C}(x)=U_{C'\;C}(x)$,  we replace 
(\ref{general relation Cp}) and (\ref{m relation}) with
 \begin{eqnarray}
 \nonumber
d_{C',C,m,n}&=&e^{-i\frac{\pi r(1-k)}{k}}e^{-\frac{\pi L_1 (2C'-1-k)}{k L_2}-\frac{2\pi m L_1}{L_2}+\frac{\pi r L_1}{k L_2}}d_{C'-r,C,m,n}\,,\quad \mbox{if}\,\, C'-r>0\,,\\
\nonumber
 d_{C',C,m,n}&=&e^{-i\frac{\pi r(1-k)}{k}}e^{-\frac{\pi L_1 (2C'-1-k)}{k L_2}-\frac{2\pi m L_1}{L_2}+\frac{\pi r L_1}{k L_2}} d_{C'_{\scriptsize\mbox{new}},C, m-1,n}\,,\\
 &&\quad\quad\quad\quad\quad\quad\quad\quad\quad\quad C'_{\scriptsize\mbox{new}}=C'-r+k\,,\quad  \mbox{if}\,\, C'-r\leq 0\,. 
\label{combined relation Cp}
\end{eqnarray}

Now we come to the solution of the difference equation (\ref{combined relation C}). This is a first-order difference equation, and thus, it yields a single solution. To this end, we substitute the following form
\begin{eqnarray}
d_{C',C,m,n}=F(C',m)e^{-\frac{\pi L_3}{\ell L_4}\left[C+S(n)\right]^2}
\end{eqnarray}
into the first equation in (\ref{combined relation C}), to obtain 
\begin{eqnarray}
S(n)=-\frac{1+(2n+1)\ell}{2}\,.
\end{eqnarray}
Thus,
\begin{eqnarray}
d_{C',C,m,n}=F(C',m)e^{-\frac{\pi L_3}{\ell L_4} \left(C-i\frac{L_4(1-\ell)}{2L_3}-\frac{1+\ell (2n+1)}{2}\right)^2}\,.
\label{solution for C}
\end{eqnarray}
It is straightforward to check that the solution (\ref{solution for C}) obeys  (\ref{combined relation C}).

On the other hand, the recurrence relation (\ref{combined relation Cp}) is a difference equation of order $r$, and thus, it should yield $r$ independent solutions. To solve it, we parameterize it as
\begin{eqnarray}
d_{C',C,m,n}=e^{-\frac{\pi L_1}{k r L_2}\left(C'+i\frac{ L_2r(1-k)}{2L_1}+S'(m)\right)^2}\,,
\label{ relation CP}
\end{eqnarray}
and, inserting into the first equation in (\ref{combined relation Cp}), we find%
\begin{eqnarray}
S'(m)=-\frac{1+k(1-2m)}{2}\,.
\label{solution for Cp}
\end{eqnarray}
We can check that (\ref{ relation CP}, \ref{solution for Cp}) satisfy   (\ref{combined relation Cp}). Combining (\ref{solution for C}) and (\ref{ relation CP}), we obtain the final answer
\begin{eqnarray}
d_{C',C,m,n}=e^{-\frac{\pi L_3}{\ell L_4} \left(C-i\frac{(1-\ell)L_4}{2L_3}-\frac{1+\ell (2n+1)}{2}\right)^2}e^{-\frac{\pi L_1}{k r L_2}\left(C'+i\frac{ r(1-k)L_2}{2L_1}-\frac{1+k(1-2m)}{2}\right)^2}\,.
\label{final solution of d}
\end{eqnarray}

Notice that $d_{C',C,m,n}\rightarrow e^{-\frac{\pi L_3 }{ L_4}\ell n^2}e^{-\frac{\pi  L_1 }{ rL_2}k m^2}$ as $n, m\rightarrow \infty$, and thus, the series (\ref{series for U}) rapidly converges. Had we not set $\lambda_{C' C\; 2}=0$ in (\ref{semi-final beta lambda}), we would have obtained a divergent series in $m,n$, and thus, no normalizable zero modes could be found. 

What is not immediately clear from (\ref{final solution of d}) is that there are $r$ independent solutions of $U_{C'C}$; this should be expected since (\ref{combined relation Cp}) is a difference equation of order $r$.  The $r$ independent solutions of $U_{C'C}$ follow from two distinct cases.
 \begin{enumerate}
 \item If $\mbox{gcd}(r,k)=r$, we can show that there are $r$ independent coefficients
 \begin{equation}
 d_{C'=1,C,m,n},\; d_{C'=2,C,m,n},\;...,d_{C'=r,C,m,n}\,,
 \end{equation}
and the sums over $m, n$ in  (\ref{series for U}) are over all integers. The $r$ independent coefficients yield $r$ independent solutions.
 \item If $\mbox{gcd}(r,k)=1$ and $r>1$, then the set of integers $m$ in  (\ref{series for U}) bifurcates into $r$ sets such that the sum over $m\in \mathbb Z$ in (\ref{series for U}) is divided into  $m_j=n_jr+n$, $n_j\in \mathbb Z$, $n=0,1,..,r-1$. These form $r$ independent orbits that correspond to $r$ independent solutions.    
\end{enumerate}
The general situation, $1<\mbox{gcd}(r,k)<r$, is a combination of both cases.

 To ease our discussion, we consider a few examples to understand the essence of each case. First, consider case 1 above, and take as an example $k=6, r=2$, where $\mbox{gcd}(6,2)=2$.  Using (\ref{combined relation Cp}), we see that the coefficients $d_{C', C,m,n}$ are related via (here we ignore $C$ and $n$ since they do not play a role. Also the arrow indicates the relations between the coefficients as we traverse the $L_1$ direction, without caring about the pre-coefficients):
\begin{eqnarray}
\nonumber
&&d_{1,m}\rightarrow d_{5,m-1}\rightarrow d_{3,m-1}\rightarrow d_{1,m-1}\rightarrow d_{5,m-2}\rightarrow ...\,,\\
&&d_{2,m}\rightarrow d_{6,m-1}\rightarrow d_{4,m-1}\rightarrow d_{2,m-1}\rightarrow d_{6,m-2}\rightarrow ...\,.
\end{eqnarray}
Each line depicts a set of coefficients, and the coefficients of line 1 and line 2 are independent in that a coefficient in line 1 cannot be reached via a coefficient in line 2 and vice versa. Notice also, for example, as we start from $d_{1,m}$ and traverse the $L_1$ direction $3$ times, we obtain the shifted $d_{1,m-1}$ by one unit. Thus, we need to sum over all integers $m$ in every line. This gives the two independent solutions. 

Next, consider case 2. For example, take $k=6, r=5$, where $\mbox{gcd}(k,r)=1$. Applying (\ref{combined relation Cp}) we find
\begin{eqnarray}
d_{1,m}\rightarrow d_{2,m-1}\rightarrow d_{3,m-2}\rightarrow d_{4,m-3}\rightarrow d_{5,m-4}\rightarrow d_{6,m-5}\rightarrow d_{1,m-5}\rightarrow d_{2,m-6}....
\end{eqnarray}
Thus, the fact that $d_{1,m}$ shifts to $d_{1,m-5}$ and $d_{2,m-1}$ to $d_{2,m-6}$, etc. means that the set of integers $m$ bifurcates into $5$ sets: $m=5m'+p$, $p=0,1,2,3,4$ and $m'\in \mathbb Z$. Thus, we obtain $5$ independent orbits corresponding to $5$ independent solutions. 

 Finally, consider the general case $1<\mbox{gcd}(r,k)<r$, and take, for example, $k=6, r=4$, where $\mbox{gcd}(6,4)=2$. Here, we find
\begin{eqnarray}
\nonumber
&&d_{1,m}\rightarrow d_{3,m-1}\rightarrow d_{5,m-2}\rightarrow d_{1,m-2}\rightarrow....\,,\\
&&d_{2,m}\rightarrow d_{4,m-1}\rightarrow d_{6,m-2}\rightarrow d_{2,m-2}\rightarrow...\,.
\end{eqnarray}
The two lines depict independent solutions. However, we also find that there are independent orbits within each line. For example, $d_{1,m}$ shifts to $d_{1,m-2}$, etc. Thus, the integers are divided into two sets, odd and even. We conclude that there are two orbits in each line, and in total, we have $4$ independent solutions, as expected.  In this general case, we find that a simple relation gives the $r$ solutions:
\begin{eqnarray}
\nonumber
r=\underbrace{\mbox{gcd}(k,r)}_{\mbox{number of vertical lines, case { (1)} }}\times \underbrace{\frac{r}{\mbox{gcd}(k,r)}}_{\mbox{independent orbits, case {(2)} }}\,.\\
\end{eqnarray}

 It is best to cast the above findings in a more effective compact notation. To this end, we define the functions:
\begin{eqnarray}
\nonumber
\tilde\Phi_{C'C}^{(p)}(x)&\equiv& e^{-\frac{\pi r x_1^2}{k L_1L_2}}e^{-\frac{\pi x_3^2}{\ell L_3L_4}} \sum_{\scriptsize m=p+\frac{rm'}{\mbox{gcd}(k,r)},\, m'\in \mathbb Z}\sum_{n'\in \mathbb Z}\\
\nonumber
&&\times e^{-\frac{\pi L_3}{\ell L_4} \left(C-i\frac{(1-\ell)L_4}{2L_3}-\frac{1+\ell (2n'+1)}{2}\right)^2}e^{-\frac{\pi L_1}{k r L_2}\left(C'+i\frac{ r(1-k)L_2}{2L_1}-\frac{1+k(1-2m)}{2}\right)^2}\\
&&\times e^{\frac{2\pi w_1 }{L_2}(m+\frac{2	C'-1-k}{2k})}e^{\frac{2\pi w_2 }{L_4}(n'-\frac{2	C-1-\ell}{2\ell})}\,,
\label{main def of Phi}
\end{eqnarray}
for $p=0,1,..., \frac{r}{\mbox{gcd}(k,r)}-1$. Thus, there are $\frac{r}{\mbox{gcd}(k,r)}$ independent solutions correspond to $\frac{r}{\mbox{gcd}(k,r)}$ independent orbits.  We can also rewrite $\tilde \Phi^{(p)}_{C',C}$ conveniently as
\begin{eqnarray}
\nonumber
\tilde \Phi_{C'C}^{(p)}(x)&=&\sum_{\scriptsize m=p+\frac{rm'}{\mbox{gcd}(k,r)},\, m'\in \mathbb Z}\sum_{n'\in \mathbb Z}\left\{e^{i\frac{2\pi x_2 }{L_2}(m+\frac{2	C'-1-k}{2k})}e^{i\frac{2\pi x_4 }{L_4}(n'-\frac{2	C-1-\ell}{2\ell})}\right.\\
\nonumber
&&\times e^{\frac{\pi r (1-k)^2L_2}{4k L_1}-i\frac{\pi (1-k)}{k}\left(C'-\frac{1+k(1-2m)}{2}\right)}\times e^{\frac{\pi(1-\ell)^2L_3}{4\ell L_4}+i\frac{\pi(1-\ell)}{\ell}\left(C-\frac{1+\ell(2n'+1)}{2}\right)}\\
&&\left.\times e^{-\frac{\pi r}{k L_1L_2}\left(x_1-\frac{L_1\left(2mk+2C'-1-k\right)}{2r}\right)^2}e^{-\frac{\pi}{\ell L_3L_4}\left(x_3-\frac{L_3\left((2n'+1)\ell-(2C-1)\right)}{2}\right)^2}\right\}\,.
\label{second main def of Phi}
\end{eqnarray}
Since the terms $ e^{\frac{\pi r (1-k)^2L_2}{4k L_1}}$ and $e^{\frac{\pi(1-\ell)^2L_3}{4\ell L_4}}$ are independent of $m,n,C,C'$, we may drop them and define the function $\Phi_{C'C}^{(p)}(x)$ as:
\begin{eqnarray}
\nonumber
 \Phi_{C'C}^{(p)}(x)&\equiv&\sum_{\scriptsize m=p+\frac{rm'}{\mbox{gcd}(k,r)},\, m'\in \mathbb Z}\sum_{n'\in \mathbb Z}\left\{e^{i\frac{2\pi x_2 }{L_2}(m+\frac{2	C'-1-k}{2k})}e^{i\frac{2\pi x_4 }{L_4}(n'-\frac{2	C-1-\ell}{2\ell})}\right.\\
\nonumber
&&\times e^{-i\frac{\pi (1-k)}{k}\left(C'-\frac{1+k(1-2m)}{2}\right)}\times e^{i\frac{\pi(1-\ell)}{\ell}\left(C-\frac{1+\ell(2n'+1)}{2}\right)}\\
&&\left.\times e^{-\frac{\pi r}{k L_1L_2}\left(x_1-\frac{L_1\left(2mk+2C'-1-k\right)}{2r}\right)^2}e^{-\frac{\pi}{\ell L_3L_4}\left(x_3-\frac{L_3\left((2n'+1)\ell-(2C-1)\right)}{2}\right)^2}\right\}\,.
\label{second main def of Phi prime}
\end{eqnarray}

The functions $\Phi_{C'C}^{(p)}(x)$ solve the equation
\begin{eqnarray}
\bar \sigma^\mu\left[\partial_\mu\Phi_{C'C}^{(p)}+ i 2\pi N A_\mu^\omega \Phi_{C'C}^{(p)}\right]=0\,,
\end{eqnarray}
 and satisfy the boundary conditions
\begin{eqnarray}
\nonumber
\Phi_{C'C}^{(p)}(x+\hat e_1 L_1)&=&e^{-i\frac{\pi r(1-k)}{k}} e^{i\frac{2\pi r x_2}{k L_2}}\Phi_{[C'-r]_k\;C}^{(p)}(x)\,,\\
\nonumber
\Phi_{C'C}^{(p)}(x+\hat e_2 L_2)&=&  e^{i \frac{2\pi (2C'-1-k)}{2k}}\Phi_{C'C}^{(p)}(x)\,,\\
\nonumber
\Phi_{C'C}^{(p)}(x+\hat e_3 L_3)&=&e^{-i\frac{\pi (1-\ell)}{\ell}}e^{i\frac{2\pi x_4}{\ell L_4}}\Phi_{C'\;[C+1]_\ell}^{(p)}(x)\,,\\
\Phi_{C'C}^{(p)}(x+\hat e_4 L_4)&=&  e^{-i \frac{2\pi (2C-1-\ell)}{2\ell}}\Phi_{C'C}^{(p)}(x)\,,
\label{the formalized way to the BCS of Phi}
\end{eqnarray}
which are the exact same boundary conditions (\ref{BCS lambda beta}). The entries with $C'=j,j+\mbox{gcd}(k,r),j+2\mbox{gcd}(k,r),..., j+k-\mbox{gcd}(k,r)$, for every $j=1,2,...,\mbox{gcd}(k,r)$, are shuffled to each other as we traverse the $L_1$ direction. Thus, the rows with $C'=1,2,...,\mbox{gcd}(k,r)$ are independent. In total, there are $\mbox{gcd}(k,r)\times \frac{r}{\mbox{gcd}(k,r)}=r$ independent solutions. In addition, $\Phi^{(p)}_{C'C}$ satisfy the cyclic properties:
\begin{eqnarray}
\nonumber
\Phi^{(p)}_{C'+k\;C}(x)=\Phi^{(p+1)}_{C'C}(x)\,,\\
\Phi^{(p)}_{C'C}(x)=\Phi^{\scriptsize \left(p+\frac{r}{\mbox{gcd}(k,r)}\right)}_{C'C}(x)\,.
\label{cyclic BC}
\end{eqnarray}
Notice the intertwining between the shift in $p$ by $1$ and $C'$ by $k$. 

We can use (\ref{cyclic BC}), noticing the intertwining between the shift in $p$ and $C'$, to write the $r$ independent zero modes of the Dirac equation as
\begin{eqnarray}
\lambda_{C'C}(x)=\sum_{p=0}^{\scriptsize \frac{r}{\mbox{gcd}(k,r)}-1}\left[\begin{array}{c} \eta^{[C'+pk]_r}\\0\end{array}\right]\Phi^{(p)}_{C'C}(x)\,,
\label{the grand final solution}
\end{eqnarray}
where $[x]_r\equiv x\, \mbox{mod}\, r$, and it is obvious that $\eta^{[C'+pk]_r}$ yields $r$ independent coefficients. This is the desired equation (\ref{lambdazeromode1}) without holonomies.

%%%%%%%%%%%%%%%%%%%%%
\subsection{Turning on holonomies}
%%%%%%%%%%%%%%%%%%%%%%%

Next, we turn on the $SU(k)$ space holonomies. In particular, the gauge field is now given by
\begin{eqnarray}
A_\mu=-\left[\hat A_\mu ^\omega +\phi_\mu\right]\omega+ H^{a'} \phi_\mu^{a'}\,,
\end{eqnarray}
where $\phi_\mu=z_\mu/L_\mu$ are the abelian holonomies, $H^{a'}$, $a'=1,2,...,k-1$ are the $k-1$ Cartan generators of the $su(k)$ algebra, and $\phi_\mu^{a'}$ are $k-1$ holonomies in every direction $\mu=1,2,3,4$. Next,  we need to compute the commutator:
\begin{eqnarray}
[ H^{a'} \phi_\mu^{a'},||\lambda||_{C'C}]=\left(H^{a'}\phi_\mu^{a'}\right)_{C'C'}\lambda_{C'C}\equiv \phi^{C'}_\mu\lambda_{C'C}\,.
\end{eqnarray}
Recalling (\ref{the first commutator}), we find it convenient to define 
\begin{eqnarray}
\hat \phi_\mu^{C'}=\phi_\mu^{C'}-2\pi N \phi_\mu\,.
\end{eqnarray}
Noticing that $A_\mu$ has to commute with the transition functions, then out of $k$ holonomies, there are at most $\mbox{gcd}(k,r)$ holonomies in every spacetime direction. Thus, we find that $\hat \phi_\mu^{C'}=\hat \phi_\mu^{C'+r}$, or we can express this fact as
\begin{eqnarray}
\hat \phi_\mu^{C'}=\hat \phi_\mu^{[C']_r}\,.
\end{eqnarray}
Using the above information in the Dirac equation $\bar\sigma^\mu D_\mu \lambda=0$, we find (compare with (\ref{semi-final beta lambda}))
\begin{eqnarray}
\nonumber
\left(\partial_3+i\hat \phi_3^{C'}+i\partial_4-\hat \phi_4^{[C']_r}+\frac{2\pi x_3}{\ell L_3 L_4}\right)\lambda_{1C',C}=0\,,\\
\left(\partial_1+i\hat \phi_1^{C'}+i\partial_2-\hat \phi_2^{[C']_r}+\frac{2\pi r x_1}{k L_1 L_2}\right)\lambda_{1C',C}=0\,.
\label{EOM with holonomies}
\end{eqnarray}
and we have set $\lambda_{C'C\; 2}=0$, as in (\ref{semi-final beta lambda}). 

Next, we use the field redefinition
\begin{eqnarray}
\lambda_{C'C\; 1}=e^{-\frac{\pi r x_1^2}{k L_1L_2}}e^{-\frac{\pi  x_3^2}{ L_3L_4}}e^{-ix_\mu\hat \phi_\mu^{[C']_r}} U_{C'C}
\end{eqnarray}
in (\ref{EOM with holonomies}) to find that $U_{C'C}$ obeys the equations
\begin{eqnarray}\label{analyticequations}
\left(\partial_1+i\partial_2\right) U_{C'C}=0\,,\quad \left(\partial_3+i\partial_4\right)U_{C'C}=0\,.
\end{eqnarray} 
These equations, as before, imply that  $U_{C'C}$ is an analytic function of $w_1\equiv x_1+ix_2$ and $w_2\equiv x_3+ix_4$. 

The BCS (\ref{BCS lambda beta}) can be rewritten in terms of the functions $U_{C'C}$:
\begin{eqnarray}
\nonumber
U_{C' C}(w_1 + L_1, w_2)&=& \gamma_k^{-r}e^{ {\pi r L_1 \over k L_2}  + {2 \pi r \over kL_2} w_1 + i L_1 \hat\phi_1^{[C']_r}  } \; U_{C'-r\; C}(w_1, w_2)\,,\\
\nonumber
 U_{C' C}(w_1+ i L_2, w_2)&=&  e^{i{   \pi \over k} (2 C' -1 - k ) + i L_2 \hat\phi_2^{[C']_r}}\; U_{C' C}(w_1,w_2)\,,\\
\nonumber
 U_{C' C}(w_1, w_2+L_3)&=&\gamma_\ell^{-1}e^{ {\pi L_3 \over \ell L_4} + {2 \pi \over \ell L_4} w_2  + i L_3 \hat\phi_3^{[C']_r}} \;U_{C'\; C+1}(w_1,w_2)\,,\\
 U_{C' C}(w_1, w_2+ i L_4)&=& \  e^{- i   {  \pi \over \ell} (2 C - \ell - 1) + i L_4 \hat\phi_4^{[C']_r}} \;  U_{C' C}(w_1, w_2)\,.
\label{BCS bar U analytic}
\end{eqnarray}
Similar to (\ref{series for U}), the transformation properties under imaginary shifts of $w_1$ by $iL_2$ and $w_2$  by $i L _4$ are satisfied by writing $U_{C'C}(w_1, w_2)$ as the phase factor \begin{equation}e^{ {w_1 \over L_2}{\pi \over k}(2 C' - 1 - k) + w_1  \hat \phi_2^{[C']_r}  -  {w_2 \over L_4}{\pi \over \ell}(2 C - \ell - 1) + w_2 \hat\phi_4^{[C']_r} }\end{equation} times an analytic function which is periodic w.r.t. these imaginary shifts. Thus, the expression for $U_{C'C}$ takes the form
\begin{eqnarray}
\bar U_{C',C}(w_1,w_2)=e^{w_1 \hat \phi_2^{[C']_r} + w_2 \hat \phi_4^{[C']_r}+ \frac{\pi w_1}{k L_2} (2C'-1-k) -\frac{\pi w_2}{\ell L_4}(2C-1-\ell)}\sum_{m,n \in \Z}d_{C',C,m,n}\;e^{2\pi m\frac{w_1}{L_2}+2\pi n \frac{w_2}{L_4}}\,,\nonumber \\
\label{series for barU}
\end{eqnarray}
which differs from (\ref{series for U}) by the prefactor $e^{w_1 \hat \phi_2^{[C']_r} + w_2 \hat \phi_4^{[C']_r}}$.

As we proceed in the absence of holonomies, our next step involves determining the coefficients $d_{C',C,m,n}$ by utilizing the first and third boundary conditions in (\ref{BCS bar U analytic}). These conditions lead to the following recurrence relations:
\begin{eqnarray}
d_{C',C,m,n}&=&e^{-i\frac{\pi r(1-k)}{k}}e^{-\frac{\pi L_1}{k L_2} (2C'-r - 1 + (2 m - 1)k) } e^{i L_1 (\hat \phi_1^{[C']_r} + i \hat \phi_2^{[C']_r})}\; d_{C'-r,C,m,n}\,,
\label{general relation Cprime with H}
\end{eqnarray}
and
\begin{eqnarray}
d_{C',C ,m,n}&=&e^{-i\frac{\pi(1-\ell)}{\ell}}e^{\frac{\pi L_3}{\ell L_4} ( 2C-(2n+1)\ell) }  e^{i L_3 (\hat \phi_3^{[C']_r} + i \hat \phi_4^{[C']_r})}\; d_{C',C+1,m,n}\,.
\label{general relation C with H}
\end{eqnarray}
We observe that (\ref{general relation Cprime with H}) and (\ref{general relation C with H}) become identical to (\ref{general relation Cp}) and (\ref{general relation C}) respectively, when we replace:
\begin{eqnarray}
\nonumber
m&\longrightarrow& m-\frac{i L_2}{2\pi}\left(\hat \phi_1^{[C']_r}+i\hat \phi_2^{[C']_r}\right)\,,\\
n&\longrightarrow& n-\frac{i L_4}{2\pi}\left(\hat \phi_3^{[C']_r}+i\hat \phi_4^{[C']_r}\right)\,,
\label{m and n replacement}
\end{eqnarray}
in (\ref{general relation Cp}) and (\ref{general relation C}). Consequently, the solution to (\ref{general relation Cprime with H}) and (\ref{general relation C with H}) is identical to (\ref{final solution of d}) after making the replacement (\ref{m and n replacement}):
\begin{eqnarray}
\nonumber
d_{C',C,m,n}&=&\\
\nonumber
&&e^{-\frac{\pi L_3}{\ell L_4} \left[C-i\frac{(1-\ell)L_4}{2L_3}-\frac{1+\ell (2n+1)}{2}+i\frac{\ell L_4}{2\pi}\left(\hat \phi_3^{[C']_r}+i\hat \phi_4^{[C']_r}\right)\right]^2}\\
\nonumber
&&\times e^{-\frac{\pi L_1}{k r L_2}\left[C'+i\frac{ r(1-k)L_2}{2L_1}-\frac{1+k(1-2m)}{2}-i \frac{kL_2}{2\pi}\left(\hat \phi_1^{[C']_r}+i \hat \phi_2^{[C']_r}\right)\right]^2}\,,\\
\label{final solution of d with H}
\end{eqnarray}
and we used the fact that $\phi_\mu^{[C']_r}=\phi_\mu^{[C'-r]_r}$.

Then, all the analyses in the absence of holonomies repeat precisely, with $\tilde\Phi^{(p)}_{C'C}(x,\hat\phi)$ now given by the expression
 \begin{eqnarray}
\nonumber
\tilde\Phi_{C'C}^{(p)}(x, \hat\phi)&\equiv&e^{-ix_\mu\hat \phi_\mu^{[C']_r}}e^{w_1\hat \phi_2^{[C'}]_r} e^{w_2\hat \phi_4^{[C']_r}} e^{-\frac{\pi r x_1^2}{k L_1L_2}}e^{-\frac{\pi x_3^2}{\ell L_3L_4}}  \sum_{\scriptsize m=p+\frac{rm'}{\mbox{gcd}(k,r)},\, m'\in \mathbb Z}\sum_{n'\in \mathbb Z}\\
\nonumber
&&e^{\frac{2\pi w_1 }{L_2}(m+\frac{2	C'-1-k}{2k})}e^{\frac{2\pi w_2 }{L_4}(n'-\frac{2	C-1-\ell}{2\ell})}\\
\nonumber
&&\times e^{-\frac{\pi L_3}{\ell L_4} \left[C-i\frac{(1-\ell)L_4}{2L_3}-\frac{1+\ell (2n'+1)}{2}+i\frac{\ell L_4}{2\pi}\left(\hat \phi_3^{[C']_r}+i\hat \phi_4^{[C']_r}\right)\right]^2}\\
&&\times e^{-\frac{\pi L_1}{k r L_2}\left[C'+i\frac{ r(1-k)L_2}{2L_1}-\frac{1+k(1-2m)}{2}-i \frac{kL_2}{2\pi}\left(\hat \phi_1^{[C']_r}+i \hat \phi_2^{[C']_r}\right)\right]^2}\,,
\label{ holonomy main def of Phi}
\end{eqnarray}
where the tilde service as a reminder that these are not precisely the functions we define in the bulk of the paper. The latter will be defined momentarily. 
Manipulating, we can  rewrite $\tilde \Phi_{C'C}^{(p)}(x, \hat\phi)$ in the more convenient form (easier form for taking derivatives)
\begin{eqnarray}
\nonumber
\tilde \Phi^{(p)}_{C'C}(x,\hat\phi)&=& \sum_{\scriptsize m=p+\frac{rm'}{\mbox{gcd}(k,r)},\, m'\in \mathbb Z}\sum_{n'\in \mathbb Z}e^{\frac{i2\pi x_2 }{L_2}(m+\frac{2	C'-1-k}{2k})}e^{\frac{i2\pi x_4 }{L_4}(n'-\frac{2	C-1-\ell}{2\ell})}\\
\nonumber
&&~~\times e^{\frac{\pi r (1-k)^2L_2}{4k L_1}-i\frac{\pi (1-k)}{k}\left(C'-\frac{1+k(1-2m)}{2}-i \frac{kL_2}{2\pi}\left(\hat \phi_1^{[C']_r}+i \hat \phi_2^{[C']_r}\right)\right)}\\
\nonumber
&&~~\times e^{\frac{\pi(1-\ell)^2L_3}{4\ell L_4}+i\frac{\pi(1-\ell)}{\ell}\left(B-\frac{1+\ell(2n'+1)}{2}+i\frac{\ell L_4}{2\pi}\left(\hat \phi_3^{[C']_r}+i\hat \phi_4^{[C']_r}\right)\right)}\\
\nonumber
&&\times e^{-\frac{\pi r}{k L_1 L_2}\left[x_1-\frac{k L_1 L_2}{2\pi r}(\hat \phi_2^{[C']_r}-i\hat \phi_1^{[C']_r})-\frac{L_1}{r}\left(km +\frac{2C'-1-k}{2}\right)\right]^2}\\
&&\times e^{-\frac{\pi }{\ell L_3 L_4}\left[x_3-\frac{\ell L_3 L_4}{2\pi }(\hat \phi_4^{[C']_r}-i\hat \phi_3^{[C']_r})-L_3\left(\ell n' -\frac{2C-1-\ell}{2}\right)\right]^2}\,.
\label{alternative form of Phi}
\end{eqnarray}

The terms $e^{\frac{\pi r (1-k)^2L_2}{4k L_1}}$, $e^{-i\frac{\pi (1-k)}{k}\left(-i \frac{kL_2}{2\pi}\left(\hat \phi_1^{[C']_r}+i \hat \phi_2^{[C']_r}\right)\right)}$, $e^{\frac{\pi(1-\ell)^2L_3}{4\ell L_4}}$, and $e^{i\frac{\pi(1-\ell)}{\ell}\left(i\frac{\ell L_4}{2\pi}\left(\hat \phi_3^{[C']_r}+i\hat \phi_4^{[C']_r}\right)\right)}$ do not explicitly depend on $C,C',m,n'$, and thus, it is convenient to drop them\footnote{One can show that they can be absorbed into the coefficients $\eta^{[C'+pk]_r}$ of the general solution of the Dirac equation, see (\ref{the grand final solution}) or (\ref{the grand final solution with H}) below. } and define the function $ \Phi^{(p)}_{C'C}(x,\hat\phi)$ as:
\begin{eqnarray}
\nonumber
 \Phi^{(p)}_{C'C}(x,\hat\phi)&\equiv& \sum_{\scriptsize m=p+\frac{rm'}{\mbox{gcd}(k,r)},\, m'\in \mathbb Z}\sum_{n'\in \mathbb Z}e^{\frac{i2\pi x_2 }{L_2}(m+\frac{2	C'-1-k}{2k})}e^{\frac{i2\pi x_4 }{L_4}(n'-\frac{2	C-1-\ell}{2\ell})}\\
\nonumber
&&~~\times e^{-i\frac{\pi (1-k)}{k}\left(C'-\frac{1+k(1-2m)}{2}\right)} e^{i\frac{\pi(1-\ell)}{\ell}\left(C-\frac{1+\ell(2n'+1)}{2}\right)}\\
\nonumber
&&\times e^{-\frac{\pi r}{k L_1 L_2}\left[x_1-\frac{k L_1 L_2}{2\pi r}(\hat \phi_2^{[C']_r}-i\hat \phi_1^{[C']_r})-\frac{L_1}{r}\left(km +\frac{2C'-1-k}{2}\right)\right]^2}\\
&&\times e^{-\frac{\pi }{\ell L_3 L_4}\left[x_3-\frac{\ell L_3 L_4}{2\pi }(\hat \phi_4^{[C']_r}-i\hat \phi_3^{[C']_r})-L_3\left(\ell n' -\frac{2C-1-\ell}{2}\right)\right]^2}\,.
\label{alternative form of Phi}
\end{eqnarray}
We may also write $\Phi^{(p)}_{C'C}(x,\hat\phi)$ in the form
\begin{eqnarray}
\nonumber
 \Phi^{(p)}_{C'C}(x,\hat\phi)&=& e^{\frac{k L_1 L_2}{2\pi r}\hat \phi_1^{[C']_r}\left(i\hat \phi_2^{[C']_r}+\hat \phi_1^{[C']_r}/2\right)}  e^{\frac{\ell L_3 L_4}{2\pi }\hat \phi_3^{[C']_r}\left(i\hat \phi_4^{[C']_r}+\hat \phi_3^{[C']_r}/2\right)}  e^{-i\hat \phi_1^{[C']_r} x_1}  e^{-i\hat \phi_3^{[C']_r} x_3}\\
\nonumber
&&\times  \sum_{\scriptsize m=p+\frac{rm'}{\mbox{gcd}(k,r)},\, m'\in \mathbb Z}\sum_{n'\in \mathbb Z} e^{i \left(\frac{ 2\pi x_2 }{L_2}+\frac{L_1 k}{r}\hat \phi_1^{[C']_r}\right)(m+\frac{2	C'-1-k}{2k})}e^{i\left(\frac{2\pi x_4 }{L_4}+\ell L_3\hat \phi_3^{[C']_r}\right)(n'-\frac{2	C-1-\ell}{2\ell})}\\
\nonumber
&&~~\times e^{-i\frac{\pi (1-k)}{k}\left(C'-\frac{1+k(1-2m)}{2}\right)} e^{i\frac{\pi(1-\ell)}{\ell}\left(C-\frac{1+\ell(2n'+1)}{2}\right)}\\
\nonumber
&&\times e^{-\frac{\pi r}{k L_1 L_2}\left[x_1-\frac{k L_1 L_2}{2\pi r}\hat \phi_2^{[C']_r}-\frac{L_1}{r}\left(km +\frac{2C'-1-k}{2}\right)\right]^2}\\
&&\times e^{-\frac{\pi }{\ell L_3 L_4}\left[x_3-\frac{\ell L_3 L_4}{2\pi }\hat \phi_4^{[C']_r}-L_3\left(\ell n' -\frac{2C-1-\ell}{2}\right)\right]^2}\,.
\label{expression of Phi that shows fractionalization}
\end{eqnarray}

Finally, the fermion zero modes are given by (compare with (\ref{the grand final solution}))
\begin{eqnarray}
\lambda_{C'C}(x)= \sum_{p=0}^{\scriptsize \frac{r}{\mbox{gcd}(k,r)}-1}\left[\begin{array}{c} \eta^{[C'+pk]_r}\\0\end{array}\right]\Phi^{(p)}_{C'C}(x,\hat\phi)\,.
\label{the grand final solution with H}
\end{eqnarray}

\section{A useful identity}
\label{appx:identity1}

Here, we evaluate the expression $I_j^{ab}$ defined in (\ref{Iab1}), $j=0,...,{\rm gcd}(k,r)-1$, repeated here
\begin{eqnarray}\label{Iab1repeated}
I^{ab}_j &=& \sum\limits_{C'=0}^{k-1} \sum\limits_{n=0}^{\frac{k}{{\rm gcd}(k,r)}-1}   \delta_{C', [j+nr]_k} \sum\limits_{p,p'=0}^{{r\over {\rm gcd}(k,r)}-1}
{{\cal C}_a^{[C'+ pk]_r} \; {\cal C}_b^{* \; [C'+ p'k]_r} \over \sqrt{V}} \int_{\mathbb T^4} \sum\limits_{B=0}^{\ell-1} \Phi^{(p)}_{C'B} \Phi^{(p') \; *}_{C'B}. ~ \end{eqnarray}
For convenience, we also repeat the expression for $\Phi^{(p)}$ (\ref{form of Phi}):
\begin{eqnarray}
\nonumber
\Phi^{(p)}_{C' B}(x,\hat\phi)&=& \sum_{\scriptsize m=p+\frac{rm'}{\mbox{gcd}(k,r)},\, m'\in \mathbb Z}~~\sum_{n'\in \mathbb Z}e^{\frac{i2\pi x_2 }{L_2}(m+\frac{2	C'-1-k}{2k})}e^{\frac{i2\pi x_4 }{L_4}(n'-\frac{2	B-1-\ell}{2\ell})}\\
\nonumber
&&~~\times e^{-i\frac{\pi (1-k)}{k}\left(C'-\frac{1+k(1-2m)}{2}\right)} e^{i\frac{\pi(1-\ell)}{\ell}\left(B-\frac{1+\ell(2n'+1)}{2}\right)}\\
\nonumber
&&~~\times \; e^{-\frac{\pi r}{k L_1 L_2}\left[x_1-\frac{k L_1 L_2}{2\pi r}(\hat \phi_2^{[C']_r}-i\hat \phi_1^{[C']_r})-\frac{L_1}{r}\left(km +\frac{2C'-1-k}{2}\right)\right]^2}\\
&&~~\times \; e^{-\frac{\pi }{\ell L_3 L_4}\left[x_3-\frac{\ell L_3 L_4}{2\pi }(\hat \phi_4^{[C']_r}-i\hat \phi_3^{[C']_r})-L_3\left(\ell n' -\frac{2B-1-\ell}{2}\right)\right]^2}\,.
\label{form of Phi 1}
\end{eqnarray}
To calculate $I^{ab}_j$, we now make a few  observations, which help evaluate (\ref{Iab1repeated}):
\begin{enumerate}
\item The integral over $x_4$ can be taken,  yielding a factor of $L_4$ and the condition $\delta_{n', \tilde n'}$, where  $n'$ is the index of summation from $\Phi^{(p)}$ and  $\tilde n'$ coming from $\Phi^{(p') \; *}$.  
\item The sum over $B = 0,..., \ell-1$ allows to extend the range of the $x_3$ integral from $-\infty, +\infty$, implying that the $\hat \phi_4$-dependence disappears.\footnote{ \label{commentfootnote} However, some factor of $\hat \phi_3$ remains which we will have to keep track of when evaluating the Gaussian integral over $x_3$.} 
\item The integral over $x_2$ can also be taken, yielding an overall factor of $L_2$ and the constraint $\delta_{m, \tilde m}$, where $m$ is from $\Phi^{(p)}$ and $\tilde m$ is from $\Phi^{(p) \; *}$. Note, in view of the definition of $m$ ($\tilde m$) in (\ref{form of Phi 1}), $m = \tilde m$ implies, recalling the range of $p, p'$, that $p=p'$ and $m' = \tilde m'$. Thus, in the end of this  step, we are left with an expression that contains only sums over $C'$, $n$,  $p$, and $m'$, and only an integral over the $x_1$  direction of $\mathbb T^4$.  
\item We also note that, for each $j$, only values of $C'$ equal to $[j+ nr]_k$ enter  the sum (\ref{Iab1repeated}) defining $I_j^{ab}$, with $n$ taking values in the range given. Now is time to recall    the relation (\ref{explicit holonomies}) defining the independent holonomies. It shows that all these  have the same $\hat\phi^{C'}_\mu$ and thus $I^{ab}_j$ only depends on the  gcd$(k,r)$ independent $\varphi_\mu^j$---as we   explicitly indicate in (\ref{intermediate1}) below.
\end{enumerate}
Explicitly performing the steps outlined in the above list, we obtain an intermediate result for (\ref{Iab1repeated}),
\begin{eqnarray} \label{intermediate1}\nonumber
I^{ab}_j 
&=&\sqrt{V} \sqrt{\ell L_4 \over 2 L_3} \; {e^{{L_1 L_2 k \over 2 \pi r} ( \varphi_1^{j})^2}}  \; {e^{{L_3 L_4 \ell \over 2 \pi} (\varphi_3^j)^2}} \\\nonumber&&\times \sum\limits_{n=0}^{\frac{k}{{\rm gcd}(k,r)}-1} \sum\limits_{p=0}^{{r\over {\rm gcd}(k,r)}-1} ({\cal C}_a {\cal C}_b^{*})^{[[j+nr]_k+ pk]_r}  \\
&&\times  \sum_{ m'\in \mathbb Z}\; \int\limits_{0}^{1} d x \; e^{- {2 \pi r L_1 \over k L_2}\left(x - {k p  + [j +  nr]_k \over r} + {1 + k \over 2 r} - {k \over {\rm gcd}(k,r)} m' - {k L_2 \over 2 \pi r L_1} \varphi_2^j\right)^2}\,,
~ \end{eqnarray}
which only contains a single integral over $x_1$,   rescaled by $L_1$ and  denoted by $x$. For brevity, we also denote 
$({\cal C}_a {\cal C}_b^{*})^A \equiv {\cal C}_a^A \; {\cal C}_b^{* \; A}$.

The next step is to rearrange the sum (\ref{intermediate1}) for $I_j^{ab}$ by grouping together terms where  the ``moduli" product $({\cal C}_a {\cal C}_b^*)^A$ has the same index. Recall that apriori  $A$ can take values in the range  $A \in {0,..., r-1}$. However, it is important to realize not all allowed values of $A$ appear in the sum defining $I_j^{ab}$ for a given $j$. One numerically finds that for any given $j$, the index $A \equiv [[j+nr]_k+ pk]_r$ takes only $r\over {\rm gcd}(k,r)$ of its possible $r$ values as $n$ and $p$ scan their possible values in the sum  in (\ref{intermediate1}).

To proceed further, 
we denote  by $S_j$ each of the gcd$(k,r)$ sets of $r\over {\rm gcd}(k,r)$ values that $A$ can take for a given $j$:\begin{eqnarray}\label{setSj}
S_j &=& \bigg\{ [[j+nr]_k+ pk]_r, {\rm for} \; n = 0,...\frac{k}{{\rm gcd}(k,r)}-1, {\rm and} \;   p = 0,...,{r\over {\rm gcd}(k,r)}-1 \bigg\}, \nonumber \\
  |S_j| &=& {r\over {\rm gcd}(k,r)}~,
\end{eqnarray}
where we stress that repeated values of $[[j+nr]_k+ pk]_r$ are identified in $S_j$ and that the set has $r/{\rm gcd}(k,r)$ elements.
 The sets $S_j$ are  straightforward to generate numerically in each case (we have used numerics extensively to obtain our final answer (\ref{IabFinal}) below).
A few examples might be useful:
\begin{eqnarray} \label{setsSexamples}
 k=5,r=4\; ({\rm gcd} (k,r)=1):&& S_0 = \{0,1,2,3 \},  \nonumber\\
 k=6,r=4\; ({\rm gcd} (k,r)=2):&& S_0 = \{0,2\}, \;S_1 = \{1,3\},  \nonumber\\
k=4, r=4 \;({\rm gcd} (k,r)=4):&& S_0=\{0\}, \;S_1 = \{1\}, \;S_2 = \{2\}, \;S_3 = \{3\}, \\
k=15, r=9 \;({\rm gcd} (k,r)=3):&& S_0 = \{0,3,6\},\; S_1 = \{1,4,7\},\; S_2 = \{2,5,8 \},\nonumber
\end{eqnarray}
 while, e.g.,  for  $k=9$, $r=9$ (gcd$(k,r)=9$), all $9$ sets $S_j$ have a single element, similar to the $k=r=4$ case above. This illustrates a general feature of the $k=r$ case, which will be important in our studies of the moduli space.

The next step is the most important to obtain our final answer. For each  different value of $A \in S_j$ that appears in $I_j^{ab}$, one also finds that $({\cal C}_a {\cal C}_b^*)^A$ is multiplied by an integral $\int\limits_{0}^1 dx$. The integral is, however,  summed over  $k \over {\rm gcd}(k,r)$ times, each time with a different constant term appearing in the exponent in the integrand, due to  the $(kp + [j+nr]_k)/r$ term. Remarkably, in each case one finds that, together with the sum over $m'$, these constant terms take precisely the values needed to extend  the range of the integration over $x$ to  the entire real line.\footnote{Admittedly, we have only numerical checks of this claim rather than an analytic proof. However, the checks are fairly easy to automate and the result is the same in each of the many cases we have studied.} Performing the Gaussian  integral over $x$, the final answer for $I_j^{ab}$ is then  remarkably simple
\begin{eqnarray} \label{IabFinal}
I_j^{ab} &=&{ \sqrt{V} \over 2} \sqrt{\ell k L_2 L_4 \over  r L_1 L_3} \; {e^{{L_1 L_2 k \over 2 \pi r} ( \varphi_1^{j})^2}}  \;   \; {e^{{L_3 L_4 \ell \over 2 \pi} (\varphi_3^j)^2}}  \sum_{A_j \in S_j} \; ({\cal C}_a {\cal C}_b^*)^{A_j} ~.
\end{eqnarray}
The complexity is, of course, hidden away in the definition of the $S_j$ sets from (\ref{setSj}).

\section{Field strength and action of the multifractional instanton}
\label{appx:strength}

Here, we compute the field strength $F_{\mu\nu}$, which we shall use to compute the action density and to verify that the action of the self-dual solution satisfies the relation $S=\frac{8\pi^2|Q|}{g^2}$. %To order $\sqrt{\Delta}$, $F_{\mu\nu}$ is written as
%
%\begin{eqnarray}
%\nonumber
%F_{\mu\nu}&=&\hat F_{\mu\nu}^\omega \omega +\left[\begin{array}{cc} 0& \sqrt \Delta {\cal F}^{(0)k\times \ell}_{\mu\nu}\\  \sqrt \Delta {\cal F}^{\dagger (0) \ell\times k}_{\mu\nu} & 0 \end{array}\right]\\
%\nonumber
%&=& \hat F_{\mu\nu}^\omega \omega +\sqrt \Delta\left[\begin{array}{cc} 0& \left( \hat D_\mu {\cal W}_\nu^{(0)k \times \ell}-\hat D_\nu {\cal W}_\mu^{(0)k \times \ell}\right)\\   \left( \hat D_\mu {\cal W}_\nu^{\dagger (0) \ell\times k }-\hat D_\nu {\cal W}_\mu^{\dagger (0) \ell\times k }\right)& 0 \end{array}\right]\,.\\
%\label{expression of F to leading order in Delta}
%\end{eqnarray}
%
The non-zero components of ${\cal F}_{\mu\nu}^{(0)}$ are %
\begin{eqnarray}
{\cal F}_{13\,C',C}^{(0)}=-i\hat D_1{\cal W}_{4C',C}+i\hat D_3 {\cal W}_{2\,C',C}\,,\quad {\cal F}_{14,C',C}^{(0)}=\hat D_1{\cal W}_{4\,C',C}+i\hat D_4 {\cal W}_{2\,C',C}\,,
\end{eqnarray}
where ${\cal W}^{(0)}_{2\,C',C}$ and ${\cal W}^{(0)}_{4\,C',C}$ are from (\ref{expressions of W2 and W4 with holonomies}).
%\begin{eqnarray}
%\nonumber
%{\cal W}^{(0)}_{2\,C',C}(x)&=&V^{-1/4} \sum_{p=0}^{\scriptsize \frac{r}{\mbox{gcd}(k,r)}-1}{\cal C}_{2}^{[C'+pk]_r}\Phi^{(p)}%_{C',C}(x,\hat\phi)\,,\\
%{\cal W}^{(0)}_{4\,C',C}(x)&=&V^{-1/4}  \sum_{p=0}^{\scriptsize \frac{r}{\mbox{gcd}(k,r)}-1}{\cal C}_{4}^{[C'+pk]_r}\Phi^{(p)}_{C',C}(x,\hat\phi)\,.
%\end{eqnarray}
% 
The covariant derivatives $\hat D_\mu$ are given by
\begin{eqnarray}
\hat D_\mu=\partial_\mu+i 2\pi N \hat A_\mu+i\hat \phi_\mu^{[C']_r}\,,
\end{eqnarray}
or in terms of the components, with $\hat \phi_\mu^{[C']_r}$ from (\ref{explicit holonomies}), 
\begin{eqnarray}
\nonumber
\hat D_1&=&\partial_1+i\hat \phi_1^{[C']_r}\,, \quad \hat D_2=\partial_2-i\frac{2\pi r x_1}{k L_1 L_2}+i\hat \phi_2^{[C']_r}\\
\hat D_3&=&\partial_3+i\hat \phi_3^{[C']_r}\,, \quad \hat D_4=\partial_4-i\frac{2\pi x_3}{\ell L_3 L_4}+i\hat \phi_4^{[C']_r}\,.
\label{hat dervatives}
\end{eqnarray}
One can check that the following identities hold
\begin{eqnarray}
i \hat D_1 \Phi^{(p)}_{C',C}=\hat D_2 \Phi^{(p)}_{C',C}\,, \quad i \hat D_3 \Phi^{(p)}_{C',C}=\hat D_4 \Phi^{(p)}_{C',C}\,.
\end{eqnarray}
Then, one finds
\begin{eqnarray}
\nonumber
-i{\cal F}_{14\,C',C}^{(0)}&=& {\cal F}_{13\,C',C}^{(0)}= iV^{-1/4} \sum_{p=0}^{\scriptsize \frac{r}{\mbox{gcd}(k,r)}-1}\left\{ -{\cal C}_{4}^{[C'+pk]_r}{\cal G}^{(p)}_{1,C',C}(x,\hat\phi)       + {\cal C}_{2}^{[C'+pk]_r}{\cal G}^{(p)}_{3,C',C}(x,\hat\phi)\right\}\,,\\
{\cal F}_{12\,C',C}^{(0)}&=& {\cal F}_{34\,C',C}^{(0)}=0\,,
\end{eqnarray}
where the functions ${\cal G}^{(p)}_{1,C',C}(x,\hat\phi)$ and ${\cal G}^{(p)}_{3,C',C}(x,\hat\phi)$ are defined as
\begin{eqnarray}
\nonumber
{\cal G}^{(p)}_{1,C',C}(x,\hat\phi)&=&\hat D_1 \Phi^{(p)}_{C',C}(x, \hat\phi) \nonumber \\
&=&-\frac{2\pi r }{k L_1 L_2}\ \sum_{\scriptsize m=p+\frac{rm'}{\mbox{gcd}(k,r)},\, m'\in \mathbb Z}\sum_{n'\in \mathbb Z}e^{\frac{i2\pi x_2 }{L_2}(m+\frac{2	C'-1-k}{2k})}e^{\frac{i2\pi x_4 }{L_4}(n'-\frac{2	C-1-\ell}{2\ell})}\\
\nonumber
&&\times e^{-i\frac{\pi (1-k)}{k}\left(C'-\frac{1+k(1-2m)}{2}\right)} e^{i\frac{\pi(1-\ell)}{\ell}\left(C-\frac{1+\ell(2n'+1)}{2}\right)}\\
\nonumber
&&\times \left( x_1-\frac{k L_1 L_2\hat \phi_2^{[C']_r}}{2\pi r} -\frac{L_1}{r}\left(km+\frac{2C'-1-k}{2}\right)\right) \\
\nonumber
&&\times e^{-\frac{\pi r}{k L_1 L_2}\left[x_1-\frac{k L_1 L_2}{2\pi r}(\hat \phi_2^{[C']_r}-i\hat \phi_1^{[C']_r})-\frac{L_1}{r}\left(km +\frac{2C'-1-k}{2}\right)\right]^2}\\
&&\times e^{-\frac{\pi }{\ell L_3 L_4}\left[x_3-\frac{\ell L_3 L_4}{2\pi }(\hat \phi_4^{[C']_r}-i\hat \phi_3^{[C']_r})-L_3\left(\ell n' -\frac{2C-1-\ell}{2}\right)\right]^2}\,,
\label{the G1 function}
\end{eqnarray}
and 
\begin{eqnarray}
\nonumber
{\cal G}^{(p)}_{3,C',C}(x,\hat\phi)&=&\hat D_3 \Phi^{(p)}_{C',C}(x, \hat\phi)\nonumber \\
&=&-\frac{2\pi }{\ell L_3 L_4}\ \sum_{\scriptsize m=p+\frac{rm'}{\mbox{gcd}(k,r)},\, m'\in \mathbb Z}\sum_{n'\in \mathbb Z}e^{\frac{i2\pi x_2 }{L_2}(m+\frac{2	C'-1-k}{2k})}e^{\frac{i2\pi x_4 }{L_4}(n'-\frac{2	C-1-\ell}{2\ell})}\\
\nonumber
&&\times e^{-i\frac{\pi (1-k)}{k}\left(C'-\frac{1+k(1-2m)}{2}\right)} e^{i\frac{\pi(1-\ell)}{\ell}\left(C-\frac{1+\ell(2n'+1)}{2}\right)}\\
\nonumber
&&\times \left( x_3-\frac{\ell L_3 L_4\hat \phi_4^{[C']_r}}{2\pi  } -L_3\left(\ell n'-\frac{2C-1-\ell}{2}\right)\right) \\
\nonumber
&&\times e^{-\frac{\pi r}{k L_1 L_2}\left[x_1-\frac{k L_1 L_2}{2\pi r }(\hat \phi_2^{[C']_r}-i\hat \phi_1^{[C']_r})-\frac{L_1}{r}\left(km +\frac{2C'-1-k}{2}\right)\right]^2}\\
&&\times e^{-\frac{\pi }{\ell L_3 L_4}\left[x_3-\frac{\ell L_3 L_4}{2\pi }(\hat \phi_4^{[C']_r}-i\hat \phi_3^{[C']_r})-L_3\left(\ell n' -\frac{2C-1-\ell}{2}\right)\right]^2}\,.
\label{the G3 function}
\end{eqnarray}
Owing to the self-duality of the solution, we also have:
\begin{eqnarray}
{\cal F}_{23\,C',C}^{(0)}={\cal F}_{14\,C',C}^{(0)}\,,\quad {\cal F}_{24\,C',C}^{(0)}&=&-{\cal F}_{13\,C',C}^{(0)}\,.
\end{eqnarray}

In the following, we   calculate the action density $\mbox{tr}\left[F_{\mu\nu}F_{\mu\nu}\right]$ of the twisted solution. Using (\ref{components of F}), the square of the field strength is
\begin{eqnarray}
\nonumber
F_{\mu\nu}F_{\mu\nu}&=&\omega^2 \left(\hat F_{\mu\nu}^\omega +F_{\mu\nu}^s\right)^2+4\pi  \left(\hat F_{\mu\nu}^\omega +F_{\mu\nu}^s\right)\left[\begin{array}{cc} \ell F_{\mu\nu}^k & {\cal F}_{\mu\nu}^{k\times\ell}\\ {\cal F}_{\mu\nu}^{\dagger \ell \times k}&  -k F_{\mu\nu}^\ell\end{array}\right]\\
&&+\left[\begin{array}{cc} F_{\mu\nu}^k F_{\mu\nu}^k+ {\cal F}_{\mu\nu}^{ k\times \ell}{\cal F}_{\mu\nu}^{\dagger \ell\times k}& F^k_{\mu\nu}{\cal F}_{\mu\nu}^{k \times \ell}+{\cal F}_{\mu\nu}^{k\times \ell}F_{\mu\nu}^\ell \\ {\cal F}_{\mu\nu}^{\dagger \ell \times k} F_{\mu\nu}^k +F_{\mu\nu}^\ell {\cal F}_{\mu\nu}^{\dagger \ell \times k}& F_{\mu\nu}^\ell F_{\mu\nu}^\ell +{\cal F}_{\mu\nu}^{\dagger \ell \times k}{\cal F}_{\mu\nu}^{k \times \ell} \end{array}\right]\,.
\end{eqnarray}
 Then, the action density  is given by the trace
\begin{eqnarray}
\nonumber
\mbox{tr}\left[F_{\mu\nu}F_{\mu\nu}\right]&=&\mbox{tr}\left[\omega^2\right]\left(\hat F_{\mu\nu}^\omega +F_{\mu\nu}^s\right)^2\\
\nonumber
&&+4\pi \ell\left(\hat F_{\mu\nu}^\omega +F_{\mu\nu}^s\right)\mbox{tr}_k\left[F_{\mu\nu}^k\right]-4\pi k \left(\hat F_{\mu\nu}^\omega +F_{\mu\nu}^s\right)\mbox{tr}_\ell\left[F_{\mu\nu}^\ell\right]\\
\nonumber
&&+ \mbox{tr}_k \left[ F_{\mu\nu}^k F_{\mu\nu}^k+ {\cal F}_{\mu\nu}^{ k\times \ell}{\cal F}_{\mu\nu}^{\dagger \ell\times k}\right]+ \mbox{tr}_\ell \left[  F_{\mu\nu}^\ell F_{\mu\nu}^\ell +{\cal F}_{\mu\nu}^{\dagger \ell\times k}{\cal F}_{\mu\nu}^{ k \times \ell}\right]\,.\\
\label{action density in full}
\end{eqnarray}
To leading order in $\Delta$ we have:
\begin{eqnarray}
\nonumber
F_{\mu\nu}^s&=&\Delta \left(\partial_\mu S_\nu^{\omega (0)}-\partial_\nu S_\mu^{\omega (0)}\right) \,,\\
\nonumber
F_{\mu\nu}^k&=&\Delta \left(\partial_\mu S_\nu^{k (0)}-\partial_\nu S_\mu^{k (0)}+i {\cal W}_\mu^{(0)k\times \ell} {\cal W}_\nu^{\dagger (0) \ell\times k}-i {\cal W}_\nu^{(0)k\times \ell} {\cal W}_\mu^{\dagger (0) \ell\times k}   \right)\,,\\
\nonumber
F_{\mu\nu}^\ell&=&\Delta \left(\partial_\mu S_\nu^{\ell (0)}-\partial_\nu S_\mu^{\ell (0)}+i {\cal W}_\mu^{\dagger(0)  \ell\times k} {\cal W}_\nu^{ (0)  k\times \ell}-i {\cal W}_\nu^{\dagger(0) \ell\times k} {\cal W}_\mu^{ (0)  k\times \ell}   \right)\,,\\
{\cal F}_{\mu\nu}^{k\times \ell}&=& \sqrt\Delta{\cal F}_{\mu\nu}^{(0)k\times \ell}=\sqrt \Delta \left(\hat D_\mu {\cal W}_\nu^{(0)k\times \ell}-\hat D_\nu {\cal W}_\mu^{(0)k\times \ell}\right)\,.
\label{leading order in field strength}
\end{eqnarray}
Substituting (\ref{leading order in field strength}) into (\ref{action density in full}), we find to ${\cal O}(\Delta)$:
\begin{eqnarray}
\nonumber
\mbox{tr}\left[F_{\mu\nu}F_{\mu\nu}\right]&=&\mbox{tr}\left[\omega^2\right]\left(\hat F_{\mu\nu}^\omega \hat F_{\mu\nu}^\omega+2\Delta(\partial_\mu {\cal S}_\nu^{\omega(0)}-\partial_\nu {\cal S}_\mu^{\omega (0)})\hat F_{\mu\nu}^\omega\right)\\
\nonumber
&&+4\pi\ell \Delta\hat F_{\mu\nu}^\omega\mbox{tr}_k\left[\partial_\mu {\cal S}_\nu^{k(0)}-\partial_\nu {\cal S}_\mu^{k (0)}\right]-4\pi k \Delta\hat F_{\mu\nu}^\omega\mbox{tr}_\ell \left[\partial_\mu {\cal S}_\nu^{\ell(0)}-\partial_\nu {\cal S}_\mu^{\ell (0)}\right]\\
\nonumber
&&+i4\pi\ell \Delta\hat F_{\mu\nu}^\omega\mbox{tr}_k\left[  {\cal W}_\mu^{(0)k\times \ell} {\cal W}_\nu^{\dagger (0) \ell\times k}- {\cal W}_\nu^{(0)k\times \ell} {\cal W}_\mu^{\dagger (0) \ell\times k}  \right]\\
\nonumber
&&-i4\pi k  \Delta\hat F_{\mu\nu}^\omega\mbox{tr}_\ell\left[   {\cal W}_\mu^{\dagger(0)  \ell\times k} {\cal W}_\nu^{ (0)  k\times \ell}-{\cal W}_\nu^{\dagger(0) \ell\times k} {\cal W}_\mu^{ (0)  k\times \ell}    \right]\\
&&+\Delta \mbox{tr}_k\left[{\cal F}_{\mu\nu}^{(0) k\times \ell}{\cal F}_{\mu\nu}^{\dagger(0) \ell\times k}\right]+\Delta \mbox{tr}_\ell \left[ {\cal F}_{\mu\nu}^{\dagger (0) \ell\times k}{\cal F}_{\mu\nu}^{(0) k \times \ell}\right]\,.
\end{eqnarray}

Then, using the trace properties $\mbox{tr}_k[{\cal S}_\mu^{(0)k}]=\mbox{tr}_\ell[{\cal S}_\mu^{(0)\ell}]=0$, along with
\begin{eqnarray}
\nonumber
 \mbox{tr}_k \left[ {\cal F}_{\mu\nu}^{(0) k\times \ell}{\cal F}_{\mu\nu}^{\dagger(0) \ell\times k}\right]&=&\mbox{tr}_\ell \left[ {\cal F}_{\mu\nu}^{\dagger (0) \ell\times k}{\cal F}_{\mu\nu}^{(0) k \times \ell}\right]\,,\\
 \nonumber
 \mbox{tr}_k\left[  {\cal W}_\mu^{(0)k\times \ell} {\cal W}_\nu^{\dagger (0) \ell\times k}- {\cal W}_\nu^{(0)k\times \ell} {\cal W}_\mu^{\dagger (0) \ell\times k}  \right]&=&-\mbox{tr}_\ell\left[   {\cal W}_\mu^{\dagger(0)  \ell\times k} {\cal W}_\nu^{ (0)  k\times \ell}-{\cal W}_\nu^{\dagger(0) \ell\times k} {\cal W}_\mu^{ (0)  k\times \ell}    \right]\,,\\
 \end{eqnarray}
we find to ${\cal O}(\Delta)$
\begin{eqnarray}
\nonumber
\mbox{tr}\left[F_{\mu\nu}F_{\mu\nu}\right]&=&\mbox{tr}\left[\omega^2\right]\left(\hat F_{\mu\nu}^\omega \hat F_{\mu\nu}^\omega+2\Delta(\partial_\mu {\cal S}_\nu^{\omega(0)}-\partial_\nu {\cal S}_\mu^{\omega (0)})\hat F_{\mu\nu}^\omega\right)\\
\nonumber
&&+i4\pi N \Delta\hat F_{\mu\nu}^\omega\mbox{tr}_k\left[  {\cal W}_\mu^{(0)k\times \ell} {\cal W}_\nu^{\dagger (0) \ell\times k}- {\cal W}_\nu^{(0)k\times \ell} {\cal W}_\mu^{\dagger (0) \ell\times k}  \right]\\
&&+2\Delta \mbox{tr}_k\left[{\cal F}_{\mu\nu}^{(0) k\times \ell}{\cal F}_{\mu\nu}^{\dagger(0) \ell\times k}\right]\,.
\end{eqnarray}

In the following, we perform the calculation of the action setting ${\cal C}_4^{[C']_r}=0$. Thus, recalling (\ref{requalkconstraints}), we are particularly interested in the cases $r=1$ and $r=k, k>1$. However, the conclusion should hold in the general case.  Keeping only the non-zero entries and using $-i{\cal F}_{14}^{(0)\bm\beta}={\cal F}_{13}^{(0)\bm\beta}$ along with the self-duality property, we arrive at
\begin{eqnarray}
\nonumber
 \mbox{tr}_k \left[ {\cal F}_{\mu\nu}^{(0) k\times \ell}{\cal F}_{\mu\nu}^{\dagger(0) \ell\times k}\right]&=&2  \mbox{tr}_k \left[ {\cal F}_{13}^{(0) k\times \ell}{\cal F}_{13}^{\dagger(0) \ell\times k}\right]+2  \mbox{tr}_k \left[ {\cal F}_{14}^{(0) k\times \ell}{\cal F}_{14}^{\dagger(0) \ell\times k}\right]\\
 \nonumber
 &+&2  \mbox{tr}_k \left[ {\cal F}_{23}^{(0) k\times \ell}{\cal F}_{23}^{\dagger(0) \ell\times k}\right]+2  \mbox{tr}_k \left[ {\cal F}_{24}^{(0) k\times \ell}{\cal F}_{24}^{\dagger(0) \ell\times k}\right]\\
 &=&8 \mbox{tr}_k \left[ {\cal F}_{13}^{(0) k\times \ell}{\cal F}_{13}^{\dagger(0) \ell\times k}\right]\,.
\end{eqnarray}
Likewise:
\begin{eqnarray}
\hat F_{\mu\nu}^\omega\mbox{tr}_k\left[  {\cal W}_\mu^{(0)k\times \ell} {\cal W}_\nu^{\dagger (0) \ell\times k}- {\cal W}_\nu^{(0)k\times \ell} {\cal W}_\mu^{\dagger (0) \ell\times k}  \right]=-4 i \hat F_{12}^\omega\mbox{tr}_k\left[{\cal W}_2^{(0)k\times\ell}{\cal W}_2^{\dagger (0)\ell \times k}\right]\,.
\end{eqnarray}
Thus, the action density is given by the expression
\begin{eqnarray}
\nonumber
\mbox{tr}\left[F_{\mu\nu}F_{\mu\nu}\right]&=&\mbox{tr}\left[\omega^2\right]\left(\hat F_{\mu\nu}^\omega \hat F_{\mu\nu}^\omega+2\Delta(\partial_\mu {\cal S}_\nu^\omega-\partial_\nu {\cal S}_\mu^\omega)\hat F_{\mu\nu}^\omega\right)\\
\nonumber
&&+16\pi N \Delta \hat F_{12}^\omega\mbox{tr}_k\left[{\cal W}_2^{(0)k\times\ell}{\cal W}_2^{\dagger (0)\ell \times k}\right]\\
&&+16\Delta\mbox{tr}_k \left[ {\cal F}_{13}^{(0) k\times \ell}{\cal F}_{13}^{\dagger(0) \ell\times k}\right]\,.
\label{long action density}
\end{eqnarray}
The action is 
\begin{eqnarray}
S=\frac{2}{g^2}\int_{\mathbb T^4}\mbox{tr}\left[F_{\mu\nu}F_{\mu\nu}\right]\,,
\end{eqnarray}
and upon integrating, the term $\partial_\mu {\cal S}_\nu^{(0)\omega}$  drops out because ${\cal S}_\nu^{(0)\omega}$ satisfies periodic boundary conditions. Thus, we finally have to ${\cal O}(\Delta)$:
\begin{eqnarray}
\nonumber
S&=&S_0+ \frac{\Delta}{2g^2}\int_{\mathbb T^4}16\pi N  \hat F_{12}^\omega\mbox{tr}_k\left[{\cal W}_2^{(0)k\times\ell}{\cal W}_2^{\dagger (0)\ell \times k}\right]
+\frac{\Delta}{2g^2}\int_{\mathbb T^4}16\mbox{tr}_k \left[ {\cal F}_{13}^{(0) k\times \ell}{\cal F}_{13}^{\dagger(0) \ell\times k}\right]\,,
\end{eqnarray}
where
\begin{eqnarray}
\nonumber
S_0&=&\frac{1}{2g^2}\int_{\mathbb T^4} \mbox{tr}\left[\omega^2\right]\left(\hat F_{\mu\nu}^\omega \hat F_{\mu\nu}^\omega\right)=\frac{1}{2g^2}\int_{\mathbb T^4} \mbox{tr}\left[\omega^2\right]\left\{2\left(\hat F_{12}^\omega \hat F_{12}^\omega+\hat F_{34}^\omega \hat F_{34}^\omega\right)\right\}\\
&=&(4\pi^2 Nk \ell)\frac{1}{g^2 N^2}\left(\frac{r^2}{k^2}\frac{L_3L_4}{L_1L_2}+\frac{1}{\ell^2}\frac{L_1 L_2}{L_3L_4}\right)\,.
\end{eqnarray}
Using the definition of $\Delta$ (\ref{deltadef}) we readily find 
\begin{eqnarray}
S_0=\frac{8\pi^2 r}{N g^2}+{\cal O}(\Delta^2)\,.
\end{eqnarray}
Then, using  $\hat F_{12}^\omega=-\frac{r}{k N L_1 L_2}$, we have
\begin{eqnarray}
\nonumber
S=S_0+\frac{\Delta}{g^2}\left(-\frac{8\pi r}{k L_1 L_2}\int_{\mathbb T^4}\mbox{tr}_k\left[{\cal W}_2^{(0)k\times\ell}{\cal W}_2^{\dagger (0)\ell \times k}\right]+8\int_{\mathbb T^4}\mbox{tr}_k \left[ {\cal F}_{13}^{(0) k\times \ell}{\cal F}_{13}^{\dagger(0) \ell\times k}\right]\right)\,.\\
\end{eqnarray}
Finally,  the remaining integrals are given by (we set all holonomies to $0$, as the final answer will not depend on them):
\begin{eqnarray}
\nonumber
\int_{\mathbb T^4}\mbox{tr}_k\left[{\cal W}_2^{(0)k\times\ell}{\cal W}_2^{\dagger (0)\ell \times k}\right]&=&\sqrt{L_1L_2L_3L_4} \sum_{C=1}^\ell\sum_{C'=1}^k |{\cal C}_2^{[C']_r}|^2\\
\nonumber
&&\sum_{\scriptsize m=p+\frac{rm'}{\mbox{gcd}(k,r)},\,m'\in \mathbb Z} \int_{0}^{1}d\tilde x_1 e^{-\frac{2\pi r L_1}{k L_2}\left(\tilde x_1-\frac{2mk+2(j+nr)-1-k}{2r}\right)^2}\\
&& \times \sum_{n' \in \mathbb Z}\int_{0}^{1}d\tilde x_3e^{-\frac{2\pi  L_3}{\ell L_4}\left(\tilde x_3-\frac{(2n'+1)\ell-(2C-1)}{2}\right)^2}\,,
\end{eqnarray}
and
\begin{eqnarray}
\nonumber
&&\int_{\mathbb T^4}\mbox{tr}_k \left[ {\cal F}_{13}^{(0) k\times \ell}{\cal F}_{13}^{\dagger(0) \ell\times k}\right]\\
\nonumber
&=&\frac{4\pi^2}{\ell^2}\sqrt{\frac{L_1L_2L_3}{L_4^3}}\sum_{C=1}^\ell\sum_{C'=1}^k |{\cal C}_2^{[C']_r}|^2\\
\nonumber
&&\sum_{\scriptsize m=p+\frac{rm'}{\mbox{gcd}(k,r)},\,m'\in \mathbb Z} \int_{0}^{1}d\tilde x_1 e^{-\frac{2\pi r L_1}{k L_2}\left(\tilde x_1-\frac{2mk+2(j+nr)-1-k}{2r}\right)^2}\\
\nonumber
&&\times\sum_{n' \in \mathbb Z}\int_{0}^{1}d\tilde x_3  \left(\tilde x_3-\frac{\left((2n'+1)\ell-(2C-1)\right)}{2} \right)^2e^{-\frac{2\pi  L_3}{\ell L_4}\left(\tilde x_3-\frac{(2n'+1)\ell-(2C-1)}{2}\right)^2}\,.\\
\end{eqnarray}
Now, collecting terms of ${\cal O}(\Delta)$ and using $r\ell L_3L_4=k L_1L_2$, thus ignoring corrections ${\cal O}(\Delta^2)$, we find:
\begin{eqnarray}
\nonumber
S&=&S_0+8\pi\sqrt{\frac{r}{\ell k}}\frac{\Delta}{g^2}\sum_{C=1}^\ell\sum_{C'=1}^k |{\cal C}_2^{[C']_r}|^2\\
\nonumber
&&\sum_{\scriptsize m=p+\frac{rm'}{\mbox{gcd}(k,r)},\,m'\in \mathbb Z} \int_{0}^{1}d\tilde x_1 e^{-\frac{2\pi r L_1}{k L_2}\left(\tilde x_1-\frac{2mk+2(j+nr)-1-k}{2r}\right)^2}\\
\nonumber
&&\times\sum_{n'}\int_{0}^{1}d\tilde x_3 \left\{-1+\frac{4\pi}{\ell}\frac{L_3}{L_4} \left(\tilde x_3-\frac{\left((2n'+1)\ell-(2C-1)\right)}{2} \right)^2\right\}e^{-\frac{2\pi  L_3}{\ell L_4}\left(\tilde x_3-\frac{(2n'+1)\ell-(2C-1)}{2}\right)^2}\,.\\
\end{eqnarray}
One can check (using Mathematica) that\footnote{One can show that (\ref{important sum integral}) is true by converting the combined infinite sum and the integral over the unit interval to an infinite integral.}:
\begin{eqnarray}
\nonumber
\sum_{C=1}^\ell \sum_{n}\int_{0}^{1}d\tilde x_3 \left\{-1+\frac{4\pi}{\ell}\frac{L_3}{L_4} \left(\tilde x_3-\frac{\left((2n+1)\ell-(2C-1)\right)}{2} \right)^2\right\}e^{-\frac{2\pi  L_3}{\ell L_4}\left(\tilde x_3-\frac{(2n+1)\ell-(2C-1)}{2}\right)^2}=0\,,\\
\label{important sum integral}
\end{eqnarray}
and thus, we conclude that, as expected
\begin{eqnarray}
S&=&S_0+{\cal O}(\Delta^2)=\frac{r}{N} \frac{8\pi^2 }{  g^2}+{\cal O}(\Delta^2)\,,
\end{eqnarray}
i.e. the action of the multifractional instanton is, to the order in $\Delta$ we are working on, equal to ${r \over N}$ times the BPST instanton action.

\section{Blow up of the gauge invariant local densities along the noncompact moduli  of the $k \ne r$ solution }
\label{appx:blowup}
To determine the gauge invariant density (\ref{field strength F34F34}), 
 we need to solve for ${\cal S}^{(0)\omega }_{\nu}$. To this end, we use (\ref{Sequations}) (or the equivalent forms (\ref{equationforSk}, \ref{equationforSell})). Acting on these equations with $\partial = \sigma^\nu\partial_\nu$ and using the identity $\sigma^\nu\bar\sigma^\mu+\sigma^\mu\bar\sigma^\nu=2\delta_{\mu\nu}$, we find the expression 
\begin{eqnarray}
\square {\cal S}^{(0)\omega}=-\frac{i}{\pi \ell k}\sigma^\nu \partial_\nu{\cal Y}\,,
\label{S omega equation}
\end{eqnarray}
where (once more, for brevity, we omit the $k \times \ell$ and $\ell\times k$ superscripts)
\begin{eqnarray}
{\cal Y}=\left[\begin{array}{cc} \mbox{tr}_k\left[{\cal W}_2^{(0)}{\cal W}_2^{\dagger(0)}-{\cal W}_4^{(0)}{\cal W}_4^{\dagger(0)}\right]&-2\mbox{tr}_k\left[{\cal W}_2^{(0)}{\cal W}_4^{\dagger(0)}\right]\\ -2\mbox{tr}_k\left[{\cal W}_4^{(0)}{\cal W}_2^{\dagger(0)}\right] & -\mbox{tr}_k\left[{\cal W}_2^{(0)}{\cal W}_2^{\dagger(0)}-{\cal W}_4^{(0)}{\cal W}_4^{\dagger(0)}\right]\end{array}\right]\,.
\end{eqnarray}
Equating the components of (\ref{S omega equation}), we arrive at the following set of equations:
\begin{eqnarray}
\nonumber
&&i \pi \ell k\square\left( {\cal S}^{(0)\omega}_4+i {\cal S}^{(0)\omega}_3\right)=\\
\nonumber
&&\left(\partial_4+i\partial_3\right)\mbox{tr}_k\left[{\cal W}_2^{(0)}{\cal W}_2^{\dagger(0)}-{\cal W}_4^{(0)}{\cal W}_4^{\dagger(0)}\right]-2\left(i\partial_1+\partial_2\right)\mbox{tr}_k\left[{\cal W}_4^{(0)}{\cal W}_2^{\dagger(0)}\right]\,,\\
\nonumber
&&i \pi \ell k\square\left( {\cal S}^{(0)\omega}_4-i {\cal S}^{(0)\omega}_3\right)=\\
\nonumber
&&-\left(\partial_4-i\partial_3\right)\mbox{tr}_k\left[{\cal W}_2^{(0)}{\cal W}_2^{\dagger(0)}-{\cal W}_4^{(0)}{\cal W}_4^{\dagger(0)}\right]-2\left(i\partial_1-\partial_2\right)\mbox{tr}_k\left[{\cal W}_2^{(0)}{\cal W}_4^{\dagger(0)}\right]\,,\\
\nonumber
&&i \pi \ell k\square\left(i {\cal S}^{(0)\omega}_1+ {\cal S}^{(0)\omega}_2\right)=\\
\nonumber
&&-2\left(\partial_4+i\partial_3\right) \mbox{tr}_k\left[{\cal W}_2^{(0)}{\cal W}_4^{\dagger(0)}\right]-\left(i\partial_1+\partial_2\right)\mbox{tr}_k\left[{\cal W}_2^{(0)}{\cal W}_2^{\dagger(0)}-{\cal W}_4^{(0)}{\cal W}_4^{\dagger(0)}\right]\,,\\
\nonumber
&&i \pi \ell k\square\left(i {\cal S}^{(0)\omega}_1- {\cal S}^{(0)\omega}_2\right)=\\
&&-2\left(\partial_4-i\partial_3\right) \mbox{tr}_k\left[{\cal W}_4^{(0)}{\cal W}_2^{\dagger(0)}\right]+\left(i\partial_1-\partial_2\right)\mbox{tr}_k\left[{\cal W}_2^{(0)}{\cal W}_2^{\dagger(0)}-{\cal W}_4^{(0)}{\cal W}_4^{\dagger(0)}\right]\,.
\label{the long expressions of different S}
\end{eqnarray}
Thus, we find
\begin{eqnarray}
\nonumber
&&\pi \ell k \Box{\cal S}_4^{(0)\omega}=\\
\nonumber
&& \partial_3\mbox{tr}_k\left[{\cal W}_2^{(0)}{\cal W}_2^{\dagger(0)}-{\cal W}_4^{(0)}{\cal W}_4^{\dagger(0)}\right]-(\partial_1-i\partial_2) \mbox{tr}_k\left[{\cal W}_4^{(0)}{\cal W}_2^{\dagger(0)}\right]-(\partial_1+i\partial_2) \mbox{tr}_k\left[{\cal W}_2^{(0)}{\cal W}_4^{\dagger(0)}\right]\,,\\
\nonumber
&&-\pi \ell k \Box{\cal S}_3^{(0)\omega}=\\
\nonumber
&& \partial_4\mbox{tr}_k\left[{\cal W}_2^{(0)}{\cal W}_2^{\dagger(0)}-{\cal W}_4^{(0)}{\cal W}_4^{\dagger(0)}\right]-(i\partial_1+\partial_2) \mbox{tr}_k\left[{\cal W}_4^{(0)}{\cal W}_2^{\dagger(0)}\right]+(i\partial_1-\partial_2) \mbox{tr}_k\left[{\cal W}_2^{(0)}{\cal W}_4^{\dagger(0)}\right]\,.\\
\end{eqnarray}
and
\begin{eqnarray}
\nonumber
\left(\partial_3{\cal S}_4^{(0)\omega}-\partial_4{\cal S}_3^{(0)\omega} \right)&=&\left(\pi \ell k \Box\right)^{-1}\left\{\left(\partial_3^2+\partial_4^2\right)\mbox{tr}_k\left[{\cal W}_2^{(0)}{\cal W}_2^{\dagger(0)}-{\cal W}_4^{(0)}{\cal W}_4^{\dagger(0)}\right]\right.\\
\nonumber
&+&\left.\left(-\partial_1\partial_3-\partial_2\partial_4+i\partial_2\partial_3-i\partial_1\partial_4\right)\mbox{tr}_k\left[{\cal W}_4^{(0)}{\cal W}_2^{\dagger(0)}\right]\right.\\
\nonumber
&+&\left.\left(-\partial_1\partial_3-\partial_2\partial_4-i\partial_2\partial_3+i\partial_1\partial_4\right)\mbox{tr}_k\left[{\cal W}_2^{(0)}{\cal W}_4^{\dagger(0)}\right]\right\}\,.\\
\label{final expression of superpositions of S}
\end{eqnarray}

We are interested in the case $r>1$ and $\mbox{gcd}(k,r)=1$. Let us consider the example $r=2,k=3$. Then, using the parameterization of (\ref{noncompact}), taking the upper sign for definiteness,
 \begin{eqnarray}
 {\cal C}_2^0=u\,,\quad  {\cal C}_2^1= u\,, \quad  {\cal C}_4^0=-iu\,, \quad  {\cal C}_4^1=  iu\,.
 \end{eqnarray}
we find\footnote{The sums over $C'$ and $C$ should be really thought of as being over $0,...,k-1$ and $0,..., \ell-1$, respectively, to be consistent with the main body of the paper. We apologize to the reader for this slight mismatch.}
\begin{eqnarray}
\nonumber
\mbox{tr}_k\left[{\cal W}_2^{(0)}{\cal W}_2^{\dagger(0)}\right]&=&u^2\sum_{C=1}^\ell \sum_{C'=1}^k \left[ \Phi_{C',C}^0+ \Phi_{C',C}^1 \right]\left[\Phi_{C',C}^{*0}+  \Phi_{C',C}^{*1} \right]\\
\nonumber
&=&u^2\sum_{C=1}^\ell \sum_{C'=1}^k | \Phi_{C',C}^0|^2+ |\Phi_{C',C}^1|+ \Phi_{C',C}^0 \Phi_{C',C}^{*1} +\Phi_{C',C}^{*0} \Phi_{C',C}^{1}\,,\\
\end{eqnarray}
\begin{eqnarray}
\nonumber
\mbox{tr}_k\left[{\cal W}_4^{(0)}{\cal W}_4^{\dagger(0)}\right]&=&u^2\sum_{C=1}^\ell \sum_{C'=1}^k \left[ \Phi_{C',C}^0- \Phi_{C',C}^1 \right]\left[\Phi_{C',C}^{*0}- \Phi_{C',C}^{*1} \right]\\\nonumber
&=&u^2\sum_{C=1}^\ell \sum_{C'=1}^k | \Phi_{C',C}^0|^2+ |\Phi_{C',C}^1|- \Phi_{C',C}^0 \Phi_{C',C}^{*1} -\Phi_{C',C}^{*0} \Phi_{C',C}^{1}\,,\\
\end{eqnarray}
and
\begin{eqnarray}
\nonumber
\mbox{tr}_k\left[{\cal W}_4^{(0)}{\cal W}_2^{\dagger(0)}\right]&=&iu^2\sum_{C=1}^\ell \left[ \left(-\Phi_{1,C}^0+\Phi_{1,C}^1\right)  \left(\Phi_{1,C}^{*0}+\Phi_{1,C}^{*1}\right)\right.\\
\nonumber
&&+\left. \left(\Phi_{2,C}^0-\Phi_{2,C}^1\right)  \left(\Phi_{2,C}^{*0}+\Phi_{2,C}^{*1}\right)\right.\\
&&+\left. \left(-\Phi_{3,C}^0+\Phi_{3,C}^1\right) \left(\Phi_{3,C}^{*0}+\Phi_{C',C}^{*1}\right)\right]\,,
\end{eqnarray}
\begin{eqnarray}
\nonumber
\mbox{tr}_k\left[{\cal W}_2^{(0)}{\cal W}_4^{\dagger(0)}\right]&=&-iu^2\sum_{C=1}^\ell \left[ \left(\Phi_{1,C}^0+\Phi_{1,C}^1\right)  \left(-\Phi_{1,C}^{*0}+\Phi_{1,C}^{*1}\right)\right.\\
\nonumber
&&+\left. \left(\Phi_{2,C}^0+\Phi_{2,C}^1\right)  \left(\Phi_{2,C}^{*0}-\Phi_{2,C}^{*1}\right)\right.\\
&&+\left. \left(\Phi_{3,C}^0+\Phi_{3,C}^1\right) \left(-\Phi_{3,C}^{*0}+\Phi_{3,C}^{*1}\right)\right]\,.
\end{eqnarray}

It is not hard, using (\ref{the formalized way to the BCS of Phi}), to check that the combinations \begin{eqnarray}\nonumber&&|\Phi_{C',C}^{0}(x)|^2, |\Phi_{C',C}^{1}(x)|^2,  \left(-\Phi_{1,C}^0+\Phi_{1,C}^1\right) (x) \left(\Phi_{1,C}^{*0}+\Phi_{1,C}^{*1}\right)(x),\\ \nonumber&& \left(\Phi_{2,C}^0-\Phi_{2,C}^1\right)(x)  \left(\Phi_{2,C}^{*0}+\Phi_{2,C}^{*1}\right)(x),   \left(-\Phi_{3,C}^0+\Phi_{3,C}^1\right)(x) \left(\Phi_{3,C}^{*0}+\Phi_{C',C}^{*1}\right)(x)\\\end{eqnarray} satisfy periodic boundary conditions.\footnote{However, the component that carries the subscript $C',C$ is sent to one with subscript $C'-r,C+1$. Nevertheless, the combinations that give the gauge invariant density are periodic. Also, from the linearity of the Fourier analysis of the Fourier-transformed components below, the superposition of the various terms makes sense. The difficulty in the analysis below is that numerical convergence is hard to achieve.}  Then, we use the Fourier transform of these combinations, namely,
\begin{eqnarray}
\nonumber
|\Phi_{C',C}^{0}(x)|^2&=&\sum_{p_\mu\in \mathbb Z} e^{i\frac{2\pi p_\mu x)\mu}{L_\mu}} {\cal X}_{0;C',C}(p)\,,\\
\nonumber
|\Phi_{C',C}^{1}(x)|^2&=&\sum_{p_\mu} e^{i\frac{2\pi p_\mu x_\mu}{L_\mu}} {\cal X}_{1;C',C}(p)\,,\\
\nonumber
\Phi_{C',C}^{0}(x)\Phi_{C',C}^{*1}(x)&=&\sum_{p_\mu} e^{i\frac{2\pi p_\mu x_\mu}{L_\mu}} {\cal X}_{2;C',C}(p)\,,\\
\nonumber
 \left(-\Phi_{1,C}^0+\Phi_{1,C}^1\right) (x) \left(\Phi_{1,C}^{*0}+\Phi_{1,C}^{*1}\right)(x)&=&\sum_{p_\mu} e^{i\frac{2\pi p_\mu x_\mu}{L_\mu}} {\cal X}_{3;C}(p)\,,\\
 \nonumber
  \left(\Phi_{2,C}^0-\Phi_{2,C}^1\right)(x)  \left(\Phi_{2,C}^{*0}+\Phi_{2,C}^{*1}\right)(x)&=&\sum_{p_\mu} e^{i\frac{2\pi p_\mu x_\mu}{L_\mu}} {\cal X}_{4;C}(p)\,,\\
   \left(-\Phi_{3,C}^0+\Phi_{3,C}^1\right)(x) \left(\Phi_{3,C}^{*0}+\Phi_{3,C}^{*1}\right)(x)&=&\sum_{p_\mu} e^{i\frac{2\pi p_\mu x_\mu}{L_\mu}} {\cal X}_{5;C}(p)\,.
\end{eqnarray}
to find
\begin{eqnarray}
\nonumber
\mbox{tr}_k\left[{\cal W}_4^{(0)}{\cal W}_4^{\dagger(0)}\right]&=&u^2\sum_{p_\mu}e^{i\frac{2\pi p_\mu x_\mu}{L_\mu}}{\cal H}(p)\\
\nonumber
&&\equiv u^2\sum_{p_\mu,C,C'}e^{i\frac{2\pi p_\mu x_\mu}{L_\mu}}\left({\cal X}_{0;C',C}(p)+{\cal X}_{1;C',C}(p)-{\cal X}_{2;C',C}(p)-{\cal X}_{2;C',C}^*(p)\right)\,.\\
\end{eqnarray}
The function ${\cal H}(p)$, the Fourier transform of $\mbox{tr}_k\left[{\cal W}_4^{(0)}{\cal W}_4^{\dagger(0)}\right]$ modulo $u^2$, will play an important role below. In addition, we find
\begin{eqnarray}
\nonumber
&&\left(\pi \ell k \Box\right)^{-1}\left\{\left(\partial_3^2+\partial_4^2\right)\mbox{tr}_k\left[{\cal W}_2^{(0)}{\cal W}_2^{\dagger(0)}-{\cal W}_4^{(0)}{\cal W}_4^{\dagger(0)}\right]\right\}\\
\nonumber
&&=\frac{2u^2}{ \pi\ell k}\sum_{p_\mu,C,C'}\frac{\left[\frac{p_3^2}{L_3^2} +\frac{p_4^2}{L_4^2} \right]e^{i\frac{2\pi p_\mu x_\mu}{L_\mu}}}{\frac{p_1^2}{L_1^2}+\frac{p_2^2}{L_2^2}+\frac{p_3^2}{L_3^2}+\frac{p_4^2}{L_4^2}}\left({\cal X}_{2;C',C}(p)+{\cal X}_{2;C',C}^*(p)\right)\,,\\
\end{eqnarray}
and
\begin{eqnarray}
\nonumber
&&\left(\pi \ell k \Box\right)^{-1}\left\{\left(-\partial_1\partial_3-\partial_2\partial_4+i\partial_2\partial_3-i\partial_1\partial_4\right)\mbox{tr}_k\left[{\cal W}_4^{(0)}{\cal W}_2^{\dagger(0)}\right]\right.\\
\nonumber
&&\left.+\left(-\partial_1\partial_3-\partial_2\partial_4-i\partial_2\partial_3+i\partial_1\partial_4\right)\mbox{tr}_k\left[{\cal W}_2^{(0)}{\cal W}_4^{\dagger(0)}\right]\right\}\\
\nonumber
&=&\frac{u^2}{\pi \ell k} \sum_{p_\mu,C}\frac{e^{i\frac{2\pi p_\mu x_\mu}{L_\mu}}}{\frac{p_1^2}{L_1^2}+\frac{p_2^2}{L_2^2}+\frac{p_3^2}{L_3^2}+\frac{p_4^2}{L_4^2}}\left\{\left(-i\frac{p_1p_3}{L_1L_3}-i\frac {p_2p_4}{L_2L_4}-\frac{p_2p_3}{L_2L_3}+\frac{p_1p_4}{L_1L_4}\right)\right.\\
\nonumber
&&\left.\times({\cal X}_{3;C}(p)+{\cal X}_{4;C}(p)+{\cal X}_{5;C}(p))\right.\\
\nonumber
&&\left. +\left(i\frac{p_1p_3}{L_1L_3}+i\frac {p_2p_4}{L_2L_4}-\frac{p_2p_3}{L_2L_3}+\frac{p_1p_4}{L_1L_4}\right)\right.\\
&&\left.\times({\cal X}_{3;C}^*(p)+{\cal X}_{4;C}^*(p)+{\cal X}_{5;C}^*(p))\right\}\,.
\end{eqnarray}

Finally, one can also define  the Fourier components of $\mbox{tr}[F_{34}F_{34}](x)$:
\begin{eqnarray}
\mbox{tr}[F_{34}F_{34}](x)=\sum_{p_\mu\in \mathbb Z}e^{i \frac{2\pi p_\mu x_\mu }{L_\mu}}{\cal Q}(p)\,.
\end{eqnarray}
Using (\ref{field strength F34F34}), we find, apart from an additive constant:
\begin{eqnarray}
\nonumber
\frac{{\cal Q}(p)}{u^2\hat F_{34}^\omega \Delta}&=&8\pi N {\cal H}(p)+\frac{4}{ \pi\ell k}\sum_{C,C'}\frac{\left[\frac{p_3^2}{L_3^2} +\frac{p_4^2}{L_4^2} \right]}{\frac{p_1^2}{L_1^2}+\frac{p_2^2}{L_2^2}+\frac{p_3^2}{L_3^2}+\frac{p_4^2}{L_4^2}}\left({\cal X}_{2;C',C}(p)+{\cal X}_{2;C',C}^*(p)\right) \\
\nonumber
&&+\frac{2}{\pi \ell k} \sum_{C}\frac{1}{\frac{p_1^2}{L_1^2}+\frac{p_2^2}{L_2^2}+\frac{p_3^2}{L_3^2}+\frac{p_4^2}{L_4^2}}\left\{\left(-i\frac{p_1p_3}{L_1L_3}-i\frac {p_2p_4}{L_2L_4}-\frac{p_2p_3}{L_2L_3}+\frac{p_1p_4}{L_1L_4}\right)\right.\\
\nonumber
&&\left.\times({\cal X}_{3;C}(p)+{\cal X}_{4;C}(p)+{\cal X}_{5;C}(p))\right.\\
\nonumber
&&\left. +\left(i\frac{p_1p_3}{L_1L_3}+i\frac {p_2p_4}{L_2L_4}-\frac{p_2p_3}{L_2L_3}+\frac{p_1p_4}{L_1L_4}\right)\right.\\
&&\left.\times({\cal X}_{3;C}^*(p)+{\cal X}_{4;C}^*(p)+{\cal X}_{5;C}^*(p))\right\}\,.
\end{eqnarray}
We need to check whether the expression on the R.H.S. vanishes for all values of $p_\mu$. The easiest check to perform is to choose $p_\mu=(0, p_2,0,0)$. With this choice, all terms vanish except ${\cal H}(p)$, the Fourier transform of $\mbox{tr}_k\left[{\cal W}_4^{(0)}{\cal W}_4^{\dagger(0)}\right]$ modulo $u^2$. One can check numerically that ${\cal H}(p)$ is non-vanishing, indicating that the gauge-invariant density $\mbox{tr}[F_{34}F_{34}](x)$ increases indefinitely as $u\rightarrow \infty$.

\section{Fermion zero modes on the deformed-$\mathbb T^4$, for  $k=r$  }
\label{appx:fermiondelta}

In this Appendix, we solve for the fermion zero modes in the background (\ref{gauge field in general}), which we rewrite in the familiar $k$/$\ell$ block matrix form, using the notation of (\ref{allowedholonomies}): 
\begin{eqnarray} \label{backgroundkr}
A_\mu = \left(\begin{array}{cc}  || \left(2 \pi \ell \; (A_\mu^\omega   - {z_\mu \over L_\mu}) + \phi_\mu^{C'}\right) \delta_{C'B'} + \epsilon^2 \; {\cal S}_{\mu \; C'B'}||& || \epsilon \;{\cal W}_{\mu \; C' B}||  \cr  ||  \epsilon\; ({\cal W}_\mu^\dagger)_{C B'} ||& ||- 2 \pi k\;  (A_\mu^\omega - {z_\mu \over L_\mu}) \delta_{CB}+  \epsilon^2 \;{\cal S}_{\mu \; CB}||\end{array} \right)~.\nonumber \\
\end{eqnarray}
Here we consider exclusively the $k=r$ case, where:
\begin{enumerate}
\item $A_\mu^\omega$ is the constant flux background $A_1^\omega = A_3^\omega = 0$, $A_2^\omega = - {  x_1 \over N   L_1 L_2}$, $A_4^\omega = - {x^3 \over N \ell L_3 L_4}$.
 \item $\phi^{C'}_\mu$ are the $r-1$ allowed holonomies in $SU(k=r)$ (thus obeying $\sum_{C'} \phi_\mu^{C'} =0$) from (\ref{allowedholonomies}) and $z_\mu$ are the holonomies in the $U(1)$-direction $\omega$, eqn.~(\ref{omega}).
 
We also recall that these are, after computing the commutator in the Weyl equation, combined into the $r$ independent $\hat \phi_\mu^{C'}$ of eqns.~(\ref{holonomy1}, \ref{explicit holonomies}) with no constraint on the trace.
 \item ${\cal W}_\mu$ is leading order  $k=r$ solution. Thus,  $ {\cal W}_3 = {\cal W}_4 = 0$ and ${\cal W}_1 = -i {\cal W}_2$, and with ${\cal W}_2$ given by  (\ref{expressions of W2 and W4 with holonomies}), with the $r$ coefficients ${\cal C}_2^A$ fixed by solving eqn.~(\ref{requalkconstraints}).
 \item The  components of ${\cal S}_\mu$ are obtained by solving (\ref{equationforSk}, \ref{equationforSell}) (recall that they obey the tracelessness condition $\sum_{C'} S_{\mu\; C' C'} + \sum_{C} S_{\mu \; C  C}=0$).
  \item Finally, to remind us of the powers of $\sqrt\Delta$ appearing in the leading order solution for ${\cal W}_\mu$ and ${\cal S}_\mu$,  we introduced a parameter $\epsilon\; (\equiv 1)$.
\end{enumerate} 
Our goal is to solve the Weyl equation $\partial_\mu\bar \sigma^\mu\lambda+i\bar\sigma^\mu[A_\mu, \lambda]=0$ in the $k=r$ background (\ref{backgroundkr}), using a series expansion in $\epsilon$, to leading order. We take $\lambda$ also in the block-diagonal form (\ref{blockform}), obeying (\ref{tracelesslambda}):
\begin{eqnarray} \label{blockform1}
\lambda &=& \left[\begin{array}{cc}||\lambda_{C'B'}|| & ||\lambda_{C'B}||\\  ||\lambda_{CB'}||& ||\lambda_{CB}||\end{array}\right]~, ~ C',B' \in \{0,...k-1\}, ~ C,B  \in \{0,...\ell-1\}~.  
\end{eqnarray} 

Newt, write the Weyl equation, using the quaternionic notation of Section \ref{sec:deforming}: $\bar\partial = \partial_\mu \bar\sigma_\mu$ and $\bar A = \bar\sigma_\mu A_\mu$ (and similar for all other vectors in (\ref{backgroundkr}), with $\bar\sigma_\mu$ defined in Footnote \ref{footnote:notation0}) and obtain the following equations for the components of $\lambda$ of (\ref{blockform1}), with a sum over repeated indices $B,B'$ implied:
\begin{eqnarray} \label{lambdaepsiloneqns}
\bar\partial \lambda_{C'D'} &=& - i \epsilon \bar\sigma_\mu(  {\cal W}_{\mu \; C'B} \lambda_{B D'} -  \lambda_{C'B} ({\cal W}^\dagger)_{\mu \; BD'}) - i  \epsilon^2 \bar\sigma_\mu ( {\cal S}_{\mu \;  C' B'} \lambda_{B' D'} - \lambda_{C'B'} {\cal S}_{\mu \;  B'D'}), \nonumber \\
\nonumber
\bar\partial \lambda_{CD} &=& - i \epsilon  \bar\sigma_\mu (({\cal W}^\dagger)_{\mu \; CB'}   \lambda_{B'D} -  \lambda_{C B'} {\cal W}_{\mu \; B'D} )- i \epsilon^2 \bar\sigma_\mu (  {\cal S}_{\mu\; C  B } \lambda_{B D } - \lambda_{C B }  {\cal S}_{\mu \; B D}),  \\
\bar\partial \lambda_{C'D} &=& -i (2 \pi N \bar{A}^\omega + \bar{\hat\phi}^{C'}) \lambda_{C' D}  - i \epsilon \bar\sigma_\mu ({\cal{W}}_{\mu \; C'B} \lambda_{BD} - \lambda_{C'B'}  {\cal{W}}_{\mu \; B'D}) \nonumber \\
&& - i \epsilon^2 \bar\sigma_\mu
 ({\cal S}_{\mu\; C'B'} \lambda_{B'D} - \lambda_{C'B}  {\cal{S}}_{\mu \; BD}), \nonumber \\
\bar\partial \lambda_{CD'} &=&  i (2 \pi N \bar{A}^\omega + \bar{\hat\phi}^{C'}) \lambda_{C D'}  - i \epsilon\bar\sigma_\mu ( ({\cal W}^\dagger)_{\mu \; CB'}  \lambda_{B'D'} - \lambda_{CB} ( {\cal W}^\dagger)_{\mu\; BD'}  )\nonumber \\
&&- i \epsilon^2 \bar\sigma_\mu 
({\cal S}_{\mu \; C B } \lambda_{B D'} - \lambda_{C B'} {\cal{S}}_{\mu \; B'D'})\,.
\end{eqnarray}

We now observe that we can consistently solve (\ref{lambdaepsiloneqns}) in a series expansion in $\epsilon$, assigning the following (leading-order only shown) $\epsilon$-scaling of the various components of $\lambda$: \begin{eqnarray}
\label{lambdaseries}
\lambda_{C'D'} &=&  \epsilon^{0} \lambda_{C'D'} + {\cal O}(\epsilon^2)\,, \nonumber \\
\lambda_{C D } &=&  \epsilon^{0} \lambda_{CD} + {\cal O}(\epsilon^2)\,,\nonumber \\
\lambda_{C' D } &=&  \epsilon \; \lambda_{C'D } + {\cal O}(\epsilon^3)\,, \nonumber\\
\lambda_{C D' } &=&  \epsilon\;  \lambda_{C D'} + {\cal O}(\epsilon^3)\,.
 \end{eqnarray}
Substituting  into (\ref{lambdaepsiloneqns}) and keeping only the leading terms in $\epsilon$ in each equation, we find the following equations for the leading order  (in $\sqrt\Delta$) fermion zero modes in the background (\ref{backgroundkr}): 
 \begin{eqnarray} \label{lambdaepsiloneqn1}\nonumber
\bar\partial \lambda_{C'D'} &=&0, \\\nonumber
\bar\partial \lambda_{CD} &=& 0,  \\\nonumber
 ( \bar\partial + i   (2 \pi N \bar{A}^\omega + \bar{\hat\phi}^{C'})) \lambda_{C'D} &=&   - i   ( {\cal{W}}_{\mu \; C'B}\; \bar\sigma_\mu \lambda_{BD} - \bar\sigma_\mu \lambda_{C'B'}  {\cal{W}}_{\mu \; B'D}), \\
 (\bar\partial- i   (2 \pi N \bar{A}^\omega + \bar{\hat\phi}^{C'})) \lambda_{CD'} &=&  - i  (   {\cal W}^*_{\mu \; B' C}    \; \bar\sigma_\mu \lambda_{B'D'} -   \bar\sigma_\mu \lambda_{CB}  {\cal W}^*_{\mu \; D'B}  ).  
\end{eqnarray}

Now, we recall that the first two equations were already solved in Section \ref{sec:undotteddiagonal}. From 
eqn.~(\ref{undotteddiagonalzeromodes}), taken with $k=r$, we have the diagonal zero mode solutions
\begin{eqnarray} \label{fermion1}
\nonumber
\lambda_{\alpha \; B'C'} &=& \delta_{B'C'}\; \theta_\alpha^{C'},\\
\lambda_{\alpha \; BC} &=& - \delta_{BC} \;{1 \over \ell} \sum_{C'} \theta_\alpha^{C'}, 
\end{eqnarray}
where we momentarily restored the spinor index $\alpha$. We first define the spinor
 \begin{equation}
 \eta^{C'} \equiv    \theta^{C'} + {1 \over \ell} \sum\limits_{B'=0}^{k-1} \theta^{B'},
 \end{equation} and then plug (\ref{fermion1}) into the last two equations in (\ref{lambdaepsiloneqn1}) to obtain:
 \begin{eqnarray} \label{lambdaepsiloneqn2}
  ( \bar\partial + i   (2 \pi N \bar{A}^\omega + \bar{\hat\phi}^{C'})) \lambda_{   C'D} &=&     i     {\cal{W}}_{\mu \; C'D} \; \bar\sigma_\mu \; \eta^{C'} , \nonumber \\
 (\bar\partial- i   (2 \pi N \bar{A}^\omega + \bar{\hat\phi}^{C'})) \lambda_{  CD'} &=&  - i    {\cal W}^*_{\mu \; D'C} \; \bar\sigma_\mu \; \eta^{D'}.
 \end{eqnarray}
We now recall that for the $k=r$ solution, $ {\cal W}_3 = {\cal W}_4 = 0$ and ${\cal W}_1 = -i {\cal W}_2$, hence
\begin{eqnarray}
 {\cal{W}}_{\mu \; C'D} \; \bar\sigma_\mu &=& (- i \bar\sigma_1 + \bar\sigma_2) {\cal W}_{2 \; C'D} = {\cal W}_{2 \; C'D} \left(\begin{array}{cc}0 & - 2 \cr 0 &0 \end{array} \right)\,,\nonumber \\
 {\cal W}^*_{\mu \; D'C} \; \bar\sigma_\mu &=& (  i \bar\sigma_1 + \bar\sigma_2) {\cal W}_{2 \; D' C}^* ={\cal W}_{2 \; D' C}^* \left(\begin{array}{cc}0 & 0\cr 2 &0 \end{array} \right)\,,
\end{eqnarray}  
 and that ${\cal W}_{2\,C',C} = V^{-1/4}{\cal C}_{2}^{C'} \Phi^{(0)}_{C',C}(x,\hat\phi)$, where ${\cal C}_2^{C'}$ is as determined   in Section \ref{sec:compactvsnoncompact}. 
 
 The equation for $\lambda_{C'D}$ then takes the form, using the derivatives from (\ref{hat dervatives}) and noting that the equations for each $C' = 1,..., r$ decouple:
\begin{eqnarray}\label{lambdazeromode}
\nonumber
(\hat D_4   - i \hat D_3 )  \lambda_{1 \; C'D} - (i \hat D_1  + \hat D_2 )\lambda_{2 \; C'D} &=& \bar\eta_2^{C'} \; \Phi^{(0)}_{2 \; C'D}\,, \\
(- i \hat D_1  +\hat D_2 )  \lambda_{1 \; C'D} + (\hat D_4  + i \hat D_3 ) \lambda_{2 \; C'D} &=&0~,
\end{eqnarray}
where we absorb various inessential constants in the redefined $\bar\eta_2^{C'}$ coefficient. The solution of these equations is given by the function ${\cal G}_{3 \; C'D}^{(0)}$ defined in (\ref{the G3 function}), explicitly
\begin{eqnarray}\label{lambdazeromode11}\nonumber
\lambda_{1 \; C'D} &= & \bar\eta_2^{C'} {\cal G}_{3 \; C'D}^{(0)} , \\
\lambda_{2 \; C'D} &=& 0~.
\end{eqnarray}
Similarly, one finds that the other zero mode is
\begin{eqnarray}\label{lambdazeromode22}\nonumber
\lambda_{1 \; CD'} &= & 0 , \\
\lambda_{2 \; CD'} &=& \bar\eta_1^{D'}   {\cal G}_{3 \; D'C}^{* \; (0)}.
\end{eqnarray}
 Thus, there are in total $2 r$ zero modes labeled by $\bar\eta_{1,2}^{C'}$, with $C'=1,...,r$. The $x$-dependence of the zero mode labeled by a given $C'$ is governed only by the holonomies $\hat\phi^{C'}_\mu$, similar to the bosonic case discussed earlier.   
 
  \bibliography{ReferencesALL.bib}
  
  \bibliographystyle{JHEP}
  \end{document}